\documentclass[%
reprint,
superscriptaddress,
 amsmath,amssymb,
 aps,
 prx,
]{revtex4-2}

\usepackage{mathrsfs}
\usepackage{gensymb}
\usepackage{graphicx}% Include figure files
\usepackage{dcolumn}% Align table columns on decimal point
\usepackage{bm}% bold math
\usepackage{lineno}
\usepackage{siunitx}
\usepackage{color}
\newcommand{\yesun}[1]{#1}

\usepackage{placeins}
\begin{document}
% BEGIN MAIN BODY (only documented layout changes)

\preprint{APS/123-QED}

\title{Scaling Quantum Networks via Phase-Stable Vacuum Beam Guide:\\ Architectural Blueprint and Benchmark}% Force line breaks with \\

\author{Yuexun Huang}
\email{yesunhuang@uchicago.edu}
\affiliation{Pritzker School of Molecular Engineering, The University of Chicago, Chicago, IL 60637, USA}
\author{Delaney Smith}
\affiliation{Pritzker School of Molecular Engineering, The University of Chicago, Chicago, IL 60637, USA}
\author{Pei Zeng}
\affiliation{Pritzker School of Molecular Engineering, The University of Chicago, Chicago, IL 60637, USA}
\author{Debayan Bandyopadhyay}
\affiliation{Pritzker School of Molecular Engineering, The University of Chicago, Chicago, IL 60637, USA}
\author{Junyu Liu}
\affiliation{Department of Computer Science, The University of Pittsburgh, Pittsburgh, PA 15260, USA}
\author{Rana X Adhikari}%
\email{rana@caltech.edu}
\affiliation{Division of Physics, Math, and Astronomy, LIGO Laboratory, California Institute of Technology, Pasadena, CA 91125 USA}
\author{Liang Jiang}
\email{liangjiang@uchicago.edu}
\affiliation{Pritzker School of Molecular Engineering, The University of Chicago, Chicago, IL 60637, USA}

\date{\today}% It is always \today, today,

\begin{abstract}
Scaling quantum networks to continental distances requires physical infrastructure capable of overcoming both exponential attenuation and severe phase decoherence. While the concept of the vacuum beam guide (VBG) has recently emerged as a promising low-loss solution, we move beyond it by proposing a \yesun{stringent and conservative} physical-layer architectural blueprint anchored by empirical data and methods from Advanced LIGO, particularly regarding its interferometric stability in dynamic environments. To evaluate its system-level viability, we benchmark this architecture against various protocols across quantum communication, metrology, and computation. Ultimately, 
we identify no fundamental technical blockers to scaling this infrastructure.
% \begin{description}
% Secondary publications and information retrieval purposes.
% You may use the \texttt{description} environment to structure your abstract;
% use the optional argument of the \verb+\item+ command to give the category of each item. 
% \end{description}
\end{abstract}

\maketitle

%\tableofcontents

\section{Introduction}
\label{sec:into}
Creating a low-loss quantum channel for global networks is challenging due to optical absorption losses that cause an exponential drop in direct quantum communication rates with distance, thereby restricting the function of fiber to tens of kilometers \cite{petrovich2025first,numkam2023loss,sakr2021hollow}. Tremendous progress has been made in extending quantum links beyond the exponential loss limits. In particular, satellite-to-ground links have pioneered landmark intercontinental entanglement distribution \cite{lu2022micius,yin2017satellite}, and emerging aerial platforms, such as highly efficient stratospheric balloon relays \cite{liu2026global}, are providing robust solutions for flexible, long-range spatial routing. Despite these remarkable advances in network coverage, unguided open-space channels remain fundamentally bounded by atmospheric coupling limits \cite{aspelmeyer2003long}. Meanwhile, the quantum repeaters currently available still lack full error correction ability, and their use results in a polynomial reduction in communication rates across long distances~\cite{muralidharan2016optimal}. \yesun{Together, these various approaches highlight that the realization of future large-scale quantum networks remains an open problem, with no single technique established as an ultimate answer for all distances and applications \cite{nstc2021quantumnetworking,illiano2022quantum}. Current architecture studies and roadmaps instead explicitly anticipate heterogeneous and hierarchical quantum  networks in which complementary technologies can serve different geographic scales and operational requirements \cite{kozlowski2023architectural,quic2025roadmap,itu2025sup98}.} \yesun {The integration of terrestrial fiber and satellite-to-ground links has already been demonstrated in quantum key distribution networks \cite{chen2021integrated}.} \yesun{However, a critical capability that is still missing from such heterogeneous architectures is a deterministic, continuous-operation, high-capacity backbone for continental-scale quantum transmission.} Crucially, such an infrastructure must simultaneously deliver ultra-high transmission capacities and guarantee the rigorous interferometric phase stability required to preserve delicate quantum coherence over continental distances. \yesun{Therefore, within this broader architectural context, we envision the vacuum beam guide (VBG) as a high-performance, long-distance backbone channel connecting regional networks, with optical fiber and free-space links providing local access \cite{huang2024vacuum}.} \yesun{This backbone role is also compatible with repeater-assisted architectures where} \yesun{theoretical studies indicate that the reduced attenuation of VBG links can significantly increase repeater spacing and lower intermediate quantum resource requirements \cite{gan2025quantum}.}

\yesun{Foundational models treated the VBG as a pure-loss bosonic channel \cite{holevo2001evaluating,takeoka2014fundamental}, demonstrating its theoretical potential to exceed tera-qubit-per-second capacities with appropriate phase-insensitive dual-rail encoding \cite{huang2024vacuum,zheng2024performance}.} While this prior work established the in-principle feasibility of an ultra-low-loss channel by extrapolating current technologies, it remained an idealized proof-of-concept. A truly universal physical layer, however, cannot rely solely on phase-insensitive protocols. In domains such as distributed quantum metrology, information is intrinsically encoded by nature into the phase of the optical field, making the channel's inherent interferometric stability the critical prerequisite to prevent catastrophic dephasing~\cite{zhang2025criteria,bartlett2007reference}. Consequently, translating the VBG from a theoretical vision into a scalable infrastructure necessitates advancing beyond fundamental loss limits to establish a concrete prototype design. \yesun{For the first time within the VBG framework, we take this step by extending its role beyond an encoding-restricted, ultra-low-loss transmission channel toward a phase-coherent, general-purpose quantum optical channel, which establishes a substantially broader physical layer that is capable of supporting a wide class of protocols, as long as the information is intrinsically carried by photons.}

Resolving the complex phase dynamics of a continent-spanning interferometer allows for the relaxation of strict constraints on encoding schemes, ultimately laying the groundwork for benchmarking the VBG's system-level viability. To this end, \yesun{we advance a new cross-field methodology that bridges gravitational wave interferometry and quantum network physics, translating the mature noise budgeting, sensing, and feedback control framework of Advanced LIGO \cite{aasi2015advanced} into continental-scale quantum optical channel design.} Crucially, rather than relying solely on idealized theoretical abstractions, we ground this cross-field framework in diverse empirical measurements from gravitational wave observatories. \yesun{We systematically evaluate the VBG’s interferometric stability through a comprehensive phase noise budget, characterized by its phase noise power spectral density (PSD), $S_\phi(f)$ \cite{riley2008handbook,robins1984phase}.} The operational success of these observatories \cite{martynov2016sensitivity,buikema2020sensitivity,capote2025advanced} provides a robust, data-driven foundation for our analysis. Furthermore, because maintaining single-photon interferometric visibility is a significantly relaxed constraint compared to LIGO’s extreme strain sensitivity, our model concludes that the VBG yields notable advantages beyond 10 Hz over the fundamental limit set by the fiber thermal noise floor. Most importantly, we show that the VBG is capable of accomplishing this unprecedented stability utilizing a considerably simpler and more scalable control system than those required for gravitational wave observatories (detailed in Supplementary Section \ref{sec:VBGControlSys}).

Beyond interferometric phase stability, evaluating the VBG as a universal physical layer requires the rigorous quantification of complementary optical properties. To provide a complete channel characterization, we also analyze its inherent polarization maintenance, transmission latency, and pulse duration limits. Equipped with this holistic physical model, we subsequently \yesun{demonstrate} the system-level viability of the VBG through quantitative performance evaluation. \yesun{We directly connect the physical channel model of the VBG to the performance limits of representative quantum applications across communication, sensing, and computation: Device Independent Quantum Key Distribution (DI-QKD) \cite{ekert1991quantum,sekatski2021device,xu2022device} as a primarily loss-limited case, Quantum Telescope (Q-Telescope) \cite{khabiboulline2019quantum,khabiboulline2019optical,gottesman2012longer} to additionally probe interferometric phase coherence, and Blind Quantum Computation (BQC) \cite{fitzsimons2017private,morimae2013blind} to further evaluate the impact of interactive communication latency.} \yesun{Additional analyses in the Supplementary Information extend this connection to a broader set of protocol--metric combinations. Together, these analyses establish a practically grounded framework for benchmarking quantum optical channels by linking physical channel properties to application-level performance limits, with the VBG serving as a concrete case study.} Our numerical results demonstrate concrete scaling advantages over terrestrial fiber networks in both transmission rate and coherence preservation. Ultimately, this quantitative analysis \yesun{indicates} that the proposed architectural blueprint for the VBG \yesun{can meet} the demanding physical-layer requirements for next-generation quantum infrastructure.

\section{Blueprint of the VBG}
\label{sec:VBGbasis}
\begin{figure*}[htbp]
\centering\includegraphics[width=\linewidth]{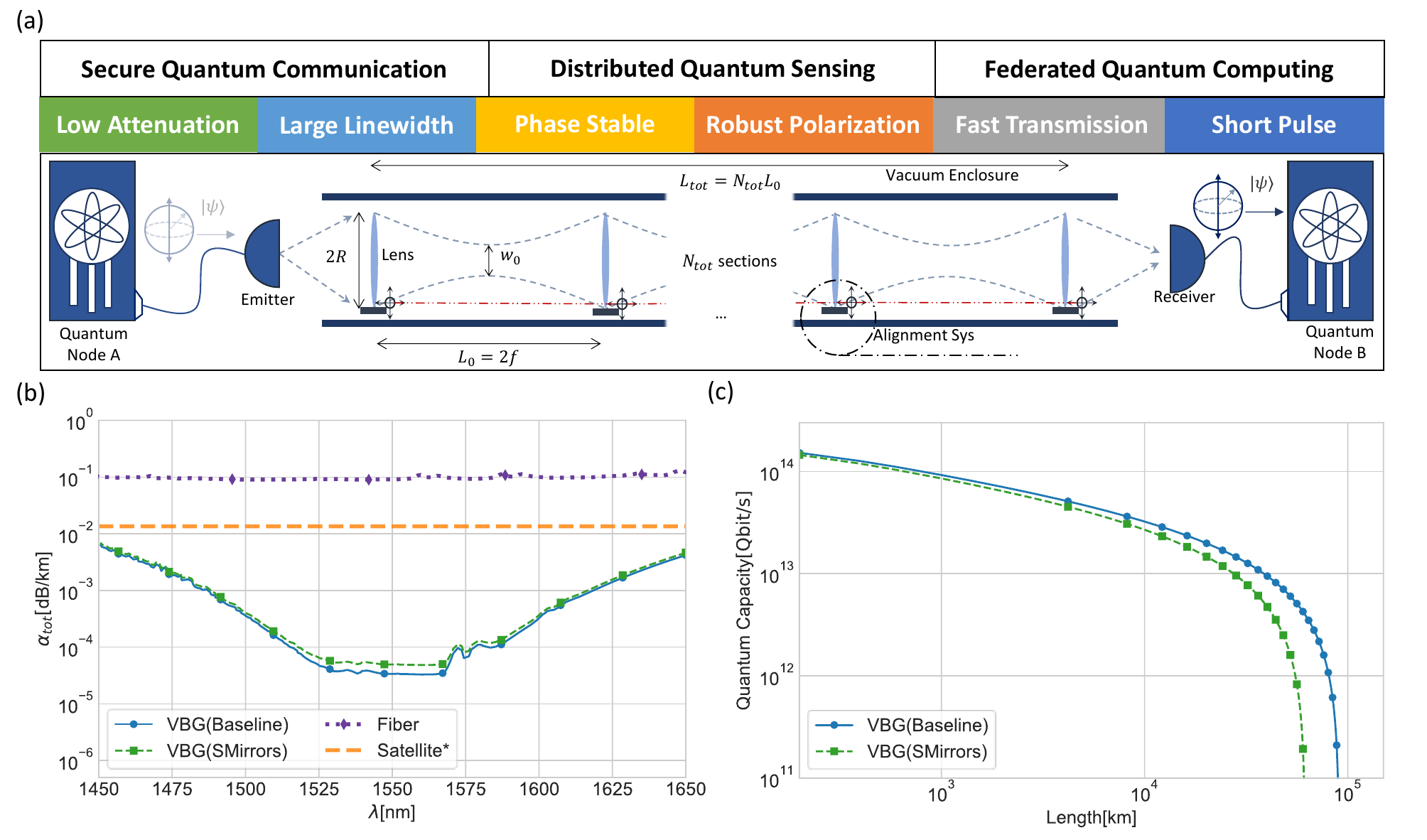}
\caption{Architectural blueprint and baseline capacity limits of the Vacuum Beam Guide (VBG).
(a) Schematic diagram of the VBG along with a representative collection of capabilities and applications. 
(b) The attenuation factor of the VBG compared to the SOTA fiber (Experimental data presented here while theoretical limit is greater than 0.01\,dB\yesun{/km} assuming practical constraints) \cite{petrovich2025first,numkam2023loss} and the satellite-to-ground link (*approximate level) \cite{lu2022micius,yin2017satellite}. 
(c) The quantum capacity of VBG under one-way protocols (Two-way protocols show similar results). 
%\rana{suggest increasing font size for the tick labels for both plots - exponents are hard to read}
(Note: Theoretical evaluations assume a relative alignment accuracy of 0.1\% for transverse displacement and 1\% for angular tilt with respect to the optical axis).
}
\label{fig:VBG_Basics}
\end{figure*}

As illustrated in Fig.~\ref{fig:VBG_Basics}(a), the physical architecture of the VBG consists of an array of precisely aligned lenses and mirrors enclosed within a continuous vacuum tube, designed to guide photon-encoded quantum information over continental baselines. The channel is composed of $N_{\mathrm{tot}}$ periodic sections of length $L_0$, utilizing lenses with a diameter of $2R$ to sustain the photonic quantum signal over a total transmission distance of $L_{\mathrm{tot}}=N_{\mathrm{tot}}L_0$. To navigate macroscopic geographic constraints and optimize construction costs, flat steering mirrors are integrated to provide both static path deflections and active alignment steering. Furthermore, to suppress atmospheric absorption and scattering, the entire propagation path operates under a moderate vacuum environment ($\lesssim$\,1\,Pa at room temperature). The parameters of this baseline blueprint are detailed in Table \ref{tab:VBGDefault} \cite{huang2024vacuum}. This configuration serves as the quantitative foundation for our subsequent physical characterization and system-level protocol benchmarks, the comprehensive results of which are summarized in Table~\ref{tab:applicationSummary}.

\begin{table}[htbp]
\caption{{Baseline Parameters of the VBG}}
  \label{tab:VBGDefault}
  \centering
\begin{tabular}{ccc}

\hline
Parameters & Symbol & Values \\
\hline
Section Length & $L_0$ & 4\,km \\
Focus Length & $f$ & 2\,km  \\
Lens Radius & $R$& 12\,cm \\
Center Wavelength & $\lambda_c$ & 1550\,nm\\
Spot Size & $w$ &  4\,cm\\
Operating Pressure & $P$ & 1\,Pa\\
Operating Temperature & $T$ & $\sim$ 300\,K\\
\hline

\end{tabular}
\end{table}

\subsection{Baseline Attenuation and Capacity Limits}
\label{subsec:Attenuation}
Because the photons propagate primarily through a vacuum, the VBG functions as an ultra-transparent channel with extremely low loss ( $\sim10^{-5}$ dB/km). 
Unlike satellite-to-ground links, a deployed VBG provides a continuous, ground-based infrastructure that inherently bypasses the $\yesun{\sim}3$ dB atmospheric loss \yesun{reported by the Micius satellite-to-ground link }\cite{lu2022micius}. The total effective attenuation of the VBG, $\alpha_{\mathrm{tot}}(\lambda)$, in the normalized unit of $\mathrm{dB/km}$, can be divided into the following four main contributions,
\begin{equation}
    \alpha_{\mathrm{tot}}(\lambda)=\alpha_{\mathrm{{lens}}}+\alpha_{\mathrm{gas}}+\alpha_{\mathrm{align}}+\frac{L_0}{L_{\mathrm{mirror}}}\alpha_{\mathrm{mirror}},
\end{equation}
where $\alpha_{\mathrm{lens}}$ denotes optical losses from absorption, scattering, diffraction, and coatings; $\alpha_{\mathrm{gas}}$ is the residual gas absorption; and $\alpha_{\mathrm{align}}$ accounts for energy leakage into higher-order transverse modes due to imperfect alignment. As established in our foundational model \cite{huang2024vacuum}, these first three terms constitute the baseline loss of the VBG, \yesun{which is approximately $4\times10^{-5}\,\mathrm{dB/km}$}. However, adapting this blueprint for continental routing necessitates steering mirrors; the angular alignment of these mirrors introduces the final term, $\alpha_{\mathrm{mirror}}$ (analyzed in detail in Supplementary Section \ref{subsec:reflectionDevice}, with $L_{\mathrm{mirror}}$ denoting the mirror spacing). As plotted in Fig.~\ref{fig:VBG_Basics}(b), numerical evaluations at the 1550 nm telecom band reveal that these active steering mirrors add only a marginal penalty ($\Delta\bar{\alpha} \approx 10^{-5}$ dB/km). Consequently, the total attenuation remains bounded at $\bar{\alpha}_{\mathrm{VBG(SMirrors)}}\approx 5 \times 10^{-5}$ dB/km, more than two orders of magnitude below that of state-of-the-art fiber and satellite links.

Beyond the center carrier wavelength, the VBG maintains this ultra-low attenuation across a broad transmission window of $\Delta\lambda \approx 50$ nm (or $\Delta\nu \approx 6$ THz). This expansive linewidth naturally supports massive time-, frequency-, and hybrid-division multiplexing. The ultimate transmission rate, $Q_k$, is bounded by integrating the pure-loss capacity per channel use, $q_k(\lambda) = \max\left[\log_2\left(\frac{\eta^{2-k}(\lambda)}{1-\eta(\lambda)}\right), 0\right]$ \cite{noh2018quantum,holevo2001evaluating,pirandola2017fundamental}, across the available frequency band~\cite{wang2022quantum},
\begin{equation}
Q_k = \int q_k(\lambda) \frac{c}{\lambda^2} d\lambda
\end{equation}
where $\eta(\lambda) = 10^{-0.1L_{\mathrm{tot}}\alpha_{\mathrm{tot}}(\lambda)}$ is the transmission efficiency and $k \in \{1,2\}$ denotes one- or two-way communication (the former being constructively realizable with appropriate encoding \cite{zheng2025performance,subramanian2025achievable}). As shown in Fig.~\ref{fig:VBG_Basics}(c), this leads to a remarkable tera-qubit-per-second channel capacity over continental scale ($10^4$\,km), quantitatively establishing the VBG's theoretical superiority over existing infrastructures. Crucially, however, while these baseline metrics establish an unprecedented throughput limit, deploying the VBG as a universal physical layer capable of supporting phase-sensitive protocols hinges on overcoming interferometric phase decoherence. This physical hurdle dictates the core focus of our subsequent phase noise budget.

\subsection{Phase Decoherence and Empirical Noise Budget}
\label{subsec:PhaseNoisePSD}

\begin{figure*}[htbp]
\centering\includegraphics[width=\linewidth]{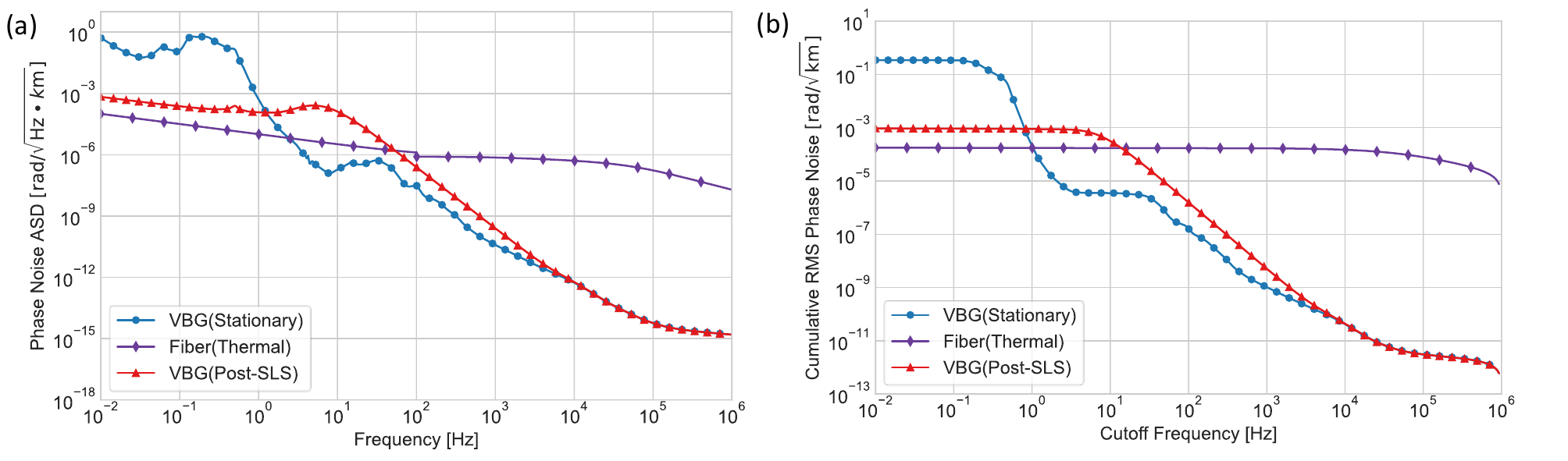}
\caption{Interferometric phase stability and cumulative noise bounds of the VBG channel in a relatively quiet, rural area.  
(a) Normalized phase noise ASD of the VBG before and after the active Section Length Stabilization (SLS) system in comparison with the fiber thermal noise floor \cite{bartolo2012thermal,wanser1992fundamental,duan2010intrinsic}. 
(b) Cumulative phase noise integrated from cutoff frequency $0.01$ Hz to 1 MHz. Comprehensive noise budgeting details are provided in Supplementary Section \ref{sec: NoiseBudgets} and Supplementary Figs. \ref{fig:Phase_PSD_APP}-\ref{fig:MTN_PSD}.
}
\label{fig:Phase_PSD}
\end{figure*}

While phase noise is conventionally regarded as secondary to attenuation due to the availability of straightforward coding mitigations like time-bin or polarization encoding, it constitutes a severe constraint for a universal global-scale quantum network.
Input/output channels linked to the VBG might implement advanced loss-counteracting coding methods, such as GKP codes \cite{gottesman2001encoding,grimsmo2021quantum,zheng2024performance,subramanian2025achievable}, 
some of which are less robust against phase noise. 
Furthermore, numerous QKD protocols utilize single-rail encoding to maximize communication distance and security \cite{lucamarini2018overcoming,sandfuchs2025security,wang2022twin,mao2021recent}, where fluctuations in the optical path length scramble the relative phase between the vacuum and single-photon components and cause direct qubit dephasing.
Most critically, the encoding in distributed quantum sensing schemes is intrinsic to the sensing entity and is deeply sensitive to phase variations \cite{nehra2024all,zang2024quantum}. 
Because converting between coding schemes would severely compromise these quantum advantages and restrict available parameter ranges, rigorously analyzing and mitigating all intrinsic phase noise sources within the VBG channel is strictly essential to maintaining its universal versatility.

As noise associated with quantum sources and detection methodologies is strictly application-dependent, our analysis isolates and evaluates the intrinsic channel-added phase noise of the VBG exclusively. Furthermore, different quantum protocols exhibit distinct sensitivities across the frequency spectrum, which means that relying solely on an aggregate root-mean-square (RMS) metric is often insufficient. Instead, we comprehensively characterize this intrinsic noise through its phase noise power spectral density (PSD), $S_\phi(f)$. This continuous spectral distribution provides the complete physical foundation necessary to calculate band-limited noise, total RMS fluctuations, or Allan deviations as dictated by specific protocol requirements. Assuming \yesun{most of the} fluctuations across individual VBG sections are random and uncorrelated, we normalize the amplitude spectral density (ASD) with the total propagation distance, yielding $\sqrt{\hat{S}_\phi}$ in units of $(\mathrm{rad}/\sqrt{\mathrm{Hz}\cdot \mathrm{km}})$. \yesun{For global correlated contributions, especially AUX-laser frequency noise, we punish it with a pessimistic coherent enhancement factor $N_{\rm coh}=10^4$ before length normalization.} To establish realistic bounds for these metrics, our evaluation incorporates a scalable prototype VBG control architecture (detailed in Supplementary Section \ref{sec:VBGControlSys} and illustrated in Supplementary Figs. \ref{fig:Reflectors}-\ref{fig:passiveIso}) comprising a Passive Isolation Platform (PIP), an Alignment Control System (ACS), and an active Section Length Stabilization (SLS) system.

The overall phase noise amplitude spectral density of the VBG under the baseline configuration is presented in Fig.~\ref{fig:Phase_PSD}(a), benchmarked against the fundamental thermal noise floor of standard optical fibers. Notably, while this theoretical limit assumes an isolated fiber spool in a pristine laboratory setting, real-world deployed fibers suffer from environmental perturbations that elevate their phase noise by several orders of magnitude \cite{bartolo2012thermal,hoghooghi2024enabling,predehl2012920,schioppo2022comparing, droste2013optical,grosche2009optical}. As detailed in the comprehensive noise budget (see Supplementary Section \ref{sec: NoiseBudgets}),
the low-frequency ASD is dominated by seismic noise filtered by the PIP, scaling approximately as $1/f^3$ in the 1 to 100 Hz band and $1/f^2$ beyond. Mid-frequency spectral features around 0.1 kHz and 10 kHz emerge from mount thermal noise and residual gas scattering, respectively, before flattening at high frequencies due to mirror thermal noise. Low-frequency drift below $10$ Hz can be mitigated by the active SLS system with the heterodyne round-trip phase noise cancellation 
technique~\cite{foreman2007remote}, at the cost of degraded high-frequency performance due to the limitations of sensing noise and \yesun{finite feedback bandwidth}. 
The cumulative noise (integrated up to $10^6$ Hz) under various lower cutoff frequencies $f_l$ is also plotted in Fig.~\ref{fig:Phase_PSD}(b). 
The integrated RMS phase noise per square-root kilometer of the VBG, before and after active subtraction, is at the level of 0.3\,rad and $1$\,mrad, respectively. Consequently, the \yesun{modeled} VBG outperforms the \yesun{typical thermal-noise limit} of optical fiber whenever the lower cutoff frequency satisfies $f_l > 10$\,Hz. While the VBG phase stability could theoretically be further enhanced by outfitting all optical components with full-scale LIGO-like active vibration isolation platforms \cite{shoemaker2009advanced,evans2009advanced,abbott2002seismic,matichard2015seismic}, such an approach may introduce prohibitive infrastructure overhead.  The proposed hybrid control architecture \yesun{provides a practical balance between stringent quantum-coherence preservation} and realistic macro-scale scalability.

\subsection{Complementary Optical Metrics}
\label{subsec:PDTD}

\begin{table*}[htbp]
    \caption{Summary of VBG's Performance Metrics and Application Requirements$^*$}
    \label{tab:applicationSummary}
    \centering
    \small
    \setlength{\tabcolsep}{3.8pt}
    \renewcommand{\arraystretch}{1.5} 

    % SOLUTION: We change 'p{0.28\textwidth}' to 'l' (letter L).
    % This removes the "p" column that crashes RevTeX.
    % We will apply the width manually inside the cells using \parbox.
    \begin{tabular}{ l c c c c c c }
        \hline\hline
        % HEADER ROW
        \textbf{Metrics} & 
        \parbox[t]{0.10\textwidth}{\centering Attenuation\\(dB/km)} & 
        \parbox[t]{0.09\textwidth}{\centering Linewidth\\(THz)} & 
        \parbox[t]{0.11\textwidth}{\centering Phase\\(rad/$\sqrt{\text{km}}$)} & 
        \parbox[t]{0.092\textwidth}{\centering Polar.\\(Prob./km)} &
        \parbox[t]{0.09\textwidth}{\centering Trans.\\Speed (c)} & 
        \parbox[t]{0.09\textwidth}{\centering Pulse\\Dur. (fs)}\\
        \hline
        
        VBG Data & $\sim\yesun{5}\times10^{-5}$ & $\sim 6$  & $\sim10^{-3}$ & $\yesun{\lesssim}1\,\mathrm{ppm}$ & $\sim1-\yesun{3}\times10^{-6}$ &$\geq300^{\yesun{\dagger}}$\\
        \hline
        
        % SECTION: Secure Quantum Communication
        \multicolumn{7}{l}{\textbf{Secure Quantum Communication}}\\
        \hline
        % MANUAL WRAPPING: We use \parbox inside the 'l' column to force the width safely
        \parbox[t]{0.28\textwidth}{DI-QKD \cite{ekert1991quantum,acin2007device,masanes2011secure,vazirani2019fully,arnon2018practical,sekatski2021device,zhang2022device,xu2022device,zapatero2023advances}
        (Sec.~\ref{subsec:DIQKD})}
        & $\checkmark$ & $\checkmark$  & $-$  & $\bigcirc$ & $-$  &$-$  \\
        
        \parbox[t]{0.28\textwidth}{QPV \cite{buhrman2014position,bluhm2022single,allerstorfer2023making}
        (SI.~\ref{subsec:QPV})}
        & $\checkmark$ & $\checkmark$  & $-$  & $\bigcirc$ & $\checkmark$ &$-$  \\
        
        \parbox[t]{0.28\textwidth}{R-QKD \cite{kravtsov2018relativistic,sandfuchs2025security}
        (SI.~\ref{subsec:GQC})}
        & $\checkmark$ & $\checkmark$  & $\checkmark$ &  $-$ & $\checkmark$ &$-$  \\
        
        \parbox[t]{0.28\textwidth}{QPQ \cite{giovannetti2008quantump}
        (SI.~\ref{subsec:GQC})}
        & $\checkmark$ & $\checkmark$  & $-$  & $-$  &$-$   & $-$ \\
        
        \parbox[t]{0.28\textwidth}{MPSD \cite{fischer2021distributing,meignant2019distributing,fan2025optimized,huang2025peer,fan2025distribution}
        (SI.~\ref{subsec:GQC})}
        & $\checkmark$ & $\checkmark$  & $-$  & $-$  &$-$   & $-$ \\
        \hline
        
        % SECTION: Distributed Quantum Sensing
        \multicolumn{7}{l}{\textbf{Distributed Quantum Sensing}}\\
        \hline
        \parbox[t]{0.28\textwidth}{QTelescope \cite{khabiboulline2019quantum,khabiboulline2019optical,gottesman2012longer}
        (Sec.~\ref{subsec:QTelescope})}
        & $\checkmark$ & $\bigcirc$  & $\checkmark$ & $-$  & $-$  &$-$  \\
        
        \parbox[t]{0.28\textwidth}{QCN \cite{giovannetti2001quantum,komar2014quantum,giovannetti2002positioning}
        (SI.~\ref{subsec:clocksyncgronization})}
        & $\checkmark$ & $\checkmark$  & $\checkmark$ & $-$ & $-$  & $\checkmark$\\
        
        \parbox[t]{0.28\textwidth}{GOG \cite{brady2021frame,giovinetti2024gingerino,di2024noise}
        (SI.~\ref{subsec:GOG})}
        & $\checkmark$ & $\checkmark$  & $\checkmark$ &$-$   & $-$  & $\checkmark$\\
        \hline
        
        % SECTION: Federated Quantum Computation
        \multicolumn{7}{l}{\textbf{Federated Quantum Computation}}\\
        \hline
        \parbox[t]{0.28\textwidth}{QDC \cite{liu2023data,liu2024quantum}
        (SI.~\ref{subsec:LDFQC})}
        & $\checkmark$ & $\checkmark$  & $-$ & $-$  & $\checkmark$ &$-$  \\
        
        \parbox[t]{0.28\textwidth}{BQC \cite{fitzsimons2017private,morimae2013blind}
        (Sec.~\ref{subsec:BDQC})}
        & $\checkmark$ & $\checkmark$  &$-$   & $-$  & $\bigcirc$ & $-$ \\
        
        \parbox[t]{0.28\textwidth}{NLQC \cite{buhrman2010nonlocality,yao1993quantum,brassard2003quantum}
        (SI.~\ref{subsec:NLQC})}
        & $\checkmark$ & $\checkmark$  & $-$  & $-$  & $-$  &$-$  \\
        \hline\hline
    \end{tabular}

    \begin{flushleft}\footnotesize
    $^*$ '$\checkmark$' means the specific application relies on a certain performance metric, while '$-$' infers that it is less relevant, and '$\bigcirc$' suggests that it may depend on a concrete protocol. SI refers to Supplementary Information. \\
    \yesun{$^\dagger$ Pessimistic Gaussian pulse duration \cite{diels2006ultrashort} residual-dispersion bound after one-time compensation of the deterministic mean GDD. The uncompensated scale is $\sim3$ ps.}
    \end{flushleft}
\end{table*}

Beyond the primary challenges of macroscopic attenuation and interferometric phase decoherence, validating the VBG as a universal physical layer necessitates rigorously quantifying its complementary optical metrics, specifically including polarization fidelity, transmission latency, and dispersion-induced pulse broadening. Analytical evaluations, detailed comprehensively 
 in the Methods Section \ref{subsec:evaluationOfCMs}, demonstrate that the VBG achieves \yesun{performance comparable to state-of-the-art optical systems} across these domains rather than merely satisfying
  baseline operational thresholds. For polarization-encoded states, the misalignment of optical reference planes introduces a worst-case error probability of approximately 1 ppm per \yesun{4 km} section, which remains \yesun{below the modeled per-section attenuation} and ensures exceptional polarization maintenance. Furthermore, the cumulative group velocity delay dictated by the ultra-pure silica substrates reduces the macroscopic transmission speed by only \yesun{3 parts per million} relative to the vacuum speed of light, vastly outperforming the inherent latency of standard optical fibers. Finally, \yesun{after one-time end-to-end compensation of the deterministic GDD}, \yesun{residual} group-velocity dispersion (GVD) along the propagation axis bounds the minimum pulse duration to roughly 270 fs over a $\yesun{\sim}10^4$-km \yesun{scale} continental baseline, \yesun{while} this temporal limit remains entirely sufficient to support ultra-dense multiplexing protocols.  
Ultimately, these results \yesun{show} that the VBG \yesun{preserves} the requisite optical \yesun{conditions} for executing versatile, large-scale quantum network architectures.

\section{Protocol Benchmarks}
\label{sec:APP}
Table~\ref{tab:applicationSummary} summarizes the transformative potential of the VBG across three key application domains.
First, the capacity for ultra-low-loss bosonic quantum state transmission enables the VBG to surpass the theoretical QKD key rate limit of the current satellite-to-ground systems by four orders of magnitude and experimental results by eight orders under the basic QKD scheme \cite{lu2022micius,huang2024vacuum}, establishing a \yesun{clear} advantage for secure quantum communication. Moreover, the unprecedented interference stability of the VBG is crucial for overcoming significant hurdles in distributed quantum sensing, where naturally encoded signals typically experience phase noise that negates quantum advantages \cite{bartlett2007reference,zhang2025criteria}. Finally, for federated quantum computing, particularly in cloud computing contexts \cite {ravi2021quantum, wehner2018quantum, devitt2016performing}, the VBG offers the critical benefit of minimizing algorithmic execution latency, an advantage derived synergistically from both its near-vacuum propagation speed and its \yesun{suppression} of loss-induced exponential retransmission delays \yesun{in sequential-retry settings}. To quantitatively demonstrate these capabilities, we execute detailed performance benchmarks for three representative protocols: Device-Independent QKD (DI-QKD) \cite{ekert1991quantum,sekatski2021device,xu2022device}, Quantum Telescope (Q-Telescope) \cite{khabiboulline2019quantum,khabiboulline2019optical,gottesman2012longer}, and Blind Quantum Computation (BQC) \cite{fitzsimons2017private,morimae2013blind}. Comprehensive illustrations and performance estimations for the remaining protocols listed in Table~\ref{tab:applicationSummary} are provided in the Supplementary Section \ref{sec:AddPro} and Figs.~\ref{fig:applicationSummary}-\ref{fig:QDC}.

\subsection{Device Independent QKD}
\label{subsec:DIQKD}

Device-independent quantum key distribution (DI-QKD) serves as a stringent benchmark for the VBG. By basing security entirely on the violation of Bell-like inequalities rather than trusted hardware \cite{ekert1991quantum}, DI-QKD necessitates exceptionally high transmission efficiency. Unfortunately, the theoretical lower bound for CHSH-based protocols \cite{larsson2001strict} is far above the coupling limit imposed by the $3$ dB atmosphere absorption using satellite-to-ground links \cite{lu2022micius}. 

To evaluate the VBG under these extreme constraints, we select the standard and simplest protocol for demonstration, which does not require any optimization of the experimental configuration but has a transmission efficiency threshold of $\eta_{\mathrm{th}}\approx90.8\%$ \cite{sekatski2021device}. 
The asymptotic secret key rate per channel use is bounded by
\begin{equation}
    r_{\mathrm{DI-QKD}}(\eta,F)\geq 1-h\left(\frac{1+\sqrt{(S/2)^2-1}}{2}\right)-\yesun{H_{IR}},
\end{equation}
where $h(x)$ is the binary entropy function, $S$ is the CHSH score, which depends on the state fidelity at generation $F$ and channel efficiency $\eta\yesun{=10^{-\alpha_{\rm dB}L/20}}$ \yesun{for each source-to-user arm}, and \yesun{$H_{IR}$ represents conditional entropy for information reconciliation.} Leveraging the massive transmission capacity of the VBG, the total \yesun{ideal
asymptotic} key rate is obtained by integrating over the continuous frequency spectrum as
\begin{equation}
    R_{\mathrm{DI-QKD}}=\int r_{\mathrm{DI-QKD}}d\nu \approx r_{\mathrm{DI-QKD}}(\eta,F)\Delta\nu,
\end{equation}
which has units of bits per second.

\begin{figure}[htbp]
\centering\includegraphics[width=\linewidth]{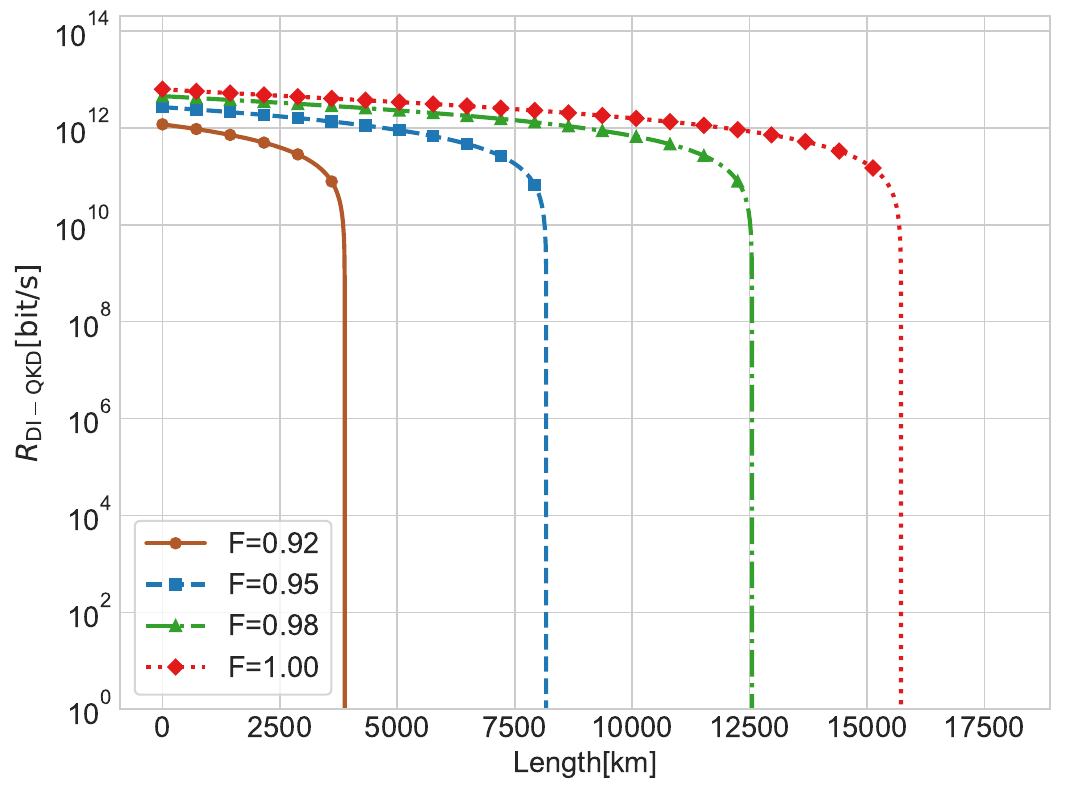}
\caption{\label{fig:QKD_DIQKD} \yesun{Ideal asymptotic channel-limited} key rate scaling for Device-Independent QKD. 
Asymptotic secret key rate $R_{\mathrm{DI-QKD}}$ as a function of continental transmission distance, 
evaluated across varying initial state fidelities $F$.}
\end{figure}

The achievable key rate \yesun{under the ideal asymptotic channel-limited assumptions} for performing the DI-QKD protocol mentioned above, with \yesun{increasing} transmission distance, is shown with different generation fidelities in Fig.~\ref{fig:QKD_DIQKD}. With the achievable fidelity $F=0.98$ \cite{liu2022toward}, the VBG can offer \yesun{a near} terabit-per-second key rate over a continental distance of $10^4\,\mathrm{km}$, which can be further improved with a more advanced protocol. As DI-QKD protocols base their security on the test of Bell-like inequalities, a practical scheme for DI-QKD also implies the feasibility of the fundamental loophole-free Bell test\cite{giustina2017significant}, \yesun{highlighting the potential of the VBG being} a transformative infrastructure for foundational quantum mechanics and unconditional cryptographic security.

\subsection{Long Base-line Quantum Telescope}
\label{subsec:QTelescope}
\begin{figure*}[htbp]
\centering\includegraphics[width=\linewidth]{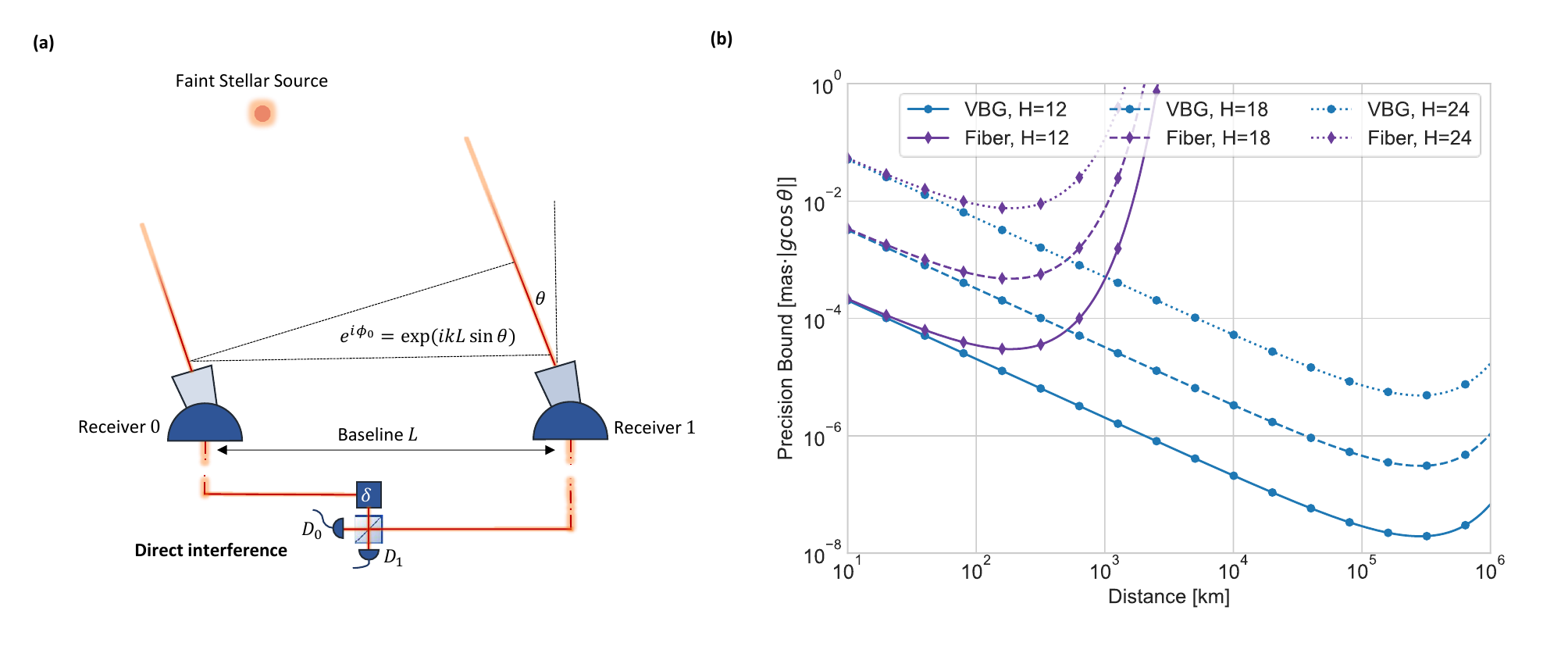}
\caption{\label{fig:QT} Interferometric precision bounds of the long-baseline Quantum Telescope. (a) The light from both telescope stations is directly interfered to eliminate the position information. $D_i$ is the photon detector. 
\yesun{(b) Channel-limited astrometric precision bound in milliarcseconds (mas) for VBG and fiber for magnitude $H=12,18,24$, with $H$ denoting the H-band apparent magnitude of the stellar object. Each measurement bin has $\epsilon_0=0.01$, and the plotted CRB uses $F_{\rm eff}(H)=N_HF_{\rm bin}(\epsilon_0)$ over the common $\tau=10$-ms atmospheric coherent period.}}
\end{figure*}

Long-baseline Quantum Telescopes (Q-Telescopes) offer a quadratic precision enhancement for the astrometry of faint stellar objects treated as weak thermal sources \cite{tsang2011quantum}. 
As illustrated in Fig.~\ref{fig:QT}(a) \cite{khabiboulline2019quantum,khabiboulline2019optical,gottesman2012longer}, incoming stellar photons from two receivers are combined via a quantum channel to generate an interference pattern. This pattern reveals the astronomical angle of arrival $\theta$ through the relative phase shift $ \phi_0=\frac{2\pi L}{\lambda}\sin{\theta}$, with $L$ being the baseline distance. Intuitively, the quantum telescope offers benefits over classical local schemes due to its ability to remove the position information of incoming photons by interference, thereby reducing uncertainty in momentum according to Heisenberg's principle. 

However, macroscopic interferometry is exquisitely sensitive to channel phase decoherence. Based on \yesun{the weak-thermal-light Fisher information~\cite{frieden2013principle,tsang2011quantum,khabiboulline2019quantum,khabiboulline2019optical}}, the fundamental lower bound for the astrometric angular variance is governed directly by the channel attenuation coefficient $\alpha=\frac{\ln(10)}{10}\alpha_{\mathrm{dB}}$ and the dephasing \yesun{coefficient} $\gamma$. Specifically, this measurement uncertainty is bounded by the exact analytical relation
\begin{equation}
(\delta\theta\cdot|g|\cos\theta)^2\geq A_H(\lambda)\times\frac{e^{(\gamma+0.5\alpha)L}}{ L^2 }, 
\end{equation}
where $|g|$ is the first-order correlation strength, and \yesun{$A_H(\lambda) = \frac{\lambda^2}{4\pi^2N_H\varepsilon_0}$ is a source-magnitude, $H$, dependent pre-factor incorporating the per measurement brightness $\varepsilon_0$ and the number of independent measurements $N_H$.} If geographic constraints are removed, the competition between exponential channel degradation and quadratic baseline enhancement yields a strict optimum for the baseline length at $L=\frac{2}{\gamma+0.5\alpha}$.

As a quantitative example, let $\tau
\approx0.01\mathrm{s}$ be the atmospheric self-coherence time \cite{lacour2014reaching}, \yesun{ $\varepsilon_0=0.01$ be the fixed brightness with the Fisher information summed over independent weak-light blocks according to the stellar photon flux, as detailed in Methods.}  Benefiting from the unprecedented interferometric stability of the VBG established in previous sections, integrating the phase noise power spectral density over the \yesun{full spectrum} yields a phenomenally low dephasing factor of $\gamma\approx 2\int \hat{S}_\phi(f)df\approx10^{-6}\mathrm{rad^2/km}$. Because this value is mathematically negligible compared to the channel attenuation, the destructive impact of phase noise is effectively erased. Consequently, the VBG enables the interferometric baseline to scale almost entirely as a function of fundamental optical loss. \yesun{The resulting channel-limited astrometric bounds for $H=12,18,24$ are shown in Fig.~\ref{fig:QT}(b).}  While fibers suffer severe precision degradation and a rapid exponential rebound at moderate distances due to overwhelming phase fluctuations, the VBG maintains a monotonic precision improvement deep into the continental scale. \yesun{At a baseline of $10^3$ km, the channel-limited CRB for the $H=24$ benchmark falls below the sub-microarcsecond scale.} Most remarkably, the optimal interferometric baseline for the VBG is pushed well beyond $10^4$ km. At this macroscopic scale, physical channel decoherence ceases to be the primary bottleneck, making the geometric curvature of the Earth the true operational limitation for ultimate astronomical resolution.

\subsection{Blind Delegated Quantum Computation}
\label{subsec:BDQC}

\begin{figure*}[htbp]
\centering\includegraphics[width=\linewidth]{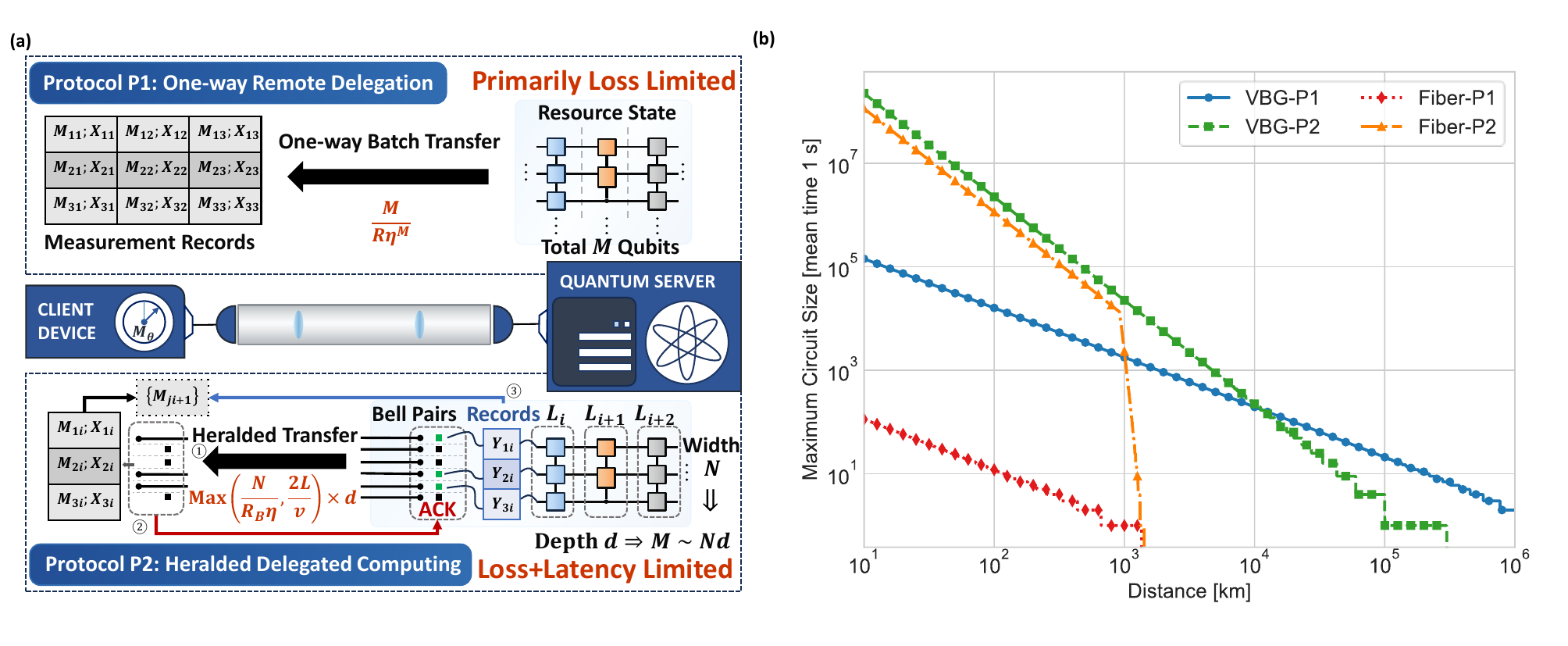}
\caption{\label{fig:BQC} \yesun{Illustration} and macroscopic scaling limits of Blind Delegated Quantum Computation. \yesun{(a) Schematic comparison of the two protocols considered here. The upper panel shows the one-way remote delegation protocol, which is primarily loss limited. The lower panel shows the heralded delegated computing protocol, where heralded Bell-pair transfer is parallel within each layer. The successive layers incur round-trip latency as the client needs to wait for the server's measurement records to determine the measurements for the next layer. (b)} The maximum circuit size assuming linear depth scaling \yesun{(we set $d=N$, hence $M=N^2$)} that can be executed with an \yesun{amortized mean execution time} of one second versus distance between server and client under \yesun{Protocol 1} (P1) and \yesun{Protocol 2} (P2) for fiber and VBG.}
\end{figure*}

Delegating quantum computation to the cloud allows users to leverage remote physical resources for executing complex algorithms. However, when these computations involve proprietary data or sensitive structures, clients require specialized protocols to conceal their tasks from the server. This specific security requirement motivates the development of blind quantum computation (BQC) \cite{fitzsimons2017private,broadbent2009universal}. Achieving BQC through measurement-based quantum computation (MBQC) necessitates the physical transmission of quantum states between the client and the server \cite{morimae2013blind}. While this approach provides profound cryptographic security, it imposes extreme physical demands on the underlying network. Consequently, BQC serves as an ideal benchmark for evaluating both the macroscopic loss and the fundamental latency of the VBG architecture. \yesun{For simplicity, we only include the protocols that do not require quantum memory on the client's side.}

Consider the one-way direct measurement protocol (P1) executing a quantum circuit with an $M$-qubit resource state under a single-qubit coupling efficiency $\eta$. \yesun{The protocol itself is extremely simple, where the client simply makes the desired measurement on the universal resource state of MBQC created and sent by the quantum server, as illustrated in the upper panel of Fig.~\ref{fig:BQC}(a)}. Because clients only perform local measurements on received states, the no-signaling theorem \yesun{guarantees} unconditional computational privacy \cite{peres2004quantum,morimae2013blind}. However, in this unconditionally secure paradigm, the worst-case success probability scales exponentially as $P_{\mathrm{success}} = \eta^M$. \yesun{Since each attempted computation consumes $M$ transmitted qubits, the corresponding amortized mean execution time at channel-use rate $R\simeq\Delta\nu$ is $\bar T_{\rm P1}=M/(R\eta^M)$.} As \yesun{shown} in Fig.~\ref{fig:BQC}\yesun{(b)}, the severe attenuation inherent \yesun{in} standard optical fibers restricts viable computations to mere 10-qubit circuits at urban scales. In stark contrast, the ultra-low loss of the VBG \yesun{directly provides huge benefits in this primarily loss-limited protocol and} facilitates complex computations involving $10^3$ qubits across thousands of kilometers, vastly exceeding current quantum computational limits. While quantum error correction \cite{barrett2010fault,xu2024constant} can eventually mitigate this exponential decay, transmission loss remains the \yesun{main} bottleneck for direct, highly secure protocols.

An alternative \yesun{entanglement-assisted, heralded} protocol (P2) \yesun{followed by interactive UBQC} \cite{fitzsimons2017private,broadbent2009universal,morimae2013blind} circumvents this exponential failure by \yesun{heralding successful entanglement distribution before the computation.} \yesun{The computation is divided into consecutive layers executed in ordered rounds, as illustrated in the lower panel of Fig.~\ref{fig:BQC}(a). At the beginning of the $i$-th round, the server distributes halves of Bell pairs to the client. The client acknowledges which pairs are successfully received and measures each received Bell-pair half in the algorithm-dependent basis $M_{ji}$ for qubit $j$. For each successful transmission, the server then couples its retained Bell-pair half to the corresponding register qubit using a CZ gate, measures the register qubit in the $X$ basis, and returns the measurement outcome to the client for it to determine the next-round measurements while also distributing Bell pairs for the subsequent round.} Although the expected number of transmitted qubits now scales only linearly as $\frac{M}{\eta}$, it fundamentally trades loss vulnerability for execution latency \yesun{due to the round-trip communication for each layer required by the protocol}. Assuming a circuit depth $d$, a layer width $N$, the minimum execution time is \yesun{bounded} by $\yesun{\bar T_{\rm P2}=}\max{\left(\frac{N}{\eta R_B},\tau_c\right)}d$, where $R_B\yesun{\approx\Delta\nu}$ is the \yesun{ideal channel-use rate for Bell-pair distribution} and $\tau_c=2L/\yesun{v}$ is the round-trip communication time, \yesun{with $v$ denoting the channel propagation speed}. This latency penalty exhibits a dual dependence on the physical channel, which standard optical fibers severely amplify through two compounding factors: a low transmission efficiency that necessitates massive heralded retransmissions; and a slow propagation speed restricted by the silica medium. The VBG architecture \yesun{substantially mitigates} both bottlenecks simultaneously. By combining a near-unity macroscopic coupling efficiency with a vacuum-level transmission speed, the VBG drastically minimizes algorithmic execution times. As demonstrated in Fig.~\ref{fig:BQC} \yesun{(b)}, while the heralded protocol avoids exponential decay at continental scales, it remains highly sensitive to communication delays, a vulnerability that becomes critically pronounced for fault-tolerant algorithms with super-linear depth scaling \cite{yamasaki2024time,tamiya2024polylog}. By addressing both the attenuation and transmission speed limits, the VBG provides \yesun{a} low-latency physical layer required to scale delegated quantum computation globally.

\section{Discussion}

In this work, we construct a comprehensive architectural blueprint for the VBG,
establishing it as a highly scalable physical channel that \yesun{can overcome}
the attenuation and decoherence bottlenecks of standard optical networks.
By \yesun{quantitatively} modeling macroscopic channel dynamics based on realistic
terrestrial topography and the empirical performance of the LIGO system \cite{abbott2009ligo,aasi2015advanced}, we demonstrate that the VBG achieves exceptional phase stability
(\yesun{$\sim10^{-3}\,\mathrm{rad/\sqrt{km}}$}) alongside negligible polarization and dispersion penalties. Through systematic protocol benchmarks, we
\yesun{demonstrate} that this \yesun{low-loss, phase-stable} physical layer unlocks
unprecedented performance regimes. Specifically, it empowers standard DI-QKD protocols
to operate over continental baselines, enables coherent signal integration for macroscopic
quantum telescopes with extreme precision, and \yesun{dramatically reduces} algorithmic execution latency for federated computations \yesun{over long distances}. \yesun{Collectively, we transform the VBG from a low-loss concept into a general-purpose quantum optical physical layer, introduce a cross-field methodology bridging gravitational-wave interferometry and quantum-network design, and establish an application-aware benchmarking framework that directly maps channel physics onto protocol performance limits.}

To transition this analytical blueprint into physical reality, detailed simulations using Finesse3 or equivalent optical tools \cite{maheshwari2024ligo,roseanalysis} are being conducted to provide a \yesun{more detailed} depiction and \yesun{further evaluate the performance} of VBG systems. Considering the projected expansion of VBG across continents, it is essential to integrate transient, significant environmental disturbances into the analysis to ensure sufficient redundancy for operational durability and longevity. The performance evaluation will proceed by implementing the system at full scale and collecting empirical data using the methods employed in LIGO \cite{nguyen2021environmental,kruk2016environmental}. Small-scale tests using former LIGO experimental tubes are also currently being pursued.

Beyond the core applications benchmarked in this study, the VBG can be integrated with all-photonic repeaters \cite{azuma2015all,gan2025quantum,fuentealba2025robust} to establish \yesun{a quantum network that is in principle scalable to arbitrary distances}, utilize its own segments as native quantum memory delay lines (evaluated in Supplementary Section \ref{sec:VBGQM} and Supplementary Figs. \ref{fig:VBGQM}–\ref{fig:VBGQM_phaseNoise}), facilitate the execution of a Bell test experiment with cognitive observers to examine the basic measurement assumption \cite{bell2004speakable}, and support quantum holographic tasks \cite{may2019quantum}. The VBG may also provide cross-disciplinary advantages, including ultra-stable optical frequency transfer (detailed in Supplementary Section \ref{sec:enhancementOfFT} and Supplementary Fig. \ref{fig:VBG_FT}), facilitating the long-range transmission of light bullets \cite{guo2021structured,chong2010airy}, directly interconnecting multiple LIGO detectors as a potential central component of the anticipated Einstein telescope \cite{kroker2014einstein}, and extending worldwide to perform large-scale geodesy measurements \cite{li2025minute}.

\section{Methods}
\label{sec:methods}
\subsection{Overview of Design}
\label{subsec:overViewStableSys}
To evaluate the macroscopic interferometric stability of the VBG, we establish a comprehensive baseline control architecture. While a full replication of the LIGO control system would yield excellent optical performance, we propose a streamlined, highly scalable design that \yesun{fulfills} our performance claims across continental distances \yesun{even under the pessimistic error model considered here}. The realistic deployment of the VBG will be built upon this prototype, allowing for further engineering optimizations.

The design of the VBG control system consists of the following three parts:
\begin{enumerate}
    \item \textbf{Passive Isolation Platform (PIP)}. 
    The nodes (lenses and mirrors) are rigidly mounted to passive isolation platforms. 
    The platforms isolate the optics from the ground motion above $\sim1$\,Hz and follow the local ground motion below 1\,Hz.
    
    \item \textbf{Alignment Control System (ACS)}. 
    The auxiliary sensing laser (AUX) beams are injected from both ends of the VBG and are co(counter)-propagated alongside the quantum information signal beam. 
    The beam position on each lens and mirror is imaged by a camera.
    All of these positions are then recorded and used to control the angle of each of the steering mirrors (and possibly also the transverse alignment of the lenses).

    \item \textbf{Section Length Stabilization (SLS)}. 
    The AUX beam reflected at each section is extracted at the steering mirror node and mixed with the AUX beam reflected (transmitted) at the previous lens (mirror) to perform section-wise, round-trip heterodyne phase detection. The detected phase (beat) serves as the feedback signal for the SLS system to drive the actuators on the mirror to perform active control of the section length. The signal can also be used for post-subtraction only for applications that do not require real-time processing.   
\end{enumerate}

Among these subsystems, the PIP and SLS are dedicated to maintaining the stringent phase stability of the VBG, whereas the ACS is essential for preserving all other performance metrics detailed in Table~\ref{tab:applicationSummary}. A comprehensive breakdown of these control subsystems, including steering mirrors, passive transfer functions, and the alignment monitoring and feedback principle, is provided in the Supplementary Section \ref{sec:VBGControlSys}.

\subsection{Evaluation of Phase Noise PSD}
To rigorously benchmark the interferometric stability of the VBG, we evaluate the phase noise power spectral density (PSD) by adapting the analytical frameworks established by Advanced LIGO \cite{aasi2015advanced}. The total open-loop PSD is modeled as the incoherent sum of dominant physical noise sources across the entire channel: $S_\phi(f) \approx S_{\mathrm{seis}} + S_{\mathrm{therm}} + S_{\mathrm{gas}}$.
\begin{enumerate}
    \item \textbf{Seismic and Alignment Noise.} Extrapolated using empirical BSSA low frequency survey data ~\cite{anthony2022seismic} and ERDC-CRREL high frequency data ~\cite{albert2017acoustic}, and converted to optical phase fluctuations via Laguerre-Gaussian mode expansion. Crucially, the high-frequency noises are filtered by the Passive Isolation Platform (PIP), which adapts a commercially available design (Supplementary Section \ref{subsec:seismicAndAlignment}).
    \item  \textbf{Residual Gas Scattering.} Modeled under thermal equilibrium conditions at 1 Pa, accounting for interactions between the propagating light and constituent gas molecules (Supplementary Section \ref{subsec:residualGas}).
    \item  \textbf{Thermal Noise.} Evaluated via the fluctuation dissipation theorem utilizing realistic material parameters and empirical measurement data anchored to the Advanced LIGO system, with mount thermal noise, mirror coating Brownian noise, and substrate thermo optic noise dominating at distinct frequency bands (Supplementary Section \ref{subsec:thermalNoise}).
    \item  \textbf{Other Noises.} Including acoustic wave noise and birefringence, analyzed quantitatively and shown to be negligible (Supplementary Section \ref{sec:polarizationNoise} \& \ref{sec:AcousticN}). 
\end{enumerate}

The resulting theoretical open-loop PSD, $S_\phi$, is subsequently processed by the Section Length Stabilization (SLS) system. The SLS system locks the section length to an ultra-stable auxiliary laser to actively suppress low-frequency length fluctuations. Because the raw physical movement of the VBG is relatively stable at high frequencies where sensor noise dominates, the feedback must be meticulously filtered. The post stabilization residual phase noise, $\hat{S}_{\phi,\yesun{\rm res}}$, is thus defined by the feedback control relation
$$\hat{S}_{\phi,\yesun{\rm res}} = S_{\mathrm{SLS}}|\mathrm{LPF}|^2 + S_\phi|\mathrm{HPF}|^2$$
where $\mathrm{LPF}$ and $\mathrm{HPF}$ are the low-pass and high-pass filter functions satisfying the \yesun{fundamental feedback} constraint $\mathrm{LPF} + \mathrm{HPF} = 1$, and $S_{\mathrm{SLS}}$ represents the total sensing noise of the SLS scheme. For a comprehensive derivation of individual noise terms, see the Supplementary Section \ref{sec: NoiseBudgets}.

\subsection{Evaluation of Complementary Optical Metrics}
\label{subsec:evaluationOfCMs}
\subsubsection{Polarization Fidelity}
In polarization-based dual-rail encoding, fidelity degradation arises from coupled amplitude and phase noise induced by optical birefringence. Because these polarization fluctuations inherently couple into the overall phase noise, the rigorous derivation of the channel's polarization evolution is integrated directly into our comprehensive noise budget (see Supplementary Section \ref{sec:polarizationNoise}). Briefly, the net effect of reference plane misalignments across sequential optical elements is modeled via Jones matrices. By averaging over the angular misalignment noise power spectrum, we can map these classical optical perturbations directly into the quantum domain, yielding the \yesun{leading} effective Kraus operators $\{K_\alpha\}$ for each VBG section:
\begin{equation}
    \{K_{\alpha\yesun{\neq0}}\}\yesun{\sim}\sqrt{p}\left\{\begin{pmatrix}
        0&1\\
        -1&0
    \end{pmatrix},\begin{pmatrix}
        0&e^{i\bar\phi}\\
        -e^{-i\bar{\phi}}&0
    \end{pmatrix}\right\},
\end{equation}
\yesun{The Jones vectors use consecutive local polarization bases.} \yesun{Here,} $\bar{\phi} \sim 10^{-2}$ rad is the average birefringence phase difference \cite{michimura2024effects}, and the error probability $p$ is obtained by integrating the noise power spectrum $S_\theta(f)$. For a worst-case root-mean-square angular fluctuation of $\delta\theta_{\mathrm{rms}} \sim 10^{-3}$ rad \cite{hirose2020characterization}, this transformation introduces an error probability of approximately 1 ppm per section. This value remains two orders of magnitude below the baseline attenuation limit, unequivocally demonstrating exceptional polarization maintenance.

\subsubsection{Transmission Speed}

The macroscopic propagation speed of a non-monochromatic signal is governed by the effective group velocity of the VBG channel. Because the beam path operates predominantly in a vacuum, the deviation from the vacuum speed of light $c$ is determined by the weighted average of the group velocities through the discrete optical elements,
\begin{equation}
    v_g=\frac{\partial\omega}{\partial k}\yesun{\approx}\frac{c}{n}\left(1+\frac{\lambda}{n} \frac{\partial n}{\partial \lambda}\right)\equiv\frac{c}{n_g}.
\end{equation}
Since most parts of the VBG are in a vacuum, the deviation from the speed of light in a vacuum arises from the weighted average of the group velocity determined by the optical elements and the vacuum velocity as $\bar{v}_g=\frac{L_{0}}{\sum_i d_i/v_{g_i}}$, where $\{d_i\}$ denotes the sections with different group velocity indices $n_g$. For our VBG system, the deviation is dominated by the ultra-pure silica substrate, which has a thickness of roughly $d_{\mathrm{substrate}}\approx2\,\mathrm{cm}$ for the worst-case consideration\cite{pinard2017mirrors} and a group velocity index of $n_g\approx1.46$ for the light centered at $1550$ nm. Therefore, the average group velocity in the VBG is calculated as
\begin{equation}
\label{eq:transmissionSpeed}
    \bar{v}_g\approx(1-\yesun{3}\times10^{-6})c,
\end{equation}
which represents only a sub-10 ppm deviation and thus can be safely neglected under most circumstances. 

\subsubsection{Dispersion and Pulse Duration}
Ultrashort optical pulses experience temporal broadening dictated by the group-velocity dispersion (GVD), defined analytically as \cite{diels2006ultrashort},
\begin{equation}
    \mathrm{GVD}\equiv\frac{\partial^2k}{\partial\omega^2}=\frac{2}{c}\left(\frac{\partial n}{\partial \omega}\right)+\frac{\omega}{c}\left(\frac{\partial^2n}{\partial\omega^2}\right).
\end{equation}
As an unchirped beam with a Gaussian profile passing through relatively thick materials with a thickness of $d_i$, the broadening of pulse duration $\tau$ can be calculated as
\yesun{
\begin{equation}
\label{eq:GDDbroadening}
    \tau'^2=\tau^2+\frac{\rm 4GDD^2}{\tau^2}
\end{equation}
}
The simulation shows that the \yesun{group-delay dispersion} (GDD$\equiv$GVD$\times d_i$) of the coating is at least two orders of magnitude smaller than that of the substrate. Therefore, the expansion effects are dominated by the silica substrate with $\mathrm{GVD}\approx-25.855 \mathrm{fs^2/mm}$ at $1550$ nm \cite{arosa2020refractive}.
\yesun{For the current $20$-mm substrate, this corresponds to a mean GDD of $-517\,\mathrm{fs^2}$ per $4$-km section. Then, for a conservative residual-dispersion benchmark, $N_{\rm sec}=10^4$ sections therefore accumulate a deterministic mean $\rm{GDD}\simeq-5.17\,\mathrm{ps^2}$. Requiring the output pulse width to remain within a factor of two of the input gives an uncompensated scale of $\sim3\,\rm ps$. }

\yesun{However, the above deterministic quadratic spectral phase is calibratable and can be compensated by a one-time end-to-end dispersion correction after the propagation. To give a calibrated bound, a linearized independent-section residual model gives
\begin{equation}
\label{eq:GVDbroadening}
    \Delta\tau=\tau'-\tau\approx\frac{2\Delta(\rm{GDD})^2}{\tau^3}\leq \frac{2\rm GDD^2}{\tau^3},\,|\Delta\tau|\ll\tau.
\end{equation}
where $\Delta(\mathrm{GDD})$ denotes the uncompensated residual GDD after removal of the deterministic mean, and is pessimistically bounded by the full section GDD. Then, suppose that the beam goes through the same $\sim 10^4$ sections with the same broadening requirement, the duration of the pulse is lower bounded by
\begin{equation}
\label{eq:GVDBroad}
N\Delta\tau<\tau_{\mathrm{GVD}}\Rightarrow\tau_{\mathrm{GVD}}\gtrsim270\,\mathrm{fs}.
\end{equation}}
For comparison, the shortest pulse duration limited by the attenuation spectrum is given by~\cite{diels2006ultrashort}
\begin{equation}
    \tau_{\mathrm{spectrum}}\geq \yesun{\frac{0.441}{\sqrt{2\ln2}\Delta\nu}\sim60\,\mathrm{fs}},
\end{equation}
which is roughly \yesun{half an} order of magnitude smaller than the value limited by the dispersion.
\subsection{Fisher Information Bound for Quantum Telescopes}
\label{subsec:QT_FI}
To quantitatively evaluate the astrometric precision of the Q-Telescope under macroscopic channel decoherence, we analyze the direct interference of incoming stellar photons. Upon detection, the incoming photon wave packet is described by the density matrix
\begin{align*}
    \rho_s &= (1-\varepsilon\eta)|00\rangle\langle00| \\&+ \frac{\eta\varepsilon}{2}\left[|01\rangle\langle01| \yesun{+|10\rangle\langle10|}\right. \\&\qquad\left. + ge^{-\frac{1}{2}\sigma_\phi^2}e^{i\phi_0}|01\rangle\langle10| + h.c.\right],
\end{align*}
where $\eta$ is the channel transmission efficiency, $\varepsilon$ represents the probability of containing a single photon from the stellar source, and $g$ denotes the first-order correlation strength. The term $\sigma_\phi^2 \equiv \gamma L$ quantifies the average phase decoherence accumulated over the baseline distance $L$. \yesun{ For $\Phi_H=\Phi_{24}10^{0.4(24-H)}$, we take
$\tau_H=\epsilon_0/\Phi_H$ and $N_H=T_{\rm atm}/\tau_H$, so that the effective phase Fisher information \cite{frieden2013principle} is $F_{\rm eff}(H)=N_H\epsilon_0\eta|g|^2e^{-\sigma_\phi^2}$ and} the fundamental lower bound for the angular uncertainty $\delta\theta$ is derived as
\begin{equation}
(\delta\theta\cdot|g|\cos\theta)^2 \geq \frac{\lambda^2}{4\pi^2 L^2}\frac{\yesun{|g|^2}}{\yesun{F_{\rm eff}(H)}} = \yesun{A_H(\lambda)}\times\frac{e^{(\gamma+0.5\alpha)L}}{ L^2 },
\end{equation}
with the source-dependent pre-factor defined as \yesun{$A_H(\lambda) = \frac{\lambda^2}{4\pi^2N_H\varepsilon_0}$}. Minimizing this uncertainty expression with respect to the baseline length $L$ directly yields the optimal baseline configuration $L = \frac{2}{\gamma+0.5\alpha}$ discussed in the main text.

\yesun{However, the co-propagating $1600$-nm AUX field introduces an additional weak-light background through anti-Stokes Raman scattering in the transmissive silica \cite{stolen1973raman,brainis2007spontaneous,agrawal2019nonlinear}, which may introduce background noise in ultra-low photon budget detection considered here. The continuous prototype $1$-W AUX operation gives an in-spectrum Raman flux of roughly $R_{\rm Raman}\sim10^3\,\mathrm{s^{-1}}$. This background will not cause a problem for the $H=12$ benchmark, whose stellar photon flux is $\sim6\times10^4\,\mathrm{s^{-1}}$. For the faint $H=24$ benchmark, an upgraded low-duty-cycle SLS operating mode with $D_{\rm AUX}\sim10^{-4}$ reduces the time-averaged Raman generation rate to $\sim0.1\,\mathrm{s^{-1}}$. Because the forward Raman photons are generated synchronously with the AUX beam, temporal detector gating rejects them from the astronomical observation window, so the principal cost is only a $\sim D_{\rm AUX}$ reduction in the science observation duty cycle rather than a continuous Raman background. The detailed estimation is given in Supplementary Section~\ref{subsec:SLS}.}

%\nocite{*}

\FloatBarrier % Keep all main-manuscript floats before the references.
\bibliography{VBGAPPs}% Produces the bibliography via BibTeX.

\begin{acknowledgments}
We thank J.C Diels, Ming Lai, Ming Li, Kaushik Seshadreesan, Allen Zang, and Haiqi Zhou for their helpful discussions. This paper describes objective technical results and analysis. This research is supported by ARO(W911NF-23-1-0077), ARO MURI (W911NF-21-1-0325), AFOSR MURI (FA9550-19-1-0399, FA9550-21-1-0209, FA9550-23-1-0338), DARPA (HR0011-24-9-0359, HR0011-24-9-0361), NSF (OMA-1936118, ERC-1941583, OMA-2137642, OSI-2326767, CCF-2312755), NTT Research, Packard Foundation (2020-71479), and the Marshall and Arlene Bennett Family Research Program. U.S. Department of Energy, Office of Science, National Quantum Information Science Research Centers and Advanced Scientific Computing Research (ASCR) program (DE-AC02-06CH11357) and Oak Ridge Leadership Computing Facility (DE-AC05-00OR22725). National Science Foundation (Grant No. DMS-1929348). University of Pittsburgh, School of Computing and Information, Department of Computer Science, Pitt Cyber, Pitt Momentum Fund, PQI Community Collaboration Awards, John C. Mascaro Faculty Scholar in Sustainability, and Cisco Research. RXA would like to acknowledge support by the Woodnext Foundation as well as the Templeton Foundation. Any subjective views or opinions that might be expressed in the paper do not necessarily represent the views of the U.S. Department of Energy or the United States Government. 
\end{acknowledgments}

\section*{Data Availability}
The empirical data used to support the physical models and theoretical analyses in this study are publicly available and appropriately referenced within the manuscript. The datasets generated during the current system-level benchmarks are available from the corresponding author upon reasonable request.

\section*{Author Contributions}
Y.H. developed the core theoretical framework, conducted the system-level benchmarks, and wrote the main manuscript. D.S. assisted in deriving the core phase-noise model. P.Z. and D.B. contributed to the analysis of the secure quantum communication protocols. J.L. evaluated the applications of federated quantum computation. R.X.A. provided expert guidance for the phase noise model and the physical control system apparatus. L.J. conceived the project, supervised the overall research, and guided the design of the quantum network architecture. All authors discussed the results and extensively reviewed the manuscript.

\section*{Competing Interests}
The authors declare no competing interests.

% END MAIN BODY

% Supplementary material: numbering and layout only.
\clearpage
\onecolumngrid
\begin{center}
\textbf{Supplementary Information of Scaling Quantum Networks via Phase-Stable Vacuum Beam Guide: Architectural Blueprint and Benchmark}
\end{center}
\twocolumngrid
\renewcommand{\thefigure}{S\arabic{figure}}
\renewcommand{\thetable}{S\arabic{table}}
\renewcommand{\theequation}{S\arabic{equation}}
\renewcommand{\thesection}{S\arabic{section}}
\setcounter{section}{0}
\setcounter{subsection}{0}
\setcounter{subsubsection}{0}
\setcounter{equation}{0}
\setcounter{figure}{0}
\setcounter{table}{0}

% BEGIN EXACT SI BODY
\section{The VBG Control System}
\label{sec:VBGControlSys}

\subsection{Steering Mirrors}
\label{subsec:reflectionDevice}
Due to the extended length of the VBG, it is necessary to deploy mirrors to deflect the beam to account for both the curvature of the Earth and man-made infrastructure above and below ground. This problem has been previously addressed in a satellite-based scheme~\cite{goswami2023satellite}. In reality, the infrastructure issue is more serious and determines the number of required steering mirrors. A conservative estimation assumes the minimum use of a steering mirror every $L_{{\mathrm{mirror}}}\sim$10\,km, which corresponds to roughly a maximum of 3 lens sections between two steering mirrors. Such a separation corresponds to a construction depth of $D\sim  \frac{L_{\mathrm{mirror}}^2}{8R_E}\sim2\,\mathrm{m}$ where $R_E\approx6000$\,km is the radius of the Earth. This result implies that the VBG system \yesun{can} be constructed on the Earth's surface to avoid digging costs.

\begin{figure*}[htbp]
\centering
\includegraphics[width=\linewidth]{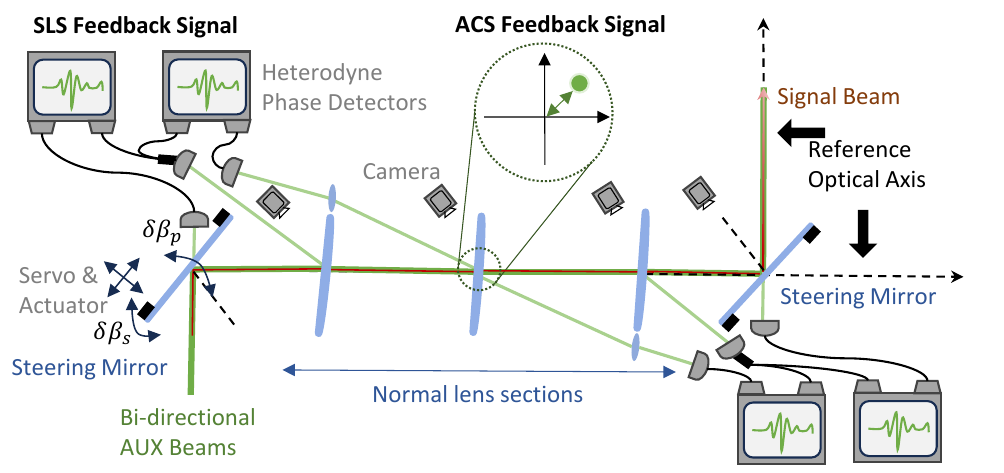}
\caption{ The illustration of the steering mirrors, alignment control system (ACS) and the section length stabilization (SLS) system.}
\label{fig:Reflectors}
\end{figure*}

These steering mirrors will be accompanied by an alignment control sensing system to ensure that the light spot on each lens is within $0.1\,\mathrm{mm}$ of the reference optical axis defined by the center of the two neighboring mirrors, as shown in Fig.~\ref{fig:Reflectors}.  

% For a realistic depth, $D$, for deploying the VBG channel, the maximum separation between two mirror stations can be estimated as
% \begin{equation}
%     L_{\mathrm{mirror}}\approx\sqrt{8R_ED},
% \end{equation}
% where $R_E\approx6000$\,km is the radius of the Earth. 
% If we plug in a reasonable depth of $D\approx30$\,m given by the common selection to build a subway, we get $L_{\mathrm{mirror}}\leq 40$\,km, which means that we need to deploy 1 mirror every 10 sections, assuming $L_{0}=4$\,km.
% However, 

The loss induced by these steering mirrors can be estimated following a similar process as that for the lenses, which gives a similar requirement for the radius, coating, and translation alignment. 
However, the loss introduced by the angular misalignment, which is negligible in the lens case, may now become a serious problem and dominate the additional mirror loss. 
In our analysis, the static centers of neighboring mirrors define the ground-truth reference optical axis. When referencing this axis, the loss associated with each lens is dominated by its transverse displacement, while angular misalignment becomes the dominant source of loss for the mirrors. 

To estimate the precision requirement, we consider a single-plane mirror with a typical incident angle of $\frac{\pi}{4}$ to deflect the beam and decompose the rotation into two independent DOFs, namely tipping $\delta\beta_s$ and tilting $\delta\beta_p$ with respect to the center point of the mirror. 
For the loss introduced by the tip, one can show by a simple argument on the solution of the Helmholtz equation and by rotating the optical axis for the plane mirror that the loss and phase shift depend only on the misalignment angle, regardless of its original orientation angle $\beta_s$ \cite{goodman2005introduction}. 
This result allows us to consider only the case $\beta_s= 0$ so that the effects of tilt angle misalignment $\delta\beta_p$ can be converted into an effective tipping angle of $\delta\beta_s^{\mathrm{eff}}\approx  \frac{\sqrt{2}}{2} \delta\beta_p$.

Suppose that the mirror is located at $z$ with respect to the beam waist. Then, using the conventional mode expansion method, 
%the additional phase gain at the plane can be expressed as
% \begin{equation}
% \label{eq:S45Mshift}
%     \exp(ik\,\delta\beta\,r\cos\phi)\approx1+\frac{i}{2}\delta\beta_s kr(e^{i\phi}+e^{-i\phi}),
% \end{equation}
% which results in a conversion factor into $E_0^1$ and $E_0^{-1}$ Laguerre-Gaussian modes with a factor
% \begin{equation}
%     M_{00\rightarrow(0,\pm 1)}\approx i\frac{w(z)}{2\sqrt{2}}k\delta\beta_s.
% \end{equation}
% with $w(z)$ being the spot size at location $z$. One notices that there is no direct first order coupling into phase noise, while 
the power leaked to the $E_0^1$ and $E_0^{-1}$ Laguerre-Gaussian modes can be calculated as,
\begin{equation}
    l_{\delta\beta_s}\approx {k^2}{w^2(z)}\delta \beta_s^2\sim \left(\frac{\delta\beta_s}{\beta_G}\right)^2,
\end{equation}
where $\beta_G=\frac{\lambda}{\pi w(0)}\sim10^{-5}$rad is the divergence angle of the Gaussian beam, which agrees with the reference~\cite{mueller2005beam,anderson1984alignment}. 
One also finds that the loss is minimized when the mirror is placed at the waist. 
Then, we have 
\begin{equation}
    l_{\delta\beta_s}\approx \frac{2\pi L_0}{\lambda}\delta\beta^2. 
\end{equation} 
For an order of magnitude estimate, plugging in $\delta\beta_s\approx0.05\,\mu$rad$\sim 0.5\%\beta_G$, one finds that $l_{\delta\beta_s}\approx40$\,ppm for the default configuration, which corresponds to an angular misalignment well above the actuator precision employed in the Advanced LIGO~\cite{barsotti2010alignment,canuel2014sub}. An intuitive way to understand such a requirement is that the $0.1$ mm displacement error corresponds to $\frac{0.1\,\mathrm{mm}}{2000\,\mathrm{m}}\approx0.05\,\mu \mathrm{rad}$ rotational precision.
In comparison, the loss induced by lens tilting is given by
\begin{equation}
    l'_{\delta\beta}\approx\frac{2 d^2}{w_0^2}\left(\frac{n-1}{n}\right)^2\delta\beta^2\sim0.05\delta\beta^2, 
\end{equation}
which implies that a tilting precision at the level of tens of milliradians will be sufficient, rendering this contribution negligible in our loss analysis. 
\subsection{Passive Isolation Platform}
\label{sec:seismic_isolation}
\begin{figure*}[htbp]
\centering
\includegraphics[width=0.8\linewidth]{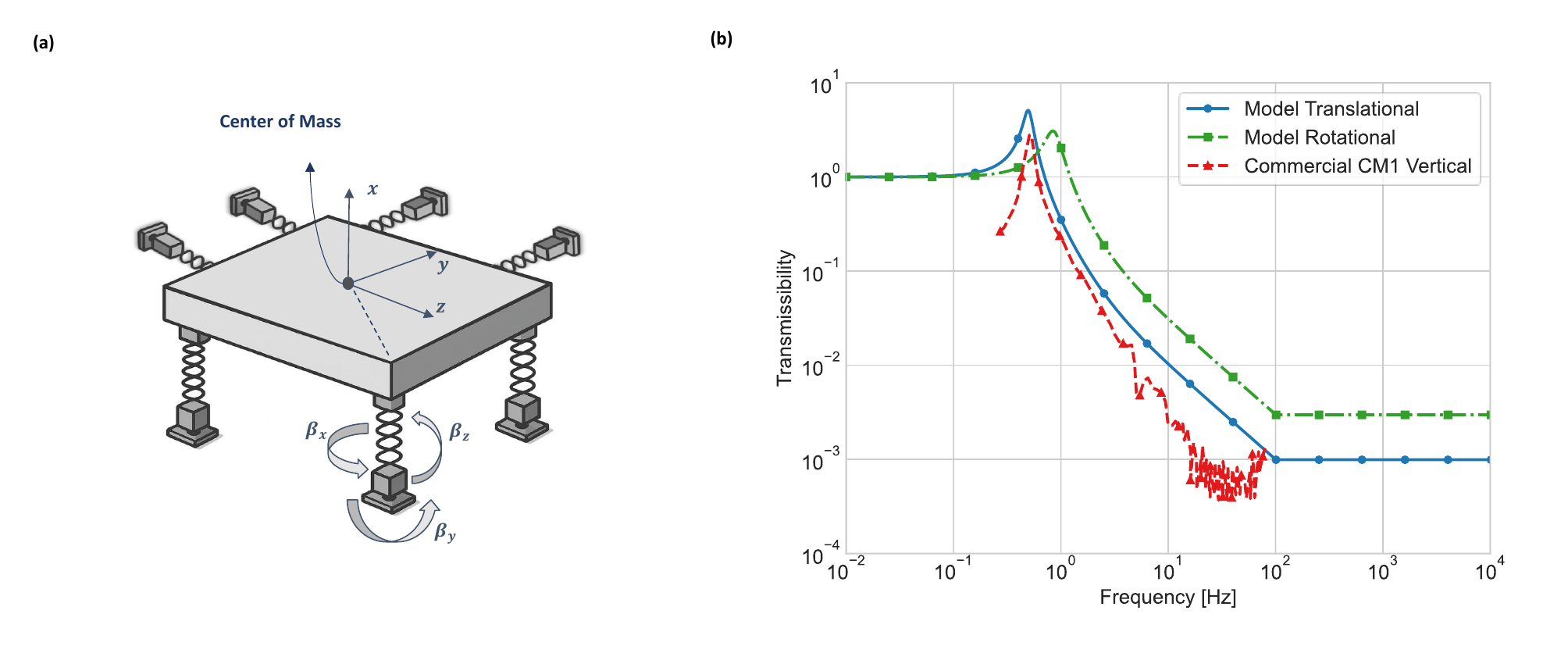}
\caption{\label{fig:passiveIso} The (a) illustration and the (b) performance of the PIP. The vertical isolation performance of the commercially available CM1 optical table \cite{MinusK_BM1_transmissibility} is also plotted for comparison.} 
\end{figure*}

The vibration isolation system we assume is based on existing commercial solutions \cite{MinusK_BM1_transmissibility}.
In this setup, the optical table acts as a rigid body isolated by pairs of parallel springs, as shown in Fig.~\ref{fig:passiveIso} (a). 
Under the simple harmonic oscillator approximation, the transmissibility can be expressed as \cite{thomas2019complementary,harris2002harris}
\begin{equation}
\label{eq:TranslationalTransmission}
    T_z(f)=\sqrt{\frac{1+\left(2\xi \frac{f}{f_n}\right)^2}{\left(1-\left(\frac{f}{f_n}\right)^2\right)^2+\left(2\xi \frac{f}{f_n}\right)^2}},
\end{equation}
where $\xi\approx0.1$, $f_n\approx0.5\,\mathrm{Hz}$ are the typical damping ratio and natural (resonant) frequency respectively for a commercial passive isolated optical table.  

The transmissibility for the ground rotations can be estimated using a similar method by substituting the mass with the moment of inertia and the force with the torque, which leads to essentially the same expression, \cite{downey2025vibration,schmitz2012mechanical}
\begin{equation}
\label{eq:RotationalTransmission}
    T_\beta(f)=\sqrt{\frac{1+\left(2\xi_\beta \frac{f}{f_{n,\beta}}\right)^2}{\left(1-\left(\frac{f}{f_{n,\beta}}\right)^2\right)^2+\left(2\xi_{\yesun{\beta}} \frac{f}{f_{n,\beta}}\right)^2}},
\end{equation}
with $\xi_{\beta}=\sqrt{3}\xi$, $f_{r,\beta}=\sqrt{3}f_r$ being the natural frequency and damping ratio for the rotation, respectively. 
The estimated transmissibility curve of the passive isolation and the corresponding measured data from a commercially available system are plotted in Fig.~\ref{fig:passiveIso} (b), demonstrating good agreement with our prediction.

However, it should be noted that the simple harmonic oscillator approximation given by Eq.(\ref{eq:TranslationalTransmission})-(\ref{eq:RotationalTransmission}) fails at frequencies above $\gtrsim100$\,Hz due to internal mode resonances of the springs and rigid bodies. 
To account for this behavior, we assume that the transmissibility becomes flat at high frequencies for a conservative estimate.

\subsection{Alignment Control System}
\label{subsec:ACS}
An auxiliary (AUX) laser at $1600\,\mathrm{nm}$ is emitted at the beginning of the VBG and co-propagates with the quantum signal beam. The spot position of the emitted auxiliary laser is monitored by cameras at each node to provide an alignment error signal for the ACS system, as shown in Fig.~\ref{fig:Reflectors}. ACS aims to eliminate static and ultra-low frequency drift, and it is expected to function at a frequency below $0.01$ Hz, implying that the measurements can be slow but must be precise. The measured displacement at each node serves as the feedback system to drive the actuators to control the rotational and translational alignment of the mirrors and lenses, so that the spots of sensing light can be locked to the centers of the mirrors and lenses within an rms error tolerance of $0.08$ mm. It is not hard to show by simple geometric optics that such a requirement corresponds to a transverse deviation of the lenses' centers from the reference axis with a maximum variance of $\delta s^2\sim(0.1\,\mathrm{mm})^2$ and a maximum mirror angular deviation variance of $\delta\beta^2\sim(0.05\,\mu\mathrm{rad})^2$.

As the required bandwidth is slow and the actuator precision is well above the requirement, the residual movement is essentially determined by the sensing noise. 
% \rana{we should cite some papers on beam spot centroid estimation and use real numbers}
Determining the spot position is necessarily a centroid estimation problem, where the noise limit is given by \cite{jia2010minimum,thompson2002precise,ober2004localization},
\begin{equation}
    \frac{\delta x_s^2}{(w/2)^2}\sim \frac{1+\frac{1}{3}(a/w)^2}{N_{ph}},
\end{equation}
where $w$ is the spot size, $a$ is the pixel size, and $N_{ph}$ is the total number of received photons in one exposure. Experimental results with advanced CCD cameras and suitable analyzing algorithms demonstrate that the above noise limit can be achieved within a factor of three, provided that the required precision is no better than the nanometer scale \cite{smith2010fast}.  For a conservative estimate, assume that the camera has $P\approx1\,\mathrm{nW}$ with $T=100\,\mathrm{ms}$ exposure time ($N_{ph}\sim10^{9}$) and a negligible pixel size ($w\gg a$). Then, the sensing noise floor is estimated to be $10^{-5}w$, which is two orders of magnitude smaller than our requirement. Thus, we expect the current design of the ACS system to work well beyond our requirement with \yesun{a peak sensing laser power of $1\,\mathrm{W}$} and a scattering power of roughly $1\,\mu\mathrm{W}$ at each lens/mirror. Notably, the most analogous setup to ours involves the alignment of the input beam in LIGO, where the control system operates at normal pressure rather than under vacuum, achieves sub-micrometer alignment precision \cite{canuel2014sub}, and provides angular sensing precision at the sub-nanoradian level. The same system can also be adapted for our purposes in future upgrades, if necessary.
% Notably, in an ultrahigh vacuum, the test masses achieved an rms of $3$ nrad \cite{barsotti2010alignment}, surpassing our requirements by two orders of magnitude. Compared to the satellite system, the satellite system has shown a dynamic tracking error of less than $0.4\,\mu$rad \cite{yin2017satellite}, despite operating in a noisier environment.
\subsection{Section Length Stabilization}
\label{subsec:SLS}
SLS is designed to suppress the phase noise below $10\,\mathrm{Hz}$ as the VBG itself is expected to be stable at high frequencies. To reduce system complexity, we propose reusing the reflected (transmitted) AUX beam generated by imperfect AR (HR) coatings, which would otherwise contribute to signal beam loss, to implement the standard round-trip phase cancellation technique that is widely employed in fiber for long-distance frequency transfer \cite{foreman2007coherent}. The AUX beam is locked to an ultra-stable laser or atomic clock at emission to provide a stable absolute length reference for the SLS.

To separate the reflected beam from the main optical path of the co-propagating AUX beam, the lenses' surfaces can be engineered such that the first surface is flat and tilted by an angle of $0.02\sim0.05\,\mathrm{mrad}$ with respect to the reference optical axis, as shown in Fig.~\ref{fig:Reflectors} \footnote{The tilting will introduce a lateral shift of the transmitted beam that leads to a dispersion separation of the AUX beam (1600 nm) and the signal beam ($\sim$1550 nm). However, the separation induced by each lens is less than $1$ nm, and thus the effects are trivial. }. This tilting will shift the spot of the reflected beam away from the center of the previous node by $20\,\mathrm{cm}$. The required tilting alignment precision for each lens is only $\sim0.01\,{\mathrm{mrad}}$, corresponding to a transverse shift of $\sim4\mathrm{cm}$, which is readily achievable by monitoring the spot location using a fraction of the reflected beam, since we can tolerate a relatively large loss in the reflected beam. 

The reflected beam is extracted at the steering mirror node \footnote{In principle, the reflected beam can also be extracted at each lens or coupled into a stabilized fiber for routing back to the mirror node. We only aim to provide one possible  solution here, while the engineering team will decide on the best strategy upon deployment.}. Then, the beam from the closest lens is split into two, with one part being mixed with the transmitted beam from the lens and the other part being mixed with the reflected beam from the second closest lens using the heterodyne phase detector \cite{de2012high}. The SLS feedback signal, the length fluctuation between the two steering mirrors, is calculated using the beat signals from the local detectors along with the results from the neighboring steering mirror node. 

Finally, the feedback signal is low-pass filtered and used to drive the local mirror actuator to ensure that the mirror is following the movement of the previous one, thus actively stabilizing the section length below $10$ Hz.

The noise budget analysis (Subsec. \ref{subsec:SLSN}) suggests that such a simple SLS design is sufficient to achieve our performance claim. However, if a more complicated scheme becomes necessary, one option is to inject an independent, ultra-stable SLS laser with a carefully selected wavelength at each steering mirror by engineering the steering mirrors to have higher transmissibility at that wavelength. Then, the multi-color arm length stabilization scheme \cite{izumi2012multicolor} can be applied to lock the section length to the laser frequency using the PDH method \cite{drever1983laser}. Alternatively, a completely separate beam path could be used for the SLS sensing laser by attaching co-moving tiny lenses and mirrors to the main optical components of the VBG, at the expense of increased system complexity.

\yesun{\textbf{Gated AUX for weak-light operation.} The co-propagating $1600$-nm AUX field can generate anti-Stokes Raman photons in the transmissive fused silica \cite{stolen1973raman,brainis2007spontaneous,agrawal2019nonlinear}. For a $10^4$-km VBG, the accumulated silica path is $L_{\rm silica}\sim50\,\mathrm{m}$. Using the fused-silica Raman response at the roughly $6$-THz AUX--signal detuning, with representative parameters, gives rise to the transmission window flux as}
\begin{equation}
\yesun{R_{\rm Raman}^{\rm CW}\sim10^3\left(\frac{P_{\rm AUX,avg}}{1\,\mathrm{W}}\right)\left(\frac{L_{\rm silica}}{50\,\mathrm{m}}\right)\,\mathrm{s^{-1}}.}
\end{equation}
\yesun{This is safely subdominant for the $H=12$ benchmark ($\sim6\times10^4\,\mathrm{s^{-1}}$), but not appropriate for the faint $H=24$ case if the SLS is operating continuously. For weak-light operation, the AUX beam can instead be time-multiplexed with science detection. As an illustrative conservative operating point, $D_{\rm AUX}=10^{-4}$ gives a time-averaged Raman generation rate of $\sim0.1\,\mathrm{s^{-1}}$. Moreover, as the forward Raman photons are generated synchronously with the AUX beam, temporal detector gating excludes them from the science window, so the principal cost is only an $O(D_{\rm AUX})$ reduction in science observation duty cycle. The associated sensing penalty remains within the conservative SLS margin with the $1$-W interpreted as the peak power rather than the continuous one.}

\section{Details of Noise Budgets}
\label{sec: NoiseBudgets}
\begin{figure*}[htbp]
\centering
\includegraphics[width=\linewidth]{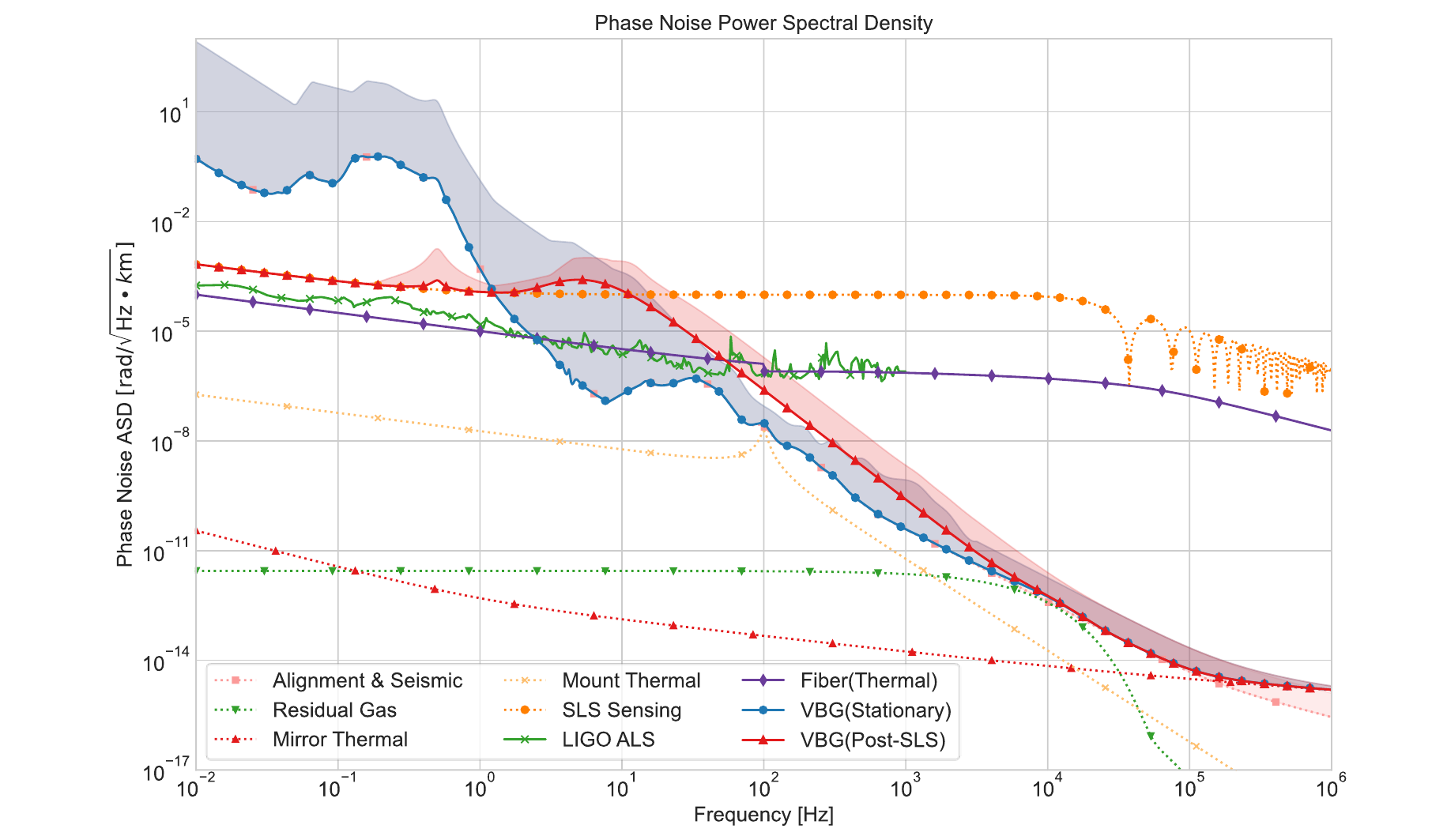}
\caption{Normalized phase noise amplitude spectral density of the VBG before and after SLS, optical fiber thermal noise \cite{bartolo2012thermal,wanser1992fundamental,duan2010intrinsic}, and the detailed noise budget. The upper edges of the shaded regions represent the corresponding noises (Alignment \& Seismic, VBG before and post SLS) under the high seismic input case.}
\label{fig:Phase_PSD_APP}
\end{figure*}

In this section, we provide a detailed description of noise budgets for estimating the phase noise power spectral density, as shown in Fig.~\ref{fig:Phase_PSD_APP}, utilizing a comparative and extrapolative method with the data and analysis from Advanced LIGO \cite{martynov2016sensitivity,buikema2020sensitivity,capote2025advanced}. 

\subsection{Seismic and Alignment Noise}
\label{subsec:seismicAndAlignment}

The seismic noise at the lens and mirror positions will introduce movement into the system and cause phase noise. Empirically, the phase noise is dominated by $k\delta z$, with $\delta z$ being the parallel movement. However, rotational and translational movement may also couple into the phase noise due to static misalignment. Here, we will present a theoretical derivation based on the Laguerre-Gaussian mode expansion, which we have used to derive the loss for cross-verification. Then, we estimate the post-isolated phase noise contribution to the final noise power spectrum based on measured and theoretical seismic movements.

\subsubsection{Lens Section}
\textbf{Parallel Movement.} The movement of the lens parallel to the laser beam will not significantly affect phase noise, as the light is merely transmitted through the lens. 
By expanding the perturbed light field in the Laguerre-Gaussian basis (LG) and collecting the imaginary part of the fundamental mode, the leading contribution from the defocus effect (Gouy phase) can be identified as,
\begin{equation}
    \delta\phi\sim  \frac{\delta z}{L_0},
\end{equation}
which is greatly suppressed by the section length $L_0$.

\textbf{Transversal Displacement.} Similarly, let $\delta s$ denote the offset of the beam and expand the light field after a perturbed lens in the original LG basis. One finds that the phase shift (piston shift) is of the second order of $\delta s$ and is estimated as
\begin{equation}
    \delta\phi\sim-k\frac{\delta s^2}{L_0}.
\end{equation}
To include its contribution to the noise power spectrum, we can replace one of the $\delta s$  with twice the expected static misalignment tolerance, $2\delta s_s\leq0.2\,\mathrm{mm}$, which corresponds to a coupling factor of $2k\frac{\delta s_s}{L_0}\sim5\times10^{-8}k$ that is expected to be negligible compared to the displacement noise introduced by the mirror. 

\textbf{Rotation.} One can expect that the phase shift is of second order in $\delta\beta_{s/p}$. Following a simple geometric argument, the leading contribution is estimated to be
\begin{equation}
    \delta\phi\approx \frac{k(n-1)d}{2n}\delta\beta_{s/p}^2,
\end{equation}
with $d\approx2\,\mathrm{cm}$ being the thickness of the lens and $n$ being the effective refractive index of the lens. Similarly, one can replace one of the $\delta\beta_{s/p}$ with twice the static tolerance value of $2\delta\beta_{s/p,s}\leq0.02\,\mathrm{mrad}$ to obtain a coupling factor of $\frac{k(n-1)d\delta\beta_{s/p,s}}{2n}\sim120\,\mu\mathrm{m/rad}$.

\subsubsection{Mirrors Section}
\textbf{Translation.} For simplicity, we can assume the incident angle to be $\ang{0}$ for an order-of-magnitude estimate. The parallel movement represents the strongest movement-to-phase coupling as
\begin{equation}
\delta\phi\sim2k\delta z.
\end{equation}
In contrast, the transversal movement ($\delta s$) does not impose any phase shift. The cross-coupling term, seen below, is greatly suppressed due to the high angular alignment control requirement for the mirror ($\sim0.05\,\mu \mathrm{rad}$).

\textbf{Rotation.}
Rotation itself only introduces pure loss, rather than a phase shift. However, it couples through the parallel movement as,
\begin{equation}
    \delta\phi\sim 2k\delta s\delta\beta,
\end{equation}
assuming the worst case scenario in which the transversal movement aligns with the rotation. With the same techniques, the coupling factor is estimated to be $200k\,\mu\mathrm{ m/rad}$. 

\subsubsection{Seismic Input}
\begin{figure*}[htbp]
\centering
\includegraphics[width=\linewidth]{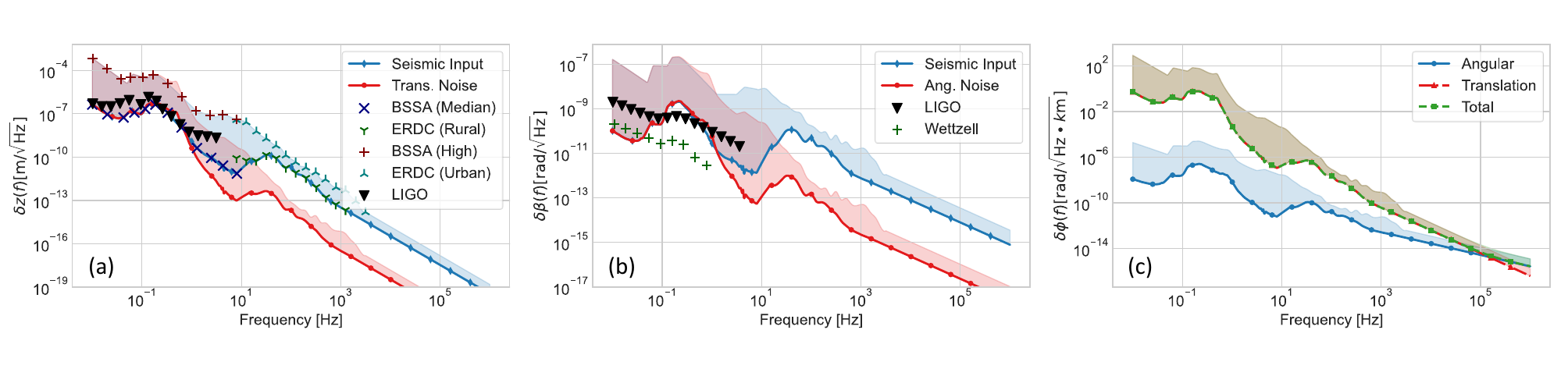}
\caption{The noise ASD of the seismic input, passive isolated platform and their contributions to the final PNASD. The upper edge of the corresponding shaded regions represents the noises under high seismic input. (a) Translation movement ASD. (b) Rotation movement ASD. (c) The expected contribution to PNASD.}
\label{fig:SeismicAAlignment} 
\end{figure*}

Generally, seismic-induced \textbf{translation} rolls off as $\delta z(f)\sim\frac{1}{f^2}$\footnote{This scaling assumption is pessimistic, as the system \yesun{is assumed to} be buried 2 meters under the ground \yesun{for the prototype deployment considered here}, which can result in exponential suppression of ground excited high-frequency seismic and acoustic noises due to dirt attenuation.}, while establishing several peaks and long-term drift at low frequencies \cite{hough2005search}. 
Therefore, to estimate the seismic-induced translation movement, we can use the low-frequency measured data from the BSSA survey~\cite{anthony2022seismic} and the high-frequency data of the urban and rural areas collected by ERDC-CRREL~\cite{albert2017acoustic}. Then, we extrapolate the spectrum to high frequencies using the aforementioned $\frac{1}{f^2}$ scaling, which is shown in Fig.~\ref{fig:SeismicAAlignment} (a) in comparison to the LIGO site data~\cite{matichard2015seismic,ross2020towards}. Specifically, the median noise data from BSSA is combined with the rural area data to define the baseline phase stability of the VBG. This baseline is then used for all subsequent performance estimates, under the assumption that the VBG will be constructed and routed through relatively quiet regions where the LIGO data is representative of this noise level. For completeness, we also combine the high-noise model (NHNM)~\cite{peterson1993observations} with the urban residential region data to provide performance estimates in noisy environments, in which the noise level is roughly two to three orders of magnitude larger in the $3-30$ Hz band. 
% \rana{Fix this section to use the USGS survey instead of LIGO.}

As for the \textbf{rotation}, it can be estimated from the translational seismic movement~\cite{brotzer2024characterizing}. 
% \rana{since we are burying the system, we may neglect the wind induced motion}

% \rana{also add a reference to some measured tilt / rotation data / rotational seismology conferences}
\nocite{sanchez2011energy}
By assuming a seismic wave propagating with displacement $z(x,t)$, we can calculate the induced tilting as
\begin{equation}
    \delta\beta(x,t)=\frac{\partial z(x,t)}{\partial x}\approx kz.
\end{equation}
Therefore, the PSD of the tilting can be calculated from the PSD of the seismic-induced translation as
\begin{equation}
    S_\beta(f)\approx k^2S_z(f)\approx \left(\frac{2\pi f}{c_s}\right)^2S_z(f),
\end{equation}
where $c_s$ is the speed of the seismic wave. At low frequencies, the seismic wave speed is determined by the hard bedrock and can reach $2\sim3$ km/s  \cite{shearer2019introduction,soomro2016phase}, whereas at higher frequencies, the wave speed is significantly slower because the waves are trapped by the shallow soil layer, with typical values ranging from $\sim100$ m/s to $\sim800$ m/s. We take a typical value of $c_{s,\mathrm{shear}}\sim300$ m/s \cite{williams1997high} \footnote{The near-surface seismic/acoustic angular noises are mainly attributed to Rayleigh waves, shear waves, and Love waves~\cite{sanchez2011energy}, among which the Rayleigh wave speed is the slowest and roughly equals $c_{s,\mathrm{Rayleigh}}\approx0.9c_{s,\mathrm{shear}}\approx270\,\mathrm{m/s}$~\cite{stein2009introduction,shearer2019introduction}.}. The tilting input data used in the following calculation, obtained from the theoretical models described above, are plotted in Fig.~\ref{fig:SeismicAAlignment} (b) and compared with the LIGO data and the Wettzell measurement data~\cite{ross2020towards,schreiber2006ring}. The LIGO data is found to be at a similar level to the low noise model and is approximately an order of magnitude larger than the Wettzell measurements, which is likely due to extra tilting induced by the wind. Because the system will be buried, we may neglect the wind induced motion, and thus our model provides a conservative estimate. In fact, the tilting induced phase noise is roughly 3 to 4 orders of magnitude smaller than the translational induced phase noise and never exceeds $10^{-5}\,\mathrm{rad}$ rms, even in the high noise case, which means its contribution is trivial and can be safely neglected.

As seen later, the performance of the VBG is acceptable even in a high-noise and unlikely environment after the engagement of the SLS system, which demonstrates the robustness of the control system design. However, the rms angular motion after the PIP system, integrated across the spectrum, in the high-noise case is approximately $0.1\,\mu\mathrm{rad}$, which is two times larger than our assumed maximum mirror angular deviation. As a result, the ACS system must operate with a faster bandwidth ($\geq 0.5\,\mathrm{Hz}$) in high-noise environments to ensure stable system performance. Alternatively, a more expensive and advanced isolation platform \cite{abbott2002seismic,giaime2001advanced} can be employed in particularly noisy areas when necessary, depending on practical engineering considerations.

\subsection{Residual Gas Noise}
\label{subsec:residualGas}
The default residual gas pressure in the VBG is $\sim$1 Pascal, which is approximately seven orders of magnitude higher than that of LIGO.
In addition to the finite transmissibility of acoustic noise, the statistical fluctuation in the effective index of refraction of the low pressure gas introduces a phase noise floor for the VBG that extends up to tens of kilohertz.

Under thermal equilibrium, interactions between gas molecules and the propagating light introduce scattering phase noise, which is dominated by forward scattering and can be estimated by \cite{weiss1989mitscattering}
\begin{equation}
    \delta\phi=\frac{(2\pi)^2\alpha\sqrt{\rho_{\#} L_0}}{\lambda\sqrt{w_0v_0}}e^{-\sqrt{2\pi} \frac{fw_0}{v_0}},
\end{equation}
where $\alpha$ is the molecular polarizability, $\rho_{\#}$ is the particle number density, $v_0$ is the thermal velocity of the gas, and $w_0$ is the average radius of the beam. And for our configuration, assuming $N_2$ is dominating, we have
\begin{equation}
    \delta \phi\sim10^{-11}\times\left(\frac{1550\,\mathrm{nm}}{\lambda}\right)^{\frac{5}{4}}\sqrt{\frac{P}{1\mathrm{Pa}}}e^{-\frac{f}{5\,\mathrm{kHz}}}\,\mathrm{rad}/\sqrt{\mathrm{Hz}}.
\end{equation}
In addition to nitrogen, hydrogen and helium should also be considered, as they have a cutoff frequency an order of magnitude larger than that of $N_2$ and may therefore dominate the residual gas scattering noise above $100\,\mathrm{kHz}$. Their contributions have also been included in the final noise budget, assuming volume ratios of $1\,\mathrm{ppm}$ and $10\,\mathrm{ppm}$, respectively.

\subsection{Thermal Noise}
\label{subsec:thermalNoise}

The thermal noise mainly comes from the rigid  mount holding the steering mirrors and the internal thermal noise of the mirrors themselves. 

% \rana{for the rigid mount, we can assume that it has a eigenfrequency of 100 Hz and a mechanical Q of 10, with a frequency independent loss angle.}

For the \textbf{mounts}, the thermal noise can be estimated by applying the fluctuation-dissipation theorem for the damped harmonic oscillator model as \cite{cumming2012design},
\begin{equation}
    \delta\phi=\sqrt{\frac{4k_BT}{2\pi m f\lambda^2}\left(\frac{f_0^2/Q}{f_0^4/Q^2+(f_0^2-f^2)^2}{}\right)},
\end{equation}
where $f_0\sim100\,\mathrm{Hz}$, $Q\sim10$, and $m\approx2\,\mathrm{kg}$ are the typical eigenfrequency, quality factor, and total mass (including mirrors/lenses), respectively, for a conservative estimate. 

As for the \textbf{mirror} noise, it may include Brownian noise and thermo-optic noise for both the substrate and the coating. Assuming that we use the same materials for the substrate and coating as LIGO, the thermal noise for the mirror will be dominated by the coating Brownian noise, which can be calculated using the fluctuation-dissipation theorem as \cite{harry2002thermal}
\begin{equation}
    \delta\phi=\frac{2\pi}{\lambda} \sqrt{\frac{2k_BTd_c}{\pi^{2}fw^2Y}\left(\frac{Y'}{Y}\phi_{\parallel}+\frac{Y}{Y'}\phi_\bot\right)},
\end{equation}
where $w$ is the beam radius at the mirror, $\phi_{\parallel}$ and $\phi_\bot$ are the parallel and orthogonal loss angles of the coating, $Y$ and $Y'$ are the Young's moduli for the substrate and the coating, and $d_c$ is the thickness of the coating. 
Using the values from LIGO, an order of magnitude estimate is found as \cite{amato2021optical} 

\begin{equation}
\label{eq:brownianNoise}
    \delta\phi_{BN,c}\sim 1\times10^{-12}\left(\frac{1550\,\mathrm{nm}}{\lambda}\right)^{3/2}\left(\frac{1\,\mathrm{Hz}}{f}\right)^{0.45} \mathrm{rad}/\sqrt{\mathrm{Hz}}.
\end{equation}

For the \textbf{lenses}, the thermal noise budgets for transmissive optics differ from those for reflective ones. Providing an exact and detailed estimate of the thermal noise is impossible due to a lack of knowledge about the material properties. Here, we only estimate the bulk transmissive noise for the substrates, which is likely the dominant noise, and adapt the values from the reflective optics case to serve as our order of magnitude estimate based on the following reasons \cite{somiya2009thermal}:
\begin{enumerate}
    \item For the Brownian noises, their values are largely determined by the loss angles, which are unknown unless a real measurement is performed. However, the mechanism of Brownian noise does not depend on whether the light is reflected or transmitted; instead, it merely depends on the material structure and properties. For coating, the noise scales as $\frac{d_c}{w^2}$, with $d_c$ being the thickness of the coating and $w$ being the waist of the beam. Considering that the AR coating is roughly an order of magnitude thinner than the HR coating being used in LIGO, it is very likely that the coating Brownian noises will be smaller in the AR case.
    
    \item For the thermal-optics noise of the coating, the PSD also scales linearly with the thickness. Therefore, if the same material is being used, it is reasonable to believe that the thermal-optic noise will be smaller in the AR case.
\end{enumerate}

For the lens' substrate, the bulk transmission thermal-optic noise can be calculated as \cite{dwyer2014radiative,benthem2009thermorefractive}
\begin{equation}
    S_x^{TO}(f)=\frac{16k_B\kappa T^2\beta^2_{eff}d}{\pi C^2\rho^2w^4\omega^2},
\end{equation}
where $\beta_{eff}=\frac{dn}{dT}+\alpha(n-1)$ is the effective thermal response, $d\sim2$cm is the thickness of the substrate, and other parameters are defined as before but for the substrate. Then, the thermal-optic noise for the substrate can be numerically estimated as
\begin{equation}
    \delta\phi_{TO,s}(f)\sim2\times10^{-12}\left(\frac{1\,\mathrm{Hz}}{f}\right)\left(\frac{1550\,\mathrm{nm}}{\lambda}\right)^2.
\end{equation}

\begin{figure*}[htbp]
\centering
\includegraphics[width=\linewidth]{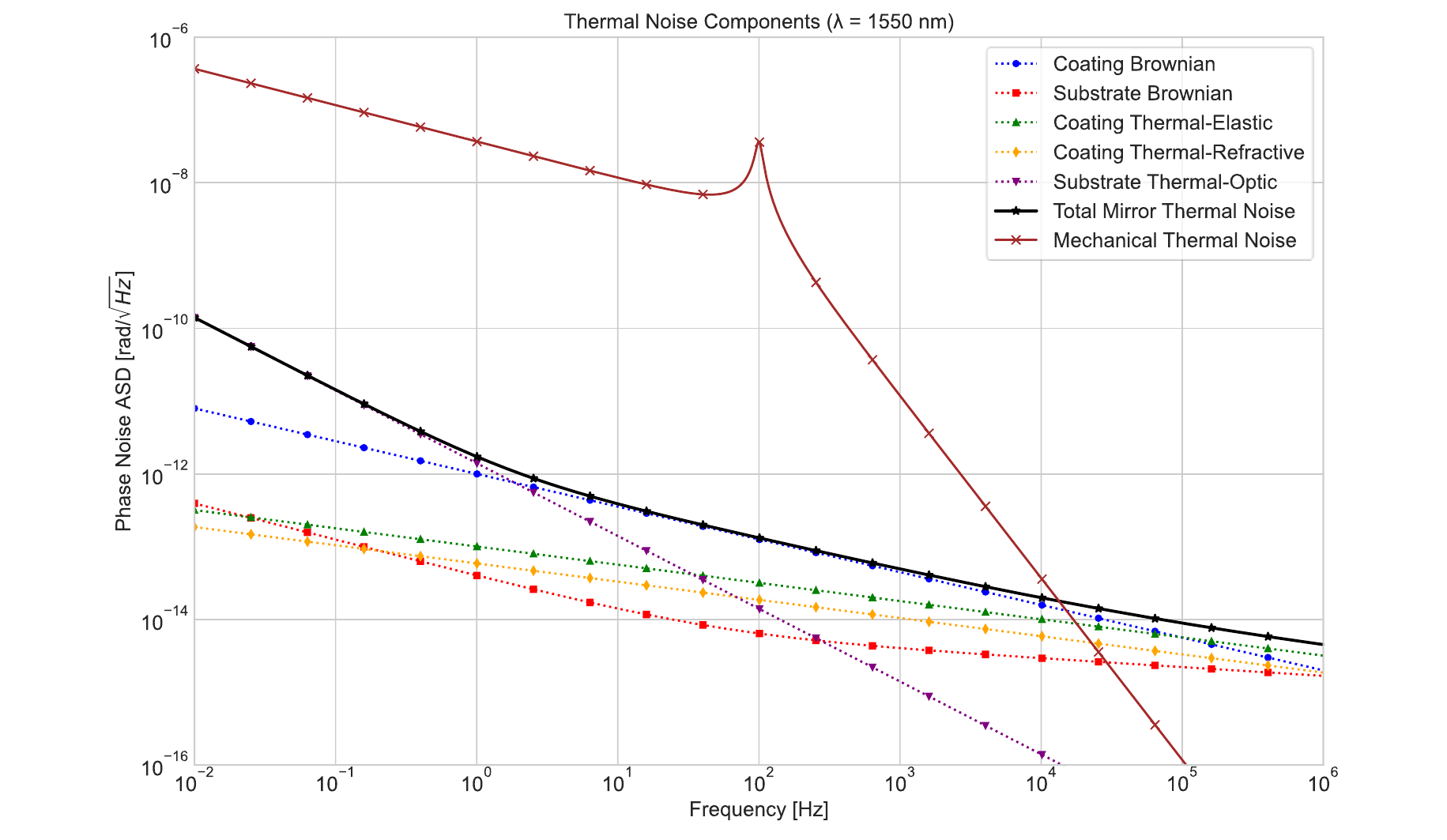}
\caption{\label{fig:MTN_PSD} Thermal noise budget order of magnitude estimate at the center carrier wavelength $\lambda=1550$ nm.}
\end{figure*}

The thermal noise budget, including all relevant components, is shown in Fig.~\ref{fig:MTN_PSD}. We observe that the substrate thermal-optic noise dominates at low frequencies, while the coating Brownian noise dominates at high frequencies. 

\subsection{Birefringence Noise}
\label{sec:polarizationNoise}
Birefringence in both the substrate and the coating not only degrades polarization stability but can also couple to phase noise when the orientation of the optic axis of an optical component is perturbed relative to the polarization of the signal light. The analysis from LIGO recently addresses this issue in a coherent way \cite{michimura2024effects}, since they have a cavity with light bouncing back and forth. We will follow a similar treatment, but we will take into account the incoherent nature of perturbation in the VBG system. 
The light transmitted in the VBG system can be decomposed into parallel and orthogonal polarizations with respect to the selected axis (e.g., the crystal axis of the first input lens), described using the Jones matrix as below
\begin{equation}
    \vec{E}=\begin{bmatrix}
    E_{\parallel}
    \\E_{\bot}
    \end{bmatrix}.
\end{equation}
Then, when the ordinary axis of the mirror/lens (including the coating) is aligned with $E_\parallel$, the transfer matrix can be represented as
\begin{equation}
    T_{m/l}=t_0\begin{pmatrix}
        1&0\\
        0&(1-t)\exp(i\phi_{m/l})
    \end{pmatrix}.
\end{equation}
In this expression, $t_0$ is the overall transmission and $t$ is the differential transmission between the two orthogonal polarization modes, which must be smaller than $100$ ppm to satisfy both the loss requirement and the limits imposed by available mirror and lens coatings. In addition, $\phi_{\mathrm{m/l}}$ is the birefringence phase shift. For a lens, it is dominated by the substrate phase shift calculated as
\begin{equation}
    \phi_{l}=kd\delta n\sim0.01\,\mathrm{rad},
\end{equation}
with $\delta n\sim10^{-7}$ for silica substrate and $d\sim0.02$ m for the lens used in the VBG. For mirrors, the birefringent phase shift is determined by the coating and is estimated to be $\phi_{m}\sim10^{-4}$ rad for the $\mathrm{Ta_2O_5/SiO_2}$ coating that we plan to use \cite{bielsa2009birefringence,kitagawa1985form}. When the optic axis of the mirror or lens is rotated by an angle $-\theta$, the transform matrix becomes
\begin{equation}
    T_{m/l}(\theta)=T_{m/l}R(\theta),\,\mathrm{with}\,R(\theta)=\begin{pmatrix}
        \cos\theta&-\sin\theta\\
        \sin\theta&\cos\theta
    \end{pmatrix},
\end{equation}
\yesun{where the Jones vector is expressed in the local polarization basis of the current optical element.} Suppose there is a perturbation $\delta\theta$. Then, to the first order, one obtains that
\begin{equation}
\label{eq:transferT}
\begin{aligned}
    T_{m/l}(\theta+\delta\theta)=&t_0\begin{pmatrix}
        1&0\\
        0&e^{i\phi_{m/l}}
    \end{pmatrix}R(\theta)\\
    -&t_0\begin{pmatrix}
        0&0\\
        0&e^{i\phi_{m/l}}
    \end{pmatrix}R(\theta)t\\
    -&t_0\begin{pmatrix}
        1&0\\
        0&e^{i\phi_{m/l}}
    \end{pmatrix}R(\theta)J\delta\theta,
\end{aligned}
\end{equation}
where 
$
    J=\begin{pmatrix}
        0&1\\
        -1&0
    \end{pmatrix}.
$
The static, unperturbed \yesun{unitary} transform can be corrected at the end by applying the inverse of the transform as
\begin{equation}
    T^r=R^\dagger(\theta)\begin{pmatrix}
        1&0\\
        0&e^{-i\phi_{m/l}}
    \end{pmatrix}.
\end{equation}
For \textbf{non-polarization encoding}, the first-order effect is the conversion of one polarization mode into its orthogonal counterpart, with a conversion factor $t_0e^{i\phi}\delta\theta$. If one polarization mode is taken as the fundamental mode, this results in a loss of $\delta\theta_{\rm rms}^2\sim1\,\mathrm{ppm}$ \yesun{per section} given $\delta\theta_{rms}\sim1\,\mathrm{mrad}$ \cite{hirose2020characterization}, which can be safely neglected, since the transverse mode conversion loss induced by misalignment is larger. The associated phase noise is only a second-order effect, with a factor $\sim i\phi_{m/l}\delta\theta_1\delta\theta_2$ due to the back-conversion from the orthogonal polarization mode. This yields a phase noise contribution of $\delta\phi\sim\phi_{m/l}\delta\theta_{rms}\delta\theta\sim10^{-5}\delta\theta$, which is negligible and dominated by other rotation misalignment coupling mechanisms. An alternative interpretation is that the light in the orthogonal mode is uncorrelated with the light in the parallel mode, so the net
effect is a reduction in the visibility of the fundamental mode. Provided that the power in the orthogonal mode is not significant (propagating through fewer than $10^6$ sections), this effect can be safely neglected.

For \textbf{polarization encoding}, birefringence can lead to phase noise between the two encoded states $|0\rangle$ and $|1\rangle$. The difference in transmission rate, $t$, may also introduce amplitude noise. However, these static and known
\yesun{unitary} birefringence effects act as a unitary for which we can
compensate, \yesun{while the polarization-dependent transmission imbalance must be treated separately.} The naive solution is to alternate the orientation of the optic axis of sequential lenses between $\ang{0}$ and $\ang{90}$, which allows for the cancelation of the transmission difference and phase shift difference between the two polarized modes. However, the sequential unknown misalignment angle $\delta\theta$ is randomized and may still introduce phase noise and amplitude noise, which is estimated here. We assume the phase difference between the two modes is $\phi_0$. Then, the input Jones matrix can be written as
\begin{equation}
    |\psi\rangle\sim\vec{E}=\begin{bmatrix}
    E_{\parallel}
    \\E_{\bot}e^{i\phi_0}
    \end{bmatrix}.
\end{equation}

We can derive the transform matrix for two consecutive lenses/mirrors as
\begin{equation}
    \begin{aligned}
    T_{1,m/l}\left(\frac{\pi}{2}+\delta\theta_2\right)&T_{2,m/l}(\delta\theta_1)\approx t_0^2\times\{(1-\bar{t})I\\
    &-\delta t_1\begin{pmatrix}
        0&0\\
        0&1
    \end{pmatrix}
    -\delta t_2\begin{pmatrix}
        1&0\\
        0&0
    \end{pmatrix}\\
    &+i(\delta\phi_1-\delta\phi_2)\begin{pmatrix}
        0&0\\
        0&1
    \end{pmatrix}\\
    &-\delta\theta_1\begin{pmatrix}
        0&1\\
        -1&0
    \end{pmatrix}-\delta\theta_2\begin{pmatrix}
        0&e^{i\bar\phi}\\
        -e^{-i\bar{\phi}}&0
    \end{pmatrix}
   \},
\end{aligned}
\end{equation}
where we set $ t_i=\bar{t}+\delta t_i,\, \phi_{m/l,i}=\bar\phi+\delta\phi_i$ for each lens/mirror to separate the relatively small statistical differences of each lens/mirror. \yesun{Again, the identity Kraus operator and the normalization factors are omitted here} \cite{kraus1983states}.

The second and third terms can be decomposed \yesun{together} into an overall attenuation channel and a \yesun{differential-amplitude} channel, while the fourth term describes dephasing due to inhomogeneous phase fluctuation, whose contribution has already been calculated in the thermal-optic noise section. The final two terms are directly associated with misalignment effects and are expected to provide the dominant contribution: the first term corresponds to a \yesun{Pauli-$Y$-type polarization error}, while the second term represents a \yesun{phase-rotated Pauli-$Y$ error induced by the birefringent phase}. By averaging over these effects, one can derive \yesun{the leading nontrivial Kraus operators} for each lens or mirror as follows,
\begin{equation}
\label{eq:Kraus}
    \frac{\delta\theta}{\sqrt{2}}\left\{\begin{pmatrix}
        0&1\\
        -1&0
    \end{pmatrix},\begin{pmatrix}
        0&e^{i\bar\phi}\\
        -e^{-i\bar{\phi}}&0
    \end{pmatrix}\right\}.
\end{equation}  

\subsection{Acoustic Wave Noise}
\label{sec:AcousticN}
Acoustic noise will introduce phase noise in the VBG.  Because the residual gas pressure is roughly 1\,Pa, there is finite acoustic transmission from the tube walls to fluctuations in the effective refractive index of the gas in the tube. Although the ambient acoustic noise levels vary substantially in the continental US, we can adopt a representative median level for our noise budgeting purposes, assuming that the majority of the VBG will be buried $\sim$2\,m underground.
% , we can also expect some acoustic attenuation in the soil (roughly 0.5\,dB/cm/kHz \cite{Oelze}).

In UHV systems, such as LIGO, acoustic waves are heavily attenuated. In contrast, for our mild vacuum conditions considered here, we must estimate the acoustic isolation quantitatively. The following analysis is performed with the simplifying assumption that $N_2$ dominates the residual gas composition.

\textbf{Attenuation.} In the continuum regime, the attenuation of the acoustic wave in gas is given by the classical Stokes-Kirchhoff formula \cite{kinsler2000fundamentals},
\begin{equation}
    \alpha_{SK}=\frac{\omega^2}{2\rho_0c_s^3}\delta_{eff},
\end{equation}
where $\yesun{\rho_0=PM/(RT)}$ is the gas density, with $M_{\yesun{\mathrm{N_2}}}\simeq28\,\mathrm{g/mol}$ being the molar mass of nitrogen, and $c_s=\sqrt{\gamma RT/M}$ is the adiabatic acoustic wave speed, with $\gamma=1.4$ being the adiabatic index. Here,
\begin{equation}
    \delta_{eff}=\frac{4}{3}\eta+\eta_B+(\gamma-1)\frac{\eta}{\mathrm{Pr}}
\end{equation}
is the effective dissipation parameter that includes the contributions from shear viscosity ($\eta\approx1.78\times10^{-5}\,\mathrm{Pa\cdot s}$), bulk viscosity ($\eta_B\approx0.8\eta$), and thermal conductivity, with $\mathrm{Pr}\approx0.72$ being the Prandtl number.

However, the continuum assumption may break down at medium vacuum for high frequency acoustic waves ($f\yesun{\gtrsim}\frac{\rho_0c_s^2}{2\pi\delta_{eff}}\sim5\,\mathrm{kHz}$). In this region, the attenuation per wavelength ($\alpha_s\lambda_s$) saturates as predicted by the super-Burnett theory ~\cite{greenspan1956propagation}, and we can use the experimentally measured value for normal air as $\alpha_{M}\lambda_s\sim 2.5$~\cite{meyer1957schallausbreitung}. Moreover, to bridge the limit and the classical Stokes-Kirchhoff prediction, we use a simple and empirical saturation function as
\begin{equation}
    \alpha_s=\frac{\alpha_{SK}}{1+\alpha_{SK}/\alpha_M}.
\end{equation}

\textbf{Phase Shift.} At low pressure, the refractive index of the residual gas scales linearly with pressure $P$ as \cite{owens1967optical}
\begin{equation}
    n(P)=1+\alpha P,
\end{equation}
where $\alpha\approx3\times10^{-9}$/Pa. Let $\delta P$ denote the pressure vibration introduced by the acoustic wave, then the refractive index changes as 
\begin{equation}
    \delta n=\alpha \delta P .
\end{equation}
Thus, the effective optical path length gain of the fundamental mode is given by
\begin{equation}
    \delta z_{\mathrm{eff}}\approx L_{\mathrm{eff}}\alpha\delta P,
\end{equation}
with $L_{\mathrm{eff}}$ denoting the effective length influenced. For a conservative upper bound, we assume that the coherence length is shorter than both the attenuation length $l_s$ and the section length $L_0$, such that 
$L_{\mathrm{eff}}$ is given by a Poisson process,
\begin{equation}
     L_{\mathrm{eff}}\sim l_c\sqrt{\frac{L_0}{l_c}}=\sqrt{L_0l_c},\,l_c=\mathrm{min}(l_s,L_0).
\end{equation}

The acoustic wave pressure, $\delta P$, excited by the tube oscillation (the seismic/acoustic translational input $z_t(f)$) can be calculated by
\begin{equation}
    \delta P(f)=Z_{\mathrm{air}}v(f)=\rho_02\pi c_sf z_t(f).
\end{equation}
For a conservative estimate, it is assumed that the tube is roughly $2$ meters beneath the ground, and the surrounding soil naturally acts as passive isolation, resulting in an attenuation calculated as \cite{stein2009introduction}
\begin{equation}
    \frac{z_t(f)}{z(f)}\equiv T_{\mathrm{dirt}}=\exp\left(-\frac{\pi f}{Q_{\mathrm{dirt}}V_s}D\right),
\end{equation}
with $D\approx 2\,m$ being the depth, $Q_{\mathrm{dirt}}\approx10$ denoting the typical quality factor of the soil \cite{kramer2024geotechnical}, and $V_s\sim300\,\mathrm{m/s}$ representing the typical shear wave speed, as mentioned earlier \footnote{In fact, the actual effective attenuation will be larger as Rayleigh waves dominate the noise power near the surface, which suffers from strong geometric decay.}. Finally, it yields a total coupling coefficient of
\begin{equation}
    g_{aw}=\yesun{\alpha}\rho_02\pi c_sf\sqrt{L_0l_c}\exp\left(-\frac{\pi f}{Q_{\mathrm{dirt}}V_s}D\right).
\end{equation}

The boundary excitation represents another coupling mechanism for the ground translational movement input $z(f)$. The coupling factor given by the boundary excitation grows linearly but is also exponentially suppressed by attenuation at high frequency, resulting in a maximum optical path length coupling factor of roughly $10^{-5}$. When one compares it to the maximum suppression factor given by PIP, which is no more than $10^{-3}$, we expect that this effect is dominated by the seismic and alignment noise.

\textbf{Mode Loss.} The acoustic wave may also introduce mode loss, which can be approximated using the Jacobi-Anger approximation as \cite{goodman2005introduction}
\begin{equation}
    l_{ac}\sim1-J_0^2(m)=\frac{m^2}{2},
\end{equation}
where $m=k\delta z_{\mathrm{eff,rms}}$ is the modulation depth. For boundary excitation, this effect is not a concern since the associated loss is dominated by the misalignment.

\subsection{SLS Noise}
\label{subsec:SLSN}
The SLS system will lock the section length between neighboring mirrors (3-section) to the linewidth of the AUX laser, which helps actively suppress length fluctuations at low frequencies. However, the rough movement of the VBG itself is relatively stable at high frequencies, where the sensor noise will probably dominate and need to be filtered out for optimized performance. The actuator noise is usually negligible compared to the sensing noise~\cite{martynov2016sensitivity}. Therefore, the PSD after stabilization can be written as,
\begin{equation}
\yesun{S_{\phi,\mathrm{res}}}=S_{\mathrm{SLS}}\mathrm{|LPF|}^2+S_\phi\mathrm{|HPF|}^2,
\end{equation}
where $\mathrm{LPF}\text{ and}\,\mathrm{HPF}$ are the low-pass and high-pass filter functions that satisfy the \yesun{fundamental feedback condition} such that $\mathrm{LPF}+\mathrm{HPF}=1$. $S_{\mathrm{SLS}}$ denotes the total sensing noise of the SLS scheme, and $S_\phi$ denotes the PSD of the open-loop phase noise. 

There are two dominant sources of the SLS sensing noise, namely the detection shot noise and the laser frequency noise, the former of which is given by \cite{francis2014weak},
\begin{equation}
\label{eq:sensingNoise}
    S_{\mathrm{SLS,Shot}}=\frac{hc}{\lambda\eta P_s},
\end{equation}
where $P_s\approx1\mu\mathrm{W}$ is the receiving power \footnote{Roughly $10$ ppm of the \yesun{$1$-W peak AUX beam} will be reflected, and we assume that $10$ percent of its power will be collected as a conservative estimate.} and $\eta\approx0.8$ is the typical detection efficiency. Estimating the frequency noise is somewhat complicated, since only two ultra-stable lasers are used for the entire system, causing the section noise to add coherently at low frequencies. For round trip detection, the average section noise is upper-bounded as
\begin{equation}
\label{eq:freqN}
    S_{\mathrm{SLS,Freq}}\approx \yesun{N_{\rm coh}}S_\nu\frac{4\sin^2(2\pi f \tau)}{f^2},
\end{equation}
where $\tau=\frac{L_0}{c}$ is the one-way delay and  $S_\nu\approx (6.4\times\frac{10^{-5}}{f}+1.4\times10^{-4})\,\mathrm{Hz^2/Hz}$ is the frequency instability of the ultra-stable laser~\cite{matei20171}. \yesun{Here, $N_{\rm coh}=10^4$ is a conservative coherent-enhancement factor for $\sim10^4\,\rm km$ scaling.} The total SLS noise is also plotted in Fig.~\ref{fig:Phase_PSD_APP} in comparison to the LIGO ALS residual movement data~\cite{cahillane2021controlling}. The latter is roughly an order of magnitude smaller than our theoretically estimated sensing noise floor, and thus we believe the current estimate is achievable.

To obtain the precise shape of the filter functions, the usual approach is to model the entire feedback control system and optimize the transfer function to achieve the lowest residual noise. However, we can approximate it with the following simple 6th-order LPF,
\begin{equation}
    \mathrm{LPF}=\frac{\tbinom{6}{3}s^3+\tbinom{6}{2}s^2+\tbinom{6}{1}s+1}{(1+s)^6},
\end{equation}
 where $s\equiv i\frac{f}{f_c}$ with cut-off frequency $f_c\sim5\,\mathrm{Hz}$, which can be optimized as long as $f_c\ll\frac{1}{2\tau}\sim10^4\,\mathrm{Hz}$ \footnote{For the case of high input noise, $f_c$ is chosen to be around $10$ Hz to further suppress the low frequency noise. The actual loop delay is mainly limited by the actuator bandwidth, rather than the signal transmission delay.}. In terms of ASD, such a filter provides $f^4$ suppression of the open-loop phase noise at low frequencies and a $1/f^3$ roll-off for the sensing noise.

The post-SLS phase noises of the VBG are plotted as the red solid curve with triangle markers in Fig.~\ref{fig:Phase_PSD_APP}, where the upper boundary of the red shaded area indicates the noise in the high seismic input case. Although the seismic input significantly increases in this case, the normalized residual rms noise after SLS is roughly at the level of $\sim5\,\mathrm{mrad}/\yesun{\sqrt{\mathrm{km}}}$, representing a degradation of less than one order of magnitude and indicating satisfactory interference stability even in relatively noisy environments.

 We remark that for applications that do not require real-time signal processing, feedback control is not necessary, and an offline post-subtracting method can be applied using the sensing data from the SLS system. This approach can result in lower noise, as the filter shape can be chosen freely without the limits of causality or stability requirements (e.g., a nearly perfect Heaviside Function).

\section{Performance of Memory Configuration}
\label{sec:VBGQM}
\begin{figure*}[htbp]
\centering
\includegraphics[width=\linewidth]{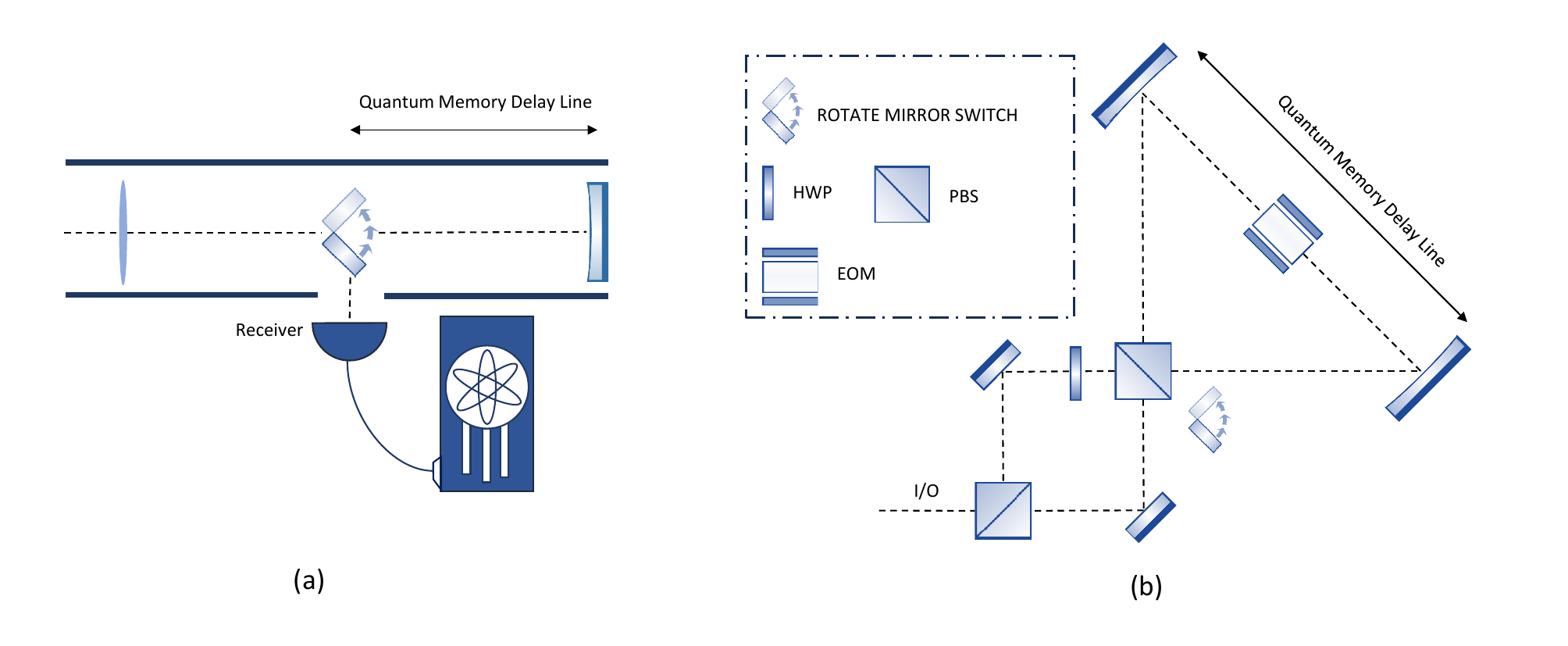}
\caption{\label{fig:VBGQM} A simple illustration of the design options for the VBGQM.  (a) A simple design utilizes a mechanical rotating mirror to perform the I/O operation to the VBGQM. The memory delay line is a section of the VBG. (b) A more complicated design with either an electro-optic modulator (EOM) or mechanical rotating mirror to perform the I/O operation. \yesun{The optional EOM/PBS route permits fast switching with an order-of-magnitude loss of $\sim1\%$ per switching cycle \cite{kaneda2019high}.} The memory delay line is also a section of the VBG, but with lenses on both sides replaced by mirrors.}
\end{figure*}

It is not surprising that the VBG can be considered as a very large, high-Q optical cavity since its initial motivation, LIGO, is itself such a system. The idea of employing an optical cavity as quantum memory has long been investigated \cite{leung2006quantum,pittman2002cyclical}, the principle of which is to use the techniques from building a Q-Switch cavity \cite{frungel2014optical,diels2006ultrashort} 
to implement the read and write operations. Employing sections of the VBG as quantum memory (VBGQM) will be particularly convenient and suitable for quantum communication protocols that require memory and are executed on the VBG. VBGQM has many advantages: a very broad bandwidth, a sufficiently long storage time, room temperature operation, and no need for conversion into matter qubits. Some particular designs for VBGQM are shown in Fig.~\ref{fig:VBGQM}. \yesun{Note that these designs are not optimal and we seek to provide a conservative stress test of the VBGQM delay-line memory.}

The performance of VBGQM can also be characterized by the attenuation and dephasing of the optical quantum state after a storage time of $T$. The decay of the probability of successful retrieval of the quantum state can be written as
\begin{equation}
    P_{\mathrm{success}}=\yesun{10^{-\alpha_{\mathrm{tot}} cT/10},}
\end{equation}
with $\alpha_{\rm tot}$ being the attenuation factor. For the memory configuration, the beam is reflected using a steering mirror. Assuming a mechanical switch is used, the angle alignment is essential and dominates the attenuation factor, as
\begin{align}
    \alpha_{\mathrm{tot}}&\approx-\frac{10}{L_0}\mathrm{log}_{10}(1-2k^2\omega_0^2\delta\beta^2)\nonumber\\
    &\sim1\times10^{-4}\,\mathrm{dB/km}\sim0.03\mathrm{dB/ms},
\end{align}
for $\delta\beta\sim0.05\,\mu$rad and $L_0=4$ km.
For an EOM/PBS switch, the loss is dominated by these optical elements, which will introduce a loss at the \yesun{$\sim1\%$} level ~\cite{kaneda2019high}, corresponding to an effective attenuation of \yesun{$\sim1 \times 10^{-2}$} dB/km.

\begin{figure}[htbp]
\centering
\includegraphics[width=\linewidth]{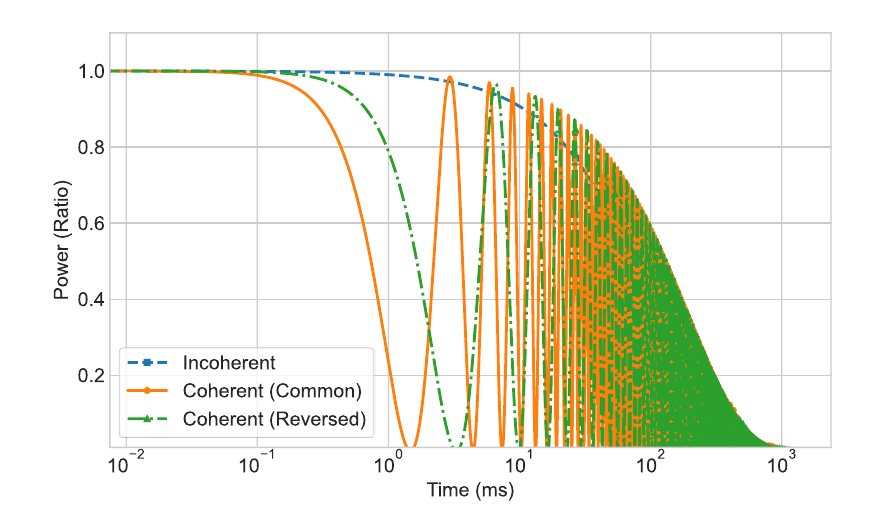}
\caption{\label{fig:VBGQM_Attenuation} Residual power ratio in the fundamental mode versus storage time of VBGQM, assuming coherent effects from common static tilting and reversed static tilting of the two mirrors by an angle $\delta\beta$, and compared with the stochastic scheme.}
\end{figure}

However, $\yesun{3\,\mathrm{dB}/{\alpha_{\mathrm{tot}}c}}\sim100\,\mathrm{ms}$ is not the actual memory lifetime of VBGQM, from which one may expect the normal VBG attenuation analysis. In the VBGQM configuration, the static misalignment of the two mirrors can give rise to coherent effects since the pulse is reflected multiple times within the same section, thereby violating the stochastic accumulation assumption. In this case, it is necessary to consider coupling to the neighboring mode $(0,\pm1)$ and to represent the beam as a vector that includes the information for both mode spaces as
\begin{equation}
    \vec{E}=\begin{pmatrix}
        E_{00}\\E_{0|\pm1|}
    \end{pmatrix}.
\end{equation}
Under this representation, the truncated conversion matrix, up to the second order of the mirror $M_i$ misalignment angle $\delta\beta$, can be derived as \cite{mueller2005beam}
\begin{equation}
M_i=\begin{pmatrix}
\sqrt{1 - \tfrac{1}{2}\,\gamma_i^2} & \tfrac{\sqrt{2}}{2}\,i\,\gamma_i \\[6pt]
\tfrac{\sqrt{2}}{2}\,i\,\gamma_i     & \yesun{\sqrt{1-\gamma_i^2}}
\end{pmatrix},\,\gamma_i\equiv k\sqrt{2} \delta\beta_iw_i, 
\end{equation}
with $w_i$ being the spot size at the mirror's location. In addition, the propagation matrix between the two mirrors can be written as
\begin{equation}
    L_{p}=\sqrt{1-l_{\mathrm{other}}}\begin{pmatrix}
1 & 0 \\[6pt]
0    & e^{i\phi_g},
\end{pmatrix}
\end{equation}
where $\phi_g$ is the Gouy phase \cite{feng2001physical,svelto2010principles} determined by the separation between the two mirrors, and $l_\mathrm{other}$ represents the additional loss that is not related to angular misalignment. \yesun{For a conservative estimate, we neglect the relative round-trip Gouy phase between the fundamental and higher-order transverse modes, corresponding to the worst-case coherent buildup of misalignment-induced mode conversion.} Therefore, the round-trip process can be described by the following product of matrices:
\begin{equation}
    C_{\mathrm{cavity}}=M_1L_{p}M_2L_p^\dagger.
\end{equation}
Assuming the configuration shown in Fig.~\ref{fig:VBGQM}(a), in which half of a standard VBG section is used for quantum memory with a plane mirror acting as the switch, the attenuation performance of the VBGQM in the presence of static angular misalignment is shown in Fig.~\ref{fig:VBGQM_Attenuation} and compared with the stochastic assumption. We observe that coherent interference shortens the effective storage time by two orders of magnitude for both the common and reversed tilting cases. One may ask whether we can convert the power in $(0,\pm1)$ mode back to the $(0,0)$ mode to avoid this effect. Unfortunately, that option would require precise knowledge of the misalignment angle and advanced spatial mode conversion techniques, the latter of which may introduce additional losses.

In addition, we note that the cycle time of the VBGQM is limited both by the length of the VBG section and the speed of the I/O switch, i.e. $\tau_{\mathrm{c}}\geq\mathrm{Max}(\tau_{\mathrm{switch}},2\frac{L}{c})$. In general, the round-trip time for a single VBG section will be approximately $\yesun{13}\,\rm{\mu s}$, which dominates the switch time regardless of whether an EOM switch or a mechanical switch is employed. The resulting trade-off between cycle time and total storage time has long been recognized as a serious limitation of this kind of cavity-based optical storage. 

We can also estimate the number of modes for frequency division multiplexing within a VBGQM using the free spectral range $\delta\nu$ for a simple cavity as 
\begin{equation}
\label{eq:modeCavity}
\#\mathrm{Modes}_{\mathrm{FDM}}
=\frac{\Delta\nu}{\delta\nu}
=\frac{2L\Delta\lambda}{\lambda^2}
\yesun{\sim 8\times10^7},
\end{equation}
which demonstrates a substantial capacity to store optical quantum signals of different frequency modes simultaneously. An alternative interpretation of Eq.~(\ref{eq:modeCavity}) is that the pulse length of the encoded bosonic information must be shorter than $L$ to keep it completely stored in the cavity. While the ability to perform time division multiplexing is limited by the switching speed if each pulse needs to be controlled independently, the maximum number of pulses that can be stored inside the VBG cavity is given by
\begin{equation}
\label{eq:VBGQMTDM}
    \#\mathrm{Modes}_{\mathrm{TDM}}=\frac{2L}{c\tau_{\mathrm{switch}}}.
\end{equation}
For a mechanical switch operating at the microsecond timescale, the storage capacity is limited to only tens of pulses. However, if all pulses are stored for the same duration, the maximum number of pulses is instead limited by the pulse width $\tau_p$. In contrast, for an EOM/PBS switch, the switching time can be comparable to or shorter than the pulse duration \cite{chai2017ultrafast}.

As for the phase noise, the situation is more subtle than in the transmission case, since the light makes repeated round trips within the same section with a round-trip time $\tau=\frac{2L}{c}$, so now the noises may add coherently instead of following a Poisson process. We adapt a similar technique to relate the phase noise PSD of the signal after $N$ storage cycles to the phase noise PSD associated with a single round trip $S_{\phi_r}\approx2S_{\phi_L}$, which represents the single section phase noise PSD of the VBG channel. This value can be calculated by multiplying Fig.~\ref{fig:Phase_PSD_APP} with $L=4 \mathrm{km}$. Let $\delta\phi_r(t)$ denote the simple round trip phase noise at time $t$, and then the total phase noise is estimated as
\begin{equation}
    \delta\phi(N)=\sum_{n=0}^{N-1} \phi_r(n\tau).
\end{equation}

In the frequency domain, the time delay acts as a phase shift factor $e^{-i2\pi fn\tau}$, after which one takes the sum to obtain the transfer function
\begin{equation}
    |H(f)|^2=\frac{\sin^2(N\pi f\tau)}{\sin^2(\pi f\tau)},
\end{equation}
and the PSD of the signal is related to the PSD of the single round-trip phase noise PSD as
\begin{equation}
    S_\phi(f)=2S_{\phi_L}(f) \frac{\sin^2(N\pi f\tau)}{\sin^2(\pi f\tau)}.
\end{equation}

\begin{figure}[htbp]
\centering
\includegraphics[width=\linewidth]{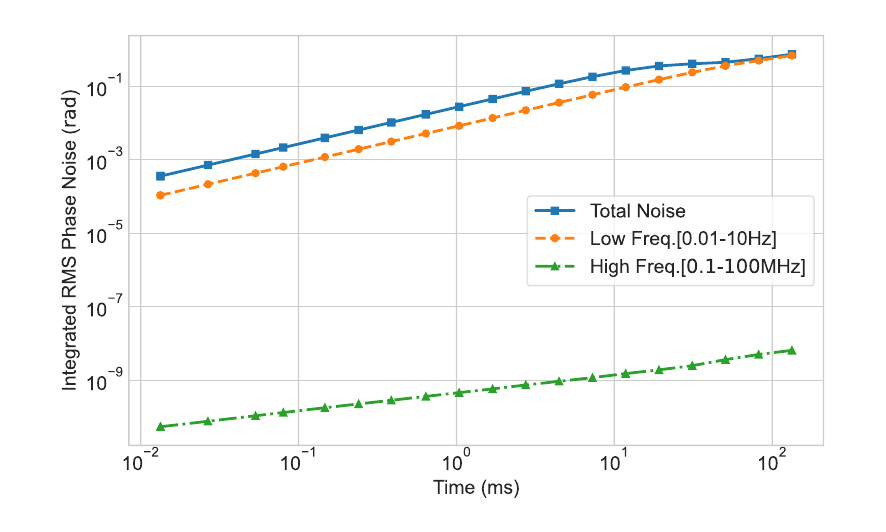}
\caption{\label{fig:VBGQM_phaseNoise} RMS phase noise of the VBG integrated over the whole phase noise PSD versus storage time after active cancellation, as well as the contributions from low and high frequencies.}
\end{figure}

This relation can also be derived by viewing the VBG cavity as an F-P etalon and calculating the final phase response. When $f\ll\frac{1}{\yesun{N}\tau}$, the pulse experiences coherent noise on each round trip. Thus, the effective transfer factor is $N^2$, corresponding to a quadratic decay. In the high-frequency region, $\yesun{f\tau\gg1}$, only the average value of this factor matters since the function oscillates quickly. Therefore, we find that
\begin{equation}
    \overline{|H^2|}\propto\frac{1}{2\pi} \int_{-\pi}^{\pi} \frac{\sin^2(Nx)}{\sin^2(x)}dx=N,
\end{equation}
which represents linear growth characteristic of a Poisson process and can be understood as the high-frequency noise at different cycles not being correlated.  Fortunately, unlike in the VBG channel case, the sensing of the error, and thus the active noise cancellation, is not limited by the round-trip delay because it is performed locally. In principle, the error can be sensed in real time and with $\tau\sim1\,\mu s$ limited by the sensing time. After cancellation, the noise is now just the sensing noise, which is dominated by frequency noise. \yesun{Because attenuation limits the useful storage depth to approximately $O(10^2)$ cycles, we set $N\sim100$ as a conservative enhancement factor for the AUX-laser frequency-noise contribution.} The RMS noise of the VBG memory channel versus storage time, obtained by integrating over the frequency range $0.01-10^{8}$ Hz, and the low and high frequency contributions, are shown in Fig.~\ref{fig:VBGQM_phaseNoise}. The observed scaling matches our previous prediction well.

\section{Enhancement of Frequency Transfer}
\label{sec:enhancementOfFT}
The investigation of frequency transfer has persisted over several decades \cite{foreman2007remote} due to its profound significance in the field of metrology. 
This endeavor has seen continual advancements in accuracy, exemplified by the implementation of extensive fiber optic linkages \cite{foreman2007coherent,newbury2007coherent,predehl2012920,schioppo2022comparing,droste2013optical} and, more recently, through atmospheric free-space connections \cite{dix2021point}. A key performance metric in such systems is the additional fractional frequency instability introduced by the transmission channel. 
This instability is commonly quantified by measuring the interference between an ultra-stable laser and its signal after a round-trip propagation through the channel. 
The results are typically reported in terms of the modified Allan deviation (MDEV) or the frequency noise PSD.

\begin{figure}[htbp]
\centering
\includegraphics[width=\linewidth]{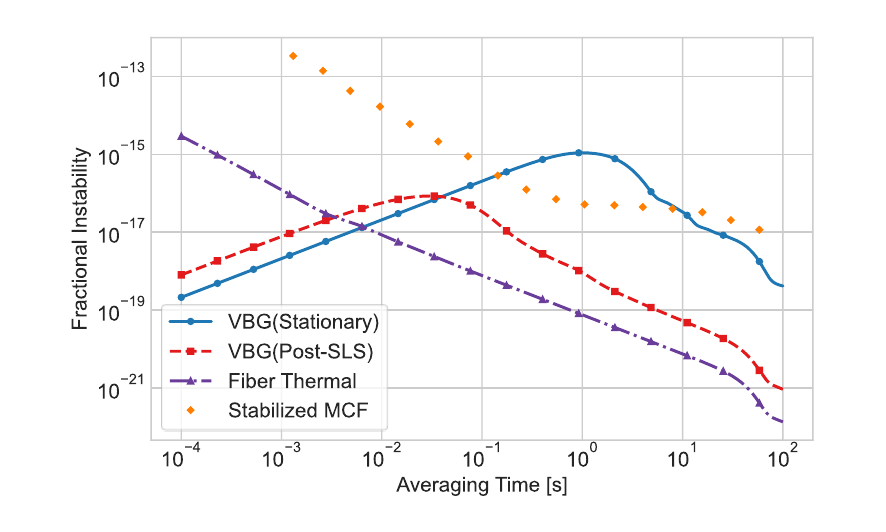}
\caption{Added fractional frequency instability for 25\,km transmission through the VBG with and without active noise cancellation, quantified by modified Allan deviation as a function of averaging time. The results are compared with the fiber thermal noise floor \cite{bartolo2012thermal,wanser1992fundamental,duan2010intrinsic} and the state-of-the-art multi-core fiber (MCF) frequency transfer experiment data \cite{hoghooghi2024enabling}.}
\label{fig:VBG_FT} 
\end{figure}

The MDEV can be calculated from the PNPSD in the continuous limit via \cite{riley2008handbook,bregni2005using}
\begin{equation}
    \sigma_y^2(\tau)=\frac{2}{\nu_0^2}\int_0^\infty S_\phi(f)\frac{\sin^6(\pi f\tau)}{(\pi\tau)^4f^2}df,
\end{equation}
with $\tau$ as the averaging time and $\nu_0$ as the carrier frequency. 
Figure~\ref{fig:VBG_FT} presents the VBG performance derived from the PNPSD, with and without SLS, alongside fiber transfer results \cite{hoghooghi2024enabling}. 
The $-1$ slope in both the VBG (SLS) and the fiber thermal floor is due to the common 1/f noise sources, as is typical in precision experiments~\cite{Ward:2007,duan2010intrinsic}.
The $-3/2$ slope in both the stabilized MCF and fiber thermal floor indicates white phase noise. 
The minor hump in both VBG curves arises from the rapid roll-off of the seismic noise. Overall, the figure illustrates that the VBG achieves superior short-term stability with significantly reduced high-frequency noise compared to fiber techniques, even in the absence of SLS. Moreover, when SLS feedforward is employed, the long-term stability approaches the fiber thermal floor within an order of magnitude. 

\section{Performance Analysis of Additional Protocols}
\label{sec:AddPro}
\begin{figure*}[htbp]
\centering\includegraphics[width=\linewidth]{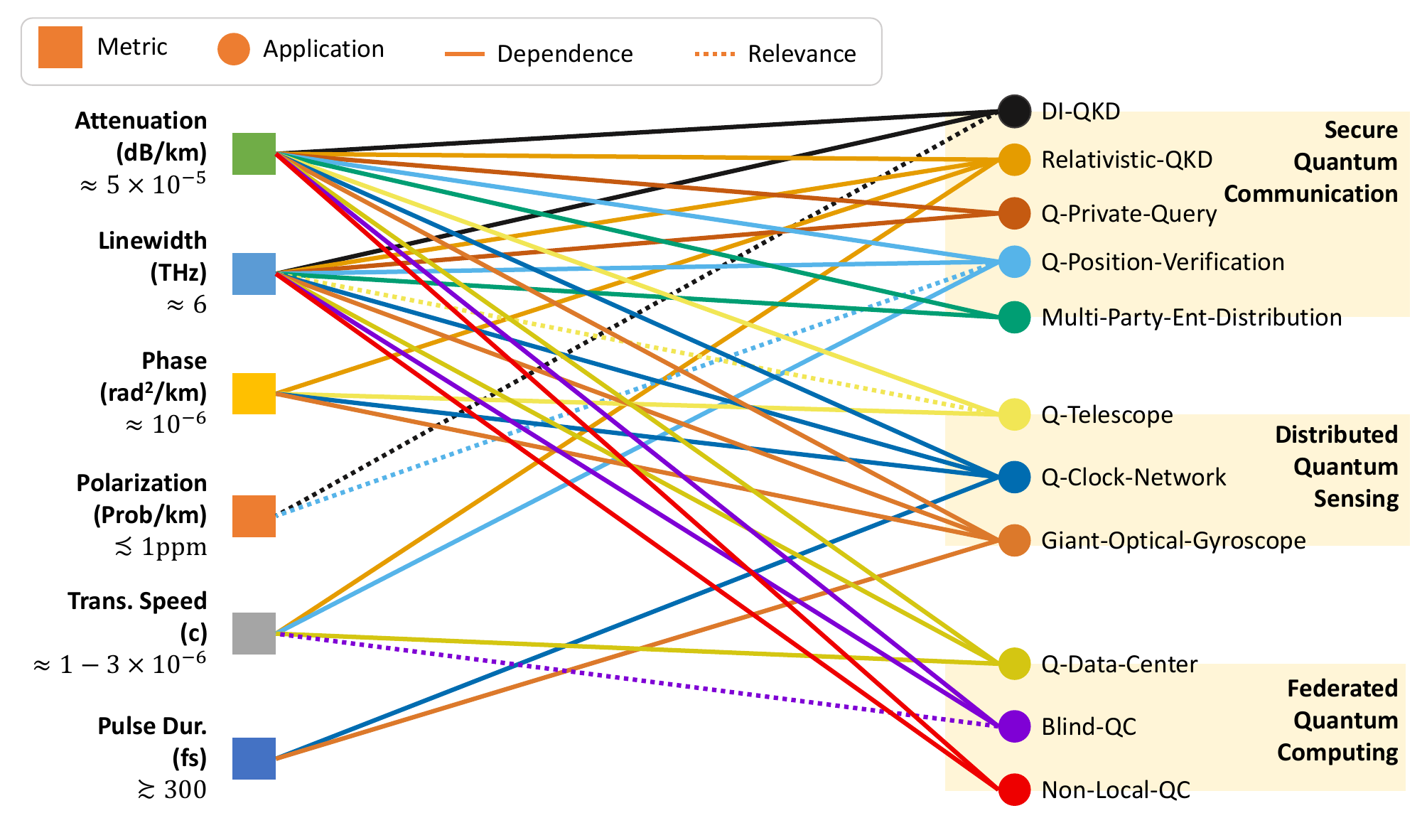}
\caption{\label{fig:applicationSummary} Summary of VBG's performance metrics and application requirements. The solid line indicates that the specific application depends on a certain performance metric, while the dotted line suggests that it may be relevant according to the concrete protocol.  Here, the applications in the category of secure quantum communication include Device-Independent Quantum Key Distribution \cite{ekert1991quantum,acin2007device,masanes2011secure,vazirani2019fully,arnon2018practical,sekatski2021device,zhang2022device,xu2022device,zapatero2023advances}
(Sec.~\ref{subsec:DIQKD}), Relativistic-QKD \cite{kravtsov2018relativistic,sandfuchs2025security}
(SI. \ref{subsec:GQC}), Quantum Private Query \cite{giovannetti2008quantump}
(SI. \ref{subsec:GQC}), Quantum Position Verification \cite{buhrman2014position,bluhm2022single,allerstorfer2023making}
(SI. \ref{subsec:QPV}), and Multi-party Entangled State Distribution \cite{fischer2021distributing,meignant2019distributing,fan2025optimized,huang2025peer,fan2025distribution}
(SI. \ref{subsec:GQC}); Distributed quantum sensing covers Quantum Telescope \cite{khabiboulline2019quantum,khabiboulline2019optical,gottesman2012longer}
(Q-Telescope, Sec.~\ref{subsec:QTelescope}), Quantum Clocks Network \cite{giovannetti2001quantum,komar2014quantum,giovannetti2002positioning}
(SI. \ref{subsec:clocksyncgronization}), and Giant Optical Gyroscope \cite{brady2021frame,giovinetti2024gingerino,di2024noise}
(SI. \ref{subsec:GOG}); and federated quantum computing consists of Quantum Data Centers \cite{liu2023data,liu2024quantum}
(SI. \ref{subsec:LDFQC}), Blind Quantum Computation \cite{fitzsimons2017private,morimae2013blind}
(Sec.~\ref{subsec:BDQC}), and Non-Local Quantum Computation \cite{buhrman2010nonlocality,yao1993quantum,brassard2003quantum}
(SI. \ref{subsec:NLQC}). } 
\end{figure*}

We summarize the key performance metrics of the VBG and the dependence of the mentioned protocols on these metrics in Fig.~\ref{fig:applicationSummary}.

\subsection{Quantum Position Verification}
\label{subsec:QPV}

\begin{figure}[htbp]
\centering\includegraphics[width=\linewidth]{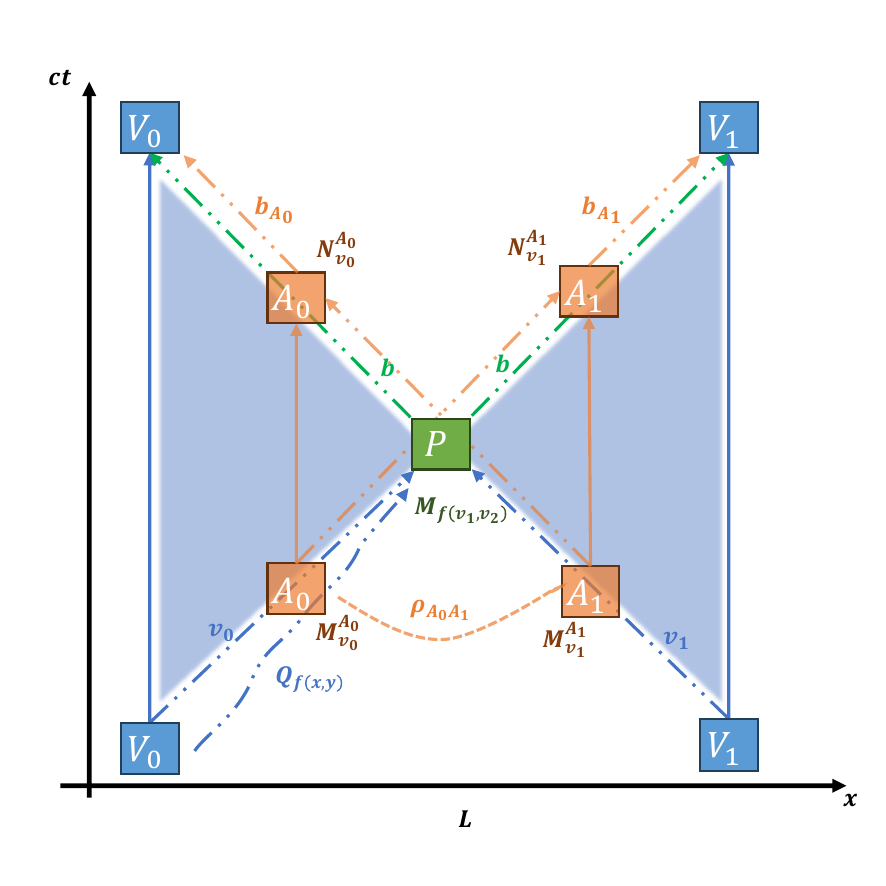}
\caption{\label{fig:QPVILLU} Illustration of the typical QPV protocol. The shaded regions represent the future light cones of both verifiers, with $P$ being the only intersection point in the ideal case. The orange boxes indicate the two possible attackers attempting to spoof the location using a pre-shared resource state with one-round simultaneous communication.}
\end{figure}

Position verification is an interesting task of significant practical importance. In many real-world scenarios, the server may want to confirm that the clients are located within the area that complies with the service contract, but the clients may not always be honest. Thus, position verification can be performed by exploiting the limitations imposed by special relativity, similar to Relativistic-QKD \cite{kravtsov2018relativistic}. For simplicity, we consider only the one-dimensional case and suppose the two verifiers, separated by $L$, want to verify that the client is indeed located at the midpoint between them. A typical method for position verification is shown by Fig.~\ref{fig:QPVILLU}, where the prover is required to compute a function whose inputs are transmitted at the speed of light from two verifiers and to return the result in a time window such that the overlap of the light cones constrains the prover's location. However, since classical information can be copied, the two attackers can easily emulate a fake position by exchanging their received inputs. It has been proven that secure classical position verification is impossible \cite{chandran2009position}.

The classical limitation can be overcome by exploiting ideas from quantum key distribution, where the quantum nature of the information inherently prohibits perfect copying. Numerous research studies have been conducted on this topic \cite{buhrman2014position,bluhm2022single,allerstorfer2023making}, and we consider a hallmark protocol that is executed as follows \cite{bluhm2022single}:
\begin{enumerate}
    \item \textbf{Random Generation.} The Verifiers, $V_0$ and $V_1$, choose two strings of random bits, $v_0 \text{ and }v_1$, of length $n$. Let $f$ be the inner product function. 
    \item \textbf{Quantum State Transmission.} If $f(v_0,v_1)=0$, $V_0$ prepares a qubit $Q$ in the $Z$ basis. Otherwise, Q is prepared in the $X$ basis.
    \item \textbf{Classical Bit Transmission.} $V_0$ and $V_1$ send $v_0$ and $v_1$ to the Prover $P$ and engineer the timing so that $Q,v_0,\text{ and }v_1$ reach the position they want to verify, $z$, at the same time.
    \item \textbf{Measurement.} If $f(v_0,v_1)=0$, the prover measures $Q$ in the $Z$ basis. Otherwise, it is measured in the $X$ basis.
    \item \textbf{Certification.} The Prover sends back the measurement results $b$ to the two verifiers immediately after receiving $Q$ at the speed of light.
    \item \textbf{Verification.} The verifiers check whether the measurement result is correct and whether the measurement result arrived in time. The above procedure is repeated for $r$ rounds to achieve the desired confidence.
\end{enumerate}
In contrast to the QKD case, verifiers in QPV cannot trust any party; consequently, attackers may also exploit entanglement to compromise the protocol. It has been shown that such attacks are always possible if two adversaries share an exponential number of entangled qubits \cite{beigi2011simplified}. \yesun{However, under reasonable computational time constraints, the latest result gives an almost-linear lower bound on the attackers' quantum-system size for the above protocol~\cite{asadi2025linear}.}

The implementation of large-scale QPV protocols is mainly limited by the attenuation in the single qubit transmission in each round, \yesun{which gives the honest prover an error probability of $p_h=(1-\eta)/2$ when a no-click round is answered with a uniformly random bit with $\eta$ denoting the single-qubit transmission probability through the VBG from the verifier to the prover.} In addition, the attackers still have a non-zero probability of success, $1-p_a$, even under the assumption of limited quantum resources, and thus set a threshold for the attenuation. Suppose we require the honest prover to succeed and the attackers to fail with the same confidence level (CL), the expected rounds $r$ can be calculated using Hoeffding's inequality as
\begin{equation}
    r\geq -\frac{2\ln(1-\mathrm{CL})}{(p_a-0.5(1-\eta))^2},\,\yesun{1-\eta<2p_a}.
\end{equation}
Dividing by the multiplexing factor $\Delta\nu$ gives the \yesun{channel-limited} minimum execution time, $\tau_{\mathrm{ex}}\approx \frac{r}{\Delta\nu}$. \yesun{For the performance benchmark, we assume a bounded-resource adversary
with a per-round error probability $p_a=0.05$. Recent results establish
strong lower bounds on the quantum resources required for sufficiently
accurate attacks on the inner-product $f$-BB84 protocol
\cite{asadi2025linear}.} However, we note that although more advanced protocols have been developed to avoid this threshold, they require additional quantum operations, such as quantum non-demolition detection \cite{allerstorfer2023making}.

\begin{figure}[htbp]
\includegraphics[width=\linewidth]{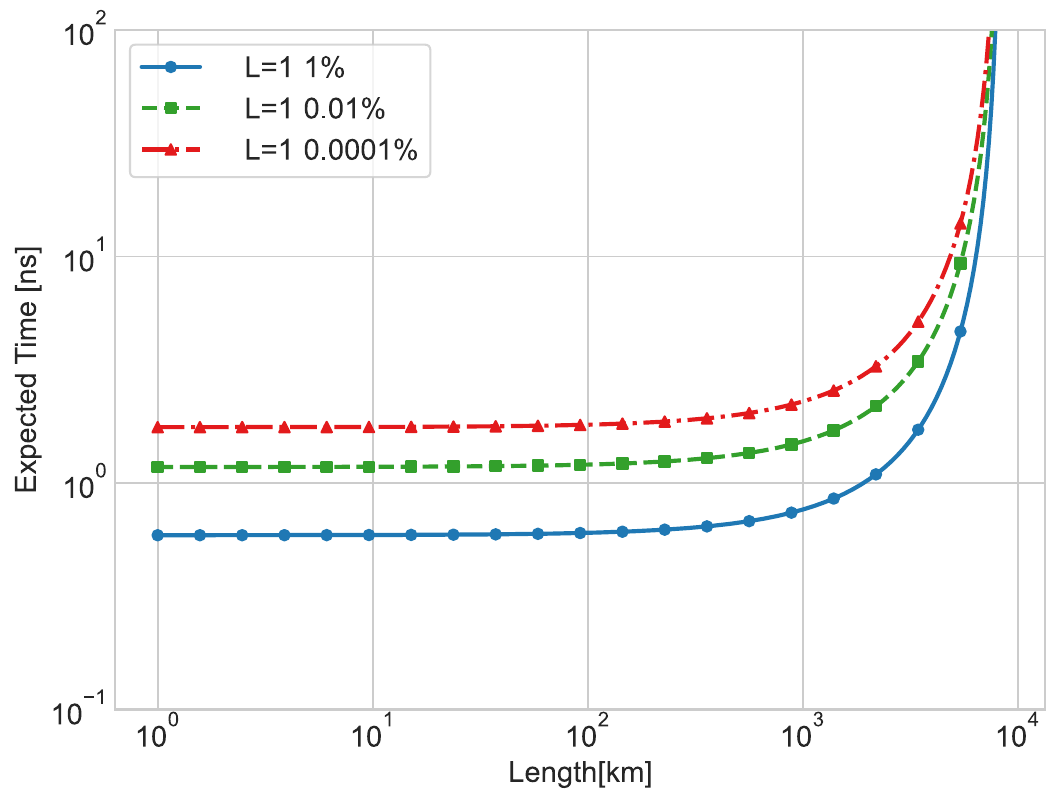}
\caption{\label{fig:QPV} Minimum execution time versus the distance between the verifiers and the honest prover under different confidence requirements.}
\end{figure}

The minimum execution time, $\tau_{\mathrm{ex}}$, versus the distance between the verifier and the honest prover, assuming the use of the VBG for transmission, is shown in Fig.~\ref{fig:QPV}. It can be concluded that when the transmission distance is relatively short, the minimum execution time is \yesun{at the nanosecond scale}, which is primarily constrained by the protocol dependent $p_a$ and \yesun{the assumed multiplexing bandwidth}. In this regime, the execution time will be dominated by the transmission time, as the total time consumption for the protocol is given by
\begin{equation}
    T_{\mathrm{tot}}=\tau_{\mathrm{ex}}+\frac{L}{\bar{v}_g},
\end{equation}
where $\bar{v}_g\approx c$ is the transmission speed of the qubit through the vacuum beam guide. Furthermore, we observe that the threshold transmission distance \yesun{approaches} $10^4$ km if the QPV is executed via the VBG. This result demonstrates that the VBG channel effectively minimizes the impact of quantum transmission losses, such that protocol execution over continental scales will ultimately be limited by the Earth's curvature and the transmission speed of the classical bit strings rather than the quantum channel itself. 

Conversely, the VBG could potentially facilitate attacks on the QPV protocol. \yesun{For the inner-product protocol, known attacks do not necessarily require exponential entanglement, and their feasibility depends on the attackers' quantum-resource and timing constraints.} Moreover, without quantum memory, attackers must transmit qubits nearly at light speed, which the VBG is capable of supporting, as demonstrated in Eq.~(\ref{eq:transmissionSpeed}) in the main text.

\subsection{General Quantum Cryptography}
\label{subsec:GQC}
The two previously discussed tasks exhibit a hard attenuation threshold in the absence of advanced quantum control techniques, highlighting the irreplaceable role of the VBG in implementing these protocols in the near term. Beyond these tasks, the VBG enables fast quantum information exchange on a global scale, providing universal advantages for general quantum cryptographic protocols involving geographically isolated parties. Although these loss-tolerant protocols may be executed without the VBG over long distances, the VBG can nevertheless provide speedups of several orders of magnitude compared to existing quantum channel techniques. We will now discuss several additional qualitative examples.

Relativistic-QKD (R-QKD) \cite{kravtsov2018relativistic,sandfuchs2025security} has emerged as a focus within the QKD field due to its ability to maintain a nonzero key rate under significant channel loss and to ensure security with minimal device assumptions through special relativity constraints. The optimal performance requires qubit trajectories to align as closely as possible with a null geodesic; otherwise, a quantum memory is necessary, and the channel must exhibit strong phase reliability. The quantum signal in the VBG travels nearly at the speed of light in a vacuum, in contrast to approximately $\frac{2}{3}c$ in optical fiber. These capabilities of the VBG significantly benefit R-QKD implementation. Additionally, the VBG phase noise after the SLS is at a level of $\yesun{10^{-3}}\,\mathrm{rad}/\sqrt{\mathrm{km}}$, and VBG-based quantum memory (VBGQM) offers millisecond-level storage without qubit conversion, as detailed in Sec.~\ref{sec:VBGQM}, thereby mitigating effective refractive index deviations and enhancing system flexibility.

Following the paradigm of QKD, the quantum private query (QPQ) protocol can be implemented using lure requests on a quantum server with quantum random access memory \cite{giovannetti2008quantump}. The randomly inserted lure request addresses, encoded in an alternative basis, enable the user to detect eavesdropping by the server with high probability. The VBG can be employed to transmit both the query and the requested data with high fidelity, which reduces the possibility of the server concealing its malicious behavior as channel loss. 

Finally, multiparty entangled state distribution (MESD) over large-scale networks is being actively explored due to its potential cryptographic advantages over pairwise Bell-pair distribution \cite{memmen2023advantage}. In particular, quantum secret-sharing protocols\cite{cleve1999share,hillery1999quantum} can be uniformly described by the construction and manipulation of graph states across a network, which form a representative class of multiparty entangled states\cite{hein2006entanglement,markham2008graph}. Advanced and efficient network layer protocols \cite{fischer2021distributing,meignant2019distributing,fan2025optimized,huang2025peer,fan2025distribution} have been proposed for the general distribution of arbitrary graph states, in which channel capacity and reliability play crucial roles in resource allocation and optimization. If future networks are built on the backbone of the VBG channel, then the network topology is approximately star-shaped, enabling the distribution of arbitrary graph states within a single time slot, significantly reducing distribution latency and the burden on quantum memory. 

\subsection{Synchronized Network of Clocks}
\label{subsec:clocksyncgronization}
\begin{figure}[htbp]
\includegraphics[width=7cm]{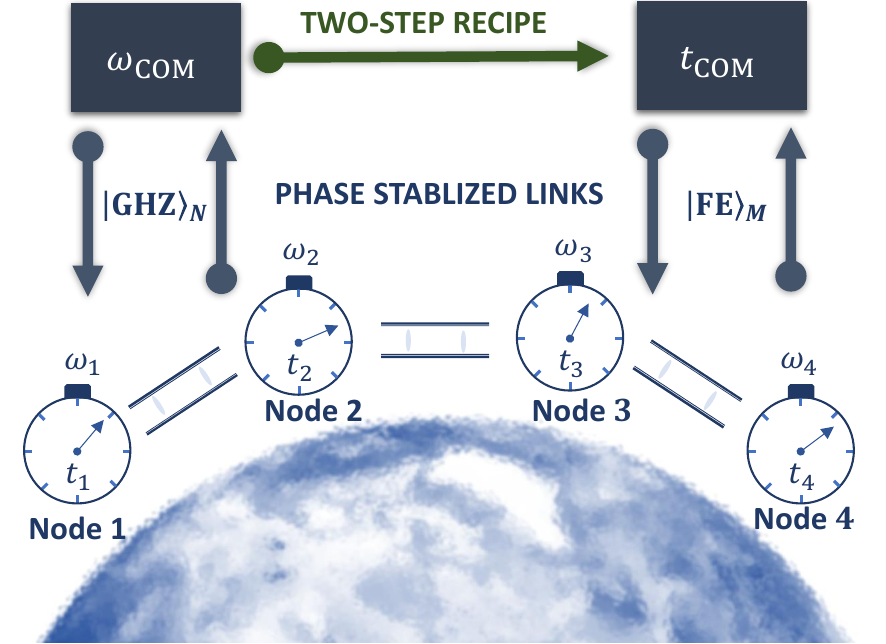}
\caption{\label{fig:QSNC} Illustration of the quantum enhanced network of synchronized clocks, which involves two key steps that benefit from different types of entangled state.}
\end{figure}

As depicted in Fig.~\ref{fig:QSNC}, constructing a highly precise synchronized quantum clock network (QCN) involves two key phases: aligning local clocks to a common ultra-stable frequency standard \cite{komar2014quantum} and achieving precise clock positioning and synchronization through the Einstein synchronization method \cite{einstein1905elektrodynamik}. Research shows that quantum protocols can enhance both of the aforementioned processes, providing a precision improvement that follows a square root scaling relative to classical techniques \cite{giovannetti2001quantum,giovannetti2002positioning}. In this framework, the deployment of the VBG can provide substantial benefits in protecting the fragile quantum signal and serve as a foundation for constructing a quantum probe of gravity \cite{covey2025probing}.

Establishing a remote frequency standard is critical for real-time sensing and many other experiments involving general relativity effects \cite{wcislo2018new,takano2016geopotential}. Quantum entanglement can be utilized to construct a global frequency standard that leverages resources from all participating nodes, with optimal scaling for short-time Allan deviation as \cite{komar2014quantum}
\begin{equation}
    \sigma_y(\tau)\sim \frac{\sqrt{\log(N)}}{\omega_0 N} \frac{1}{\tau},
\end{equation}
where $N$ is the total number of qubits involved in the GHZ state, $|\mathrm{GHZ}\rangle_N$ , and $\omega_0$ is the frequency reference of the atomic clock. An intuitive explanation is that these established entanglements effectively bring the distributed local resources of each node together at the logical level and allow for the collective operations required for coherent quantum enhancement. This protocol requires a multiqubit GHZ state of extremely high fidelity ($\sim0.99$), and thus highly transparent channels, such as the VBG, as any loss of photons destroys the GHZ state. In addition, establishing such a quantum clock network also requires robust frequency transfer between nodes to synthesize the global standard oscillator signal and implement feedback control. This task is purely classical and well-studied \cite{foreman2007remote}, though the VBG may also provide potential enhancement due to its lower intrinsic phase noise compared to fiber when above $1$ Hz, which was discussed in more detail in Sec. \ref{sec:enhancementOfFT}. 

Once the frequency of each node is locked to a common standard, the nodes can perform Einstein's protocol to eliminate the offsets of each clock. For simplicity, we assume a case involving two nodes, Alice and Bob, separated by a distance $L$ while remaining in the same global inertial frame. The synchronization procedure works as follows:
\begin{enumerate}
    \item Alice sends a pulse to Bob at $t_a^s$, as recorded by Alice's clock.
    \item Upon receiving that pulse, Bob records the time $t_b$ \yesun{on Bob's clock} and reflects the pulse (or sends another pulse immediately).
    \item Alice receives the reflective pulse and records the time as $t_a^e$.
    \item \yesun{The measured magnitude of the clock offset is}
    $$
    \yesun{\Delta t_{ab}=\left|t_b-\frac{t_a^s+t_a^e}{2}\right|.}
    $$
    The above procedure is repeated and executed in $M$ parallel modes to achieve the desired precision.
\end{enumerate}

The synchronization precision relies on the measurement accuracy of $t_b$ and $t_a^e$. To streamline performance evaluation, we focus solely on the one-way transfer from Alice to Bob, assessing the variance of $t_b$ as a positioning method, with an analogous analysis applicable for synchronization.

Using entanglement across $M$ parallel modes can enhance measurement precision by $\sqrt{M}$. The frequency entangled (FE) state is assumed to be in the following form \cite{giovannetti2001quantum,giovannetti2002positioning},
\begin{equation}
    |\mathrm{FE}\rangle_{M}=\int d\omega \phi(\omega)\otimes_{i=1}^M |1_\omega\rangle_i,
\end{equation}
where $|1_\omega\rangle_i$ is the single photon state at frequency $\Omega_i+\omega$ and $\phi(\omega)$ is the spectral function shared by $M$ modes, with $\omega$ defined as the detuning with respect to the carrier frequency $\Omega_i$. Then, the variance of the measurement is given by
\begin{equation}
    \yesun{\Delta t^2_{\mathrm{en}}=\frac{\tau^2}{4M^2},}
\end{equation}
\yesun{where $\tau$ is the Gaussian pulse-duration parameter used throughout this work; for $g(t)\propto e^{-(t/\tau)^2}$, the single-pulse arrival-time variance is $\tau^2/4$.} The envelope function is
\begin{equation}
    g(t)=\frac{1}{\sqrt{2\pi}}\int d\omega \phi (\omega)e^{-i\omega t}.
\end{equation}
However, attenuation in the transmission channel can seriously impact the performance of the entangled state. Suppose that any one of the photons from the $M$ modes is lost. In that case, the state will be projected onto a diagonal density matrix in the $|\omega\rangle$ basis, which is essentially a continuous wave instead of a pulse, and thus destroys any information regarding the arrival time \cite{giovannetti2002positioning}. When the protocol is repeated for rounds $r\gg1$, the asymptotic variance for the arrival time of each round can be calculated as
\begin{equation}
    \yesun{\Delta t^2_{\mathrm{en}}\simeq\frac{\tau^2}{4\eta^M M^2},}
\end{equation}
\yesun{where $\eta$ is the single-photon transmission probability over the distance $L$.} In the classical case, the information associated with any mode whose photon is not lost is unaffected, giving
\begin{equation}
    \yesun{\Delta t^2_{\mathrm{cl}}\simeq\frac{\tau^2}{4\eta M}.}
\end{equation}
The different scaling with $\eta$ gives rise to a threshold value of the \yesun{transmission probability}, below which the classical scheme is more advantageous. In addition, the channel introduces phase noise $\sigma_\phi^2$, as before, and the broadening of the envelope due to GVD is estimated in Sec.~\ref{subsec:PDTD} in the main text. \yesun{The phase noise contributes a common timing variance $\sigma_{t,\phi}^2=\lambda^2\sigma_\phi^2/(4\pi^2v_g^2)$, which is numerically negligible. For GVD, the deterministic mean GDD compensation model is used as in the main text. However, it shall be noted that as all photons in the FE state share the same detuning $\omega$, the residual-GVD term does not receive the $1/M^2$ enhancement. Therefore, the total arrival time uncertainty can be calculated as}
\yesun{
\begin{align}
\Delta t_{\mathrm{en}}^2&\simeq\frac{1}{\eta^M}
\left(\frac{\tau^2}{4M^2}+\frac{N_{\rm sec}\sigma_D^2}{\tau^2}\right),\\
\Delta t_{\mathrm{cl}}^2&\simeq\frac{1}{\eta M}
\left(\frac{\tau^2}{4}+\frac{N_{\rm sec}\sigma_D^2}{\tau^2}\right).
\end{align}}
\yesun{where $N_{\rm sec}=L/L_0$ is the number of VBG sections.}

\begin{figure}[htbp]
\includegraphics[width=\linewidth]{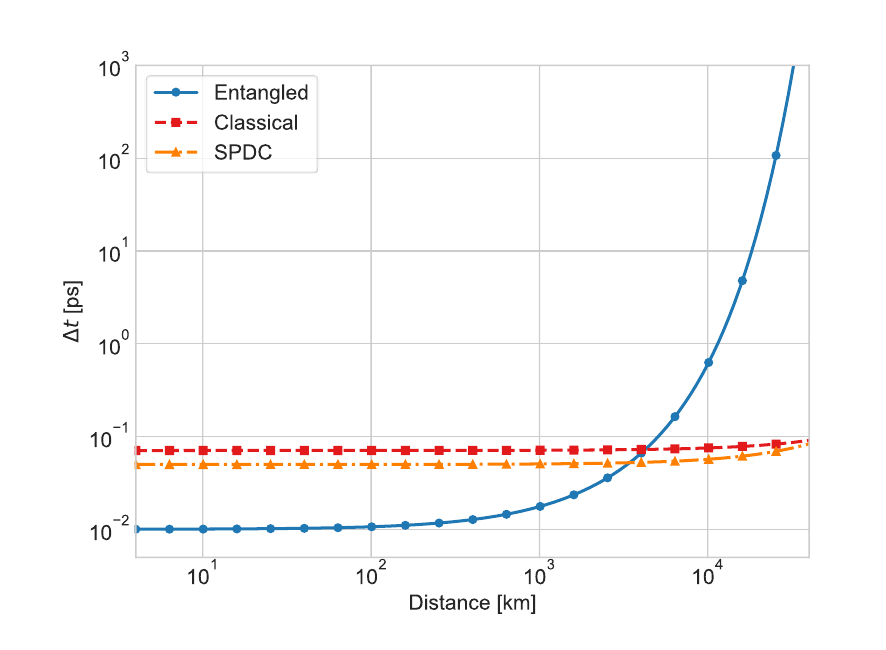}
\caption{\label{fig:QCS} Arrival time \yesun{uncertainty} versus various separation lengths under classical, $M$-entangled state, and SPDC ($M=2$) schemes. For the SPDC state, it is a partially entangled state separated into $M/2$ tensor components, with each component containing two entangled photons.}
\end{figure}

For performance estimation, we set $\tau=1\text{ ps}$ with $g(t)$ taken to be a Gaussian envelope function. The phase noise is calculated using the PNPSD as before, and \yesun{we use $M=50$ as a representative order-of-magnitude frequency-multiplexing benchmark and $\sigma_D\sim5\times10^2\,\mathrm{fs^2}$ per $4$-km section as a conservative residual-GDD scale after deterministic compensation.} The synchronization precision achieved using the VBG at different separation distances $L$ is shown in Fig.~\ref{fig:QCS}. The quantum scheme provides approximately an order of magnitude improvement over the classical scheme over a distance range of $10^3$ km, while the performance degrades significantly across \yesun{$4\times10^3$} km due to \yesun{the all-photon transmission penalty}. In addition, the SPDC scheme with pairwise entanglement demonstrates a longer threshold distance but achieves only a $\sqrt{2}$ enhancement.

\subsection{Giant Optical Gyroscope}
\label{subsec:GOG}
The VBG itself is a large-scale optical platform, and it is therefore well suited for measurements that benefit from scaling up, such as the realization of a giant optical gyroscope (GOG). A quantum optical gyroscope has recently been proposed for applications in seismology and for probing the interplay between quantum physics and general relativity \cite{brady2021frame}. In this scheme, an indistinguishable photon pair is emitted into a large ring cavity attached to the Earth's surface, with one photon propagating clockwise and the other counterclockwise. The Earth's rotation, the speed of which is denoted as $\omega_E$, induces a Sagnac effect \cite{sagnac1913preuve} that introduces a relative time delay $\delta\tau_{\mathrm{sag}}$ between the two photons that is proportional to the area size of the ring cavity, which can be written as
\begin{equation}
  \tau=\mathcal{F}A,  
\end{equation}
with $\mathcal{F}$ being the sensitive factor and $A$ being the area size. This delay $\tau$ shifts the HOM dip, from which the time delay can be determined. In addition, general relativity predicts a correction to the locally observed angular rotation rate, known as the frame-dragging effect \cite{thirring1918wirkung}, 
\begin{equation}
    \Delta\omega_{\mathrm{FD}}=\frac{2GI}{c^{\yesun{2}}R_E^3}\omega_E\sim10^{-9}\omega_E,
\end{equation}
where $G$ is the gravitational constant, $I$ is the moment of inertia, and $R_E$ is the Earth's radius. This corresponds to a sensitivity factor of $\mathcal{F}_{\mathrm{FD}}\sim10^{-24}\,\mathrm{s/km^2}$, implying that an area of $10^6\mathrm{km^2}$ can only generate a time delay at the attosecond level, which lies barely within the demonstrated sensitivity of the HOM-dip measurements \cite{lyons2018attosecond}. 

Such a measurement necessarily requires ultra-transparent transmission channels to preserve the photons during propagation, a regime in which the VBG offers a natural solution. At the same time, this measurement also requires a relative precision on the order of $10^{-9}$, since it can be shown that the frame-dragging correction cannot be separated from the ordinary angular rotation as long as the gyroscope is located on the Earth's surface. Although the gyroscope itself benefits from common-mode noise rejection, this still imposes stringent requirements for delay line scanning, requiring micrometer-level accuracy over a cavity of thousand-kilometer scale. Furthermore, additional suppression of pulse broadening due to the lens-induced GVD is required to reduce the width of the HOM-dip.

An alternative solution is the use of the active gyroscope technique, namely the intra-cavity phase interferometer (IPI) \cite{arissian2014intracavity,scully1981proposed}, which measures the resonant frequency shift rather than the time delay itself, although it operates as a classical measurement technique. Despite significant efforts toward constructing the first ground-based gyroscope capable of measuring the frame-dragging effect, current implementations remain two to three orders of magnitude away from the required precision \cite{giovinetti2024gingerino,di2024noise}. Beyond continuous wave measurement, the use of correlated frequency combs may further scale the system and eliminate the dead band arising due to cross-scattering, since the gain medium need not fill the entire cavity \cite{zhu2022phase}. The precision of the measurement can also be improved using squeezed light \cite{diels2023sub}, further expanding the potential of VBG channels.

\subsection{Long-Distance Distributed Fault-Tolerant Quantum Computing}
\label{subsec:LDFQC}

\begin{figure}[htbp]
\includegraphics[width=\linewidth]{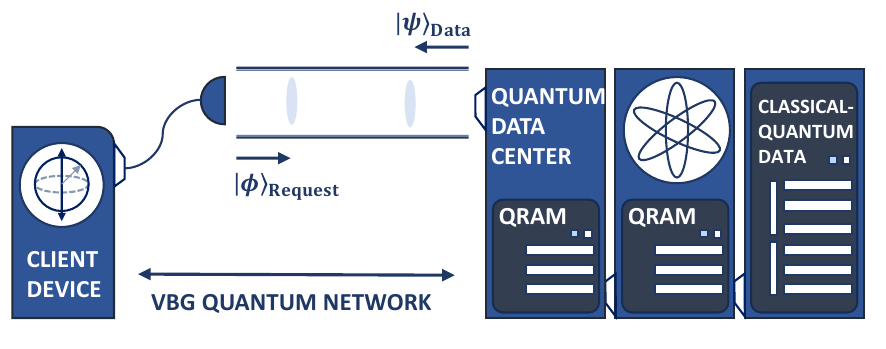}
\caption{\label{fig:QDC} Illustration of the interaction between a quantum data center (QDC) and a client's processor via a VBG-based quantum network. }
\end{figure}

VBG technologies facilitate the development of long-range, distributed, fault-resistant quantum computing through the realization of quantum data centers (QDC), which minimally include a QRAM and quantum network interface, as illustrated in Fig.~\ref{fig:QDC}. The extremely low attenuation rate of the VBG is vital for preserving coherence and reducing information loss across lengthy quantum channels \cite{liu2023data,gan2025quantum,fuentealba2025robust}. In such architectures, the VBG can enable high quality distribution of $T$-gates, support remote fault-tolerant quantum computing, and allow remote distillation of magic states through high-fidelity entangling gates over long distances. These capabilities are essential for scalable quantum architectures based on modular and distributed subsystems. Such configurations have been shown to offer significant advantages in terms of data-size scaling and the efficiency of outsourcing computational tasks when fast quantum memory is available \cite{liu2023data,giovannetti2008quantum}. For example, high-fidelity quantum teleportation not only reduces failure rates and improves Bell pair quality by suppressing attenuation but also enhances the robustness of quantum information transfer across distant nodes. This improvement directly impacts the \emph{delay time} in distributed quantum computing architectures \cite{liu2023data,liu2024quantum}, where the delay of quantum communication plays a key role in determining the performance and architecture of surface code magic state distillation schemes. 

Reduced communication latency and higher gate fidelities enabled by the VBG will lead to shorter delay times and more cost-effective magic state distillation protocols. For a transmission distance $d$, the decay of the photonic intensity ratio will be $10^{-d \alpha_{\text{att}}/10}$. \yesun{Assuming a simple sequential-retry model,} in the absence of photonic loss, the delay time will be $d/c$. In the presence of photonic loss, however, the \yesun{expected} delay time will be amplified by $\frac{d}{c} \times 10^{d \alpha_{\text{att}}/10}$. Thus, the \yesun{loss-induced relative delay factor} comparing the VBG with optical fiber is approximately $10^{-d \alpha_{\text{att,fiber}}/10}$ for $\alpha_{\text{att,fiber}} \gg \alpha_{\text{att,VBG}}$. The advantage, therefore, increases exponentially with distance. \yesun{Independently of this retry advantage, the near-vacuum propagation speed of the VBG also reduces the bare propagation delay compared to standard optical fiber.} For instance, for $d \approx 500\,\text{km}$, we have a $31\%$ reduction in delay time. Such savings, according to Figure 3 of \cite{liu2024quantum}, \yesun{can} switch the optimal magic state distillation protocol and yield qubit savings of around $20\%$. Consequently, these improvements will enhance the overall throughput and scalability of distributed quantum computing and significantly improve the efficiency and reliability of quantum task outsourcing.

\subsection{Non-Local Computation Tasks}
\label{subsec:NLQC}
Robust entanglement establishment is also essential for enabling non-local computation tasks, in which quantum mechanics offers a fundamental advantage in reducing communication costs. In the QPV section, it was mentioned that hackers may possess an exponential amount of quantum entanglement \cite{beigi2011simplified}, which can be modeled as a non-local computation task executed with one round of communication.

The general framework of non-local quantum computation (NLQC) was first proposed and pioneered by Yao \textit{et al.} to investigate communication complexity in a quantum regime \cite{buhrman2010nonlocality,yao1993quantum,brassard2003quantum}. In this setting, Alice and Bob receive local inputs $x$ and $y$ (classical or quantum) and are required to compute a function $f(x,y)$ by producing outputs $a$ and $b$ such that $f(x,y)=a\oplus b$. In the ideal case where the parties share a non-local box that is capable of computing $f(x,y)=x\wedge y$ nearly perfectly (with a probability higher than $90.8\%$), it is shown that every function can be distributively computed with just one bit of information exchange \cite{brassard2006limit,popescu1994quantum}. This threshold is closely related to the CHSH game, in which a quantum strategy is capable of achieving an optimal success probability of approximately $85\%$ \cite{cirel1980quantum}. Although this probability is insufficient to trivialize communication complexity, it still demonstrates the advantages of nonlocal quantum computation. 

Beyond its role in attacks on QPV, non-local computation has concrete applications in coordinated decision-making through quantum telepathy, which can be exploited to establish an efficient stock market strategy beyond the classical limit \cite{ding2024coordinating}. However, if the entangled pair is distributed via a direct photonic connection, the coupling efficiency $\eta$ will seriously affect the performance of the protocol and establish a threshold below which the quantum strategy cannot develop any advantages. In the simplest case in which each party receives a single input bit, the achievable distance in standard optical fiber is limited to only $10$ kilometers, rendering the protocol impractical for latency-sensitive applications. However, if the VBG is used for distribution, the distance can be scaled to more than $30000$ km, allowing HKEX and NYSE traders to carry out a quantum-enhanced coordinated trading strategy with a \yesun{potential operational advantage under the idealized model}.

\section{Limitations and Future Work}
The results presented in this work are primarily based on first-principles analysis. While as much real measurement data is incorporated as possible and assumptions are adapted conservatively, these results are still subject to limitations, which we discuss in this section.

\textbf{Construction Route.} One of the major challenges in deploying such a large scale quantum infrastructure ($10^4$ km) is finding a suitable place to construct it. Regardless of the civil and legal issues, to maintain the advanced phase stability of the VBG, it must be located and routed across a relatively quiet rural area when possible, rather than in noisy urban centers. However, this may not always be possible, and it might be inevitable to have sections going across areas with active geological or human activities (e.g., rail lines). In that case, the on-site seismic and acoustic input measurement must be performed to assess real translational and rotational motion input, so that the engineers can decide whether to make a detour or to deploy a more expensive seismic isolation platform based on budget concerns. In addition, as mentioned in Sec.~\ref{subsec:reflectionDevice}, we will need steering mirrors to deflect the direction of the channel to avoid man-made structures. We believe the current assumption of placing one mirror every three sections is conservative, and we anticipate geographic tomography being performed prior to the construction for an optimized route plan, similar to the approach utilized by the 40-km LIGO project~\cite{effler2023roadmap, datrier2025site}.

\textbf{Ultra-low Frequency Drift.} In this work, the noise spectrum under consideration extends down to $f_l\sim0.01$ Hz, which corresponds to an observation time of $100$ s that is assumed to be sufficient for most applications. As we can see from Fig. \ref{fig:Phase_PSD}(b) in the main text, the RMS noise appears to saturate when the lower cut-off frequency is below $1$ Hz. The lower frequency noise is dominated by low-frequency drifts due to geophysical and geodetic movement \cite{webb1998broadband,nishida2017ambient}, as well as flicker noise in the SLS sensing devices. The integrated contributions of these nonstationary noise sources diverge as $\sim\ln(1/f_l)$ with $f_l$ approaching zero \cite{allan2012should}. To remove these near-DC noises, either calibration against a reference or differential measurement schemes that reject the common-mode noise \cite{lenoir2013predicting} are required. Since the performance PSD is evaluated down to $0.01\mathrm{Hz}$ in this work, such mitigation procedures are expected to be conducted on a timescale around $100$ s, which is feasible given that the maximum signal delay allowed in the VBG is no more than $1$ s. These processes can be integrated into the ACS system, which operates on the same timescale and already partially suppresses these effects. Further decreasing the lower cutoff frequency is also possible and will not significantly change the rms noise due to the typical logarithmic scaling. The detailed analysis and budgets for these ultra-low frequency noises are beyond the scope of this work, and we look forward to future investigations for the confirmation of our cut-off frequency assumption. 

\textbf{Residual Gas Evacuation.}
In the analysis of the vacuum tube, a dry residual air condition is assumed, with the volume fractions of the constituent gases to be approximately, 
\begin{align}
    &\mathrm{N_2:O_2:Ar:H_2O:CO_2:He:H_2}\nonumber\\
    &\approx\mathrm{0.78:0.21:0.009:0.001:0.0004:10^{-5}:10^{-6}}.
\end{align}

Importantly, water molecules exhibit several strong absorption lines in the telecom band. Therefore, maintaining a dry air condition is crucial for the fundamental performance of the VBG. In LIGO, water molecules form strong hydrogen bonds with the metal oxide layers on the inner surfaces of the stainless steel tubes and thus require substantial effort to remove beyond simple pumping \cite{ligo_vacuum}. The solution adopted by LIGO involves baking the vacuum tubes for weeks prior to operation (a one-time procedure) and maintaining cryopumping with liquid nitrogen. However, for our 1 Pascal medium vacuum requirement with a $0.1\%$ water volume ratio, the desired conditions are expected to be readily achievable using inexpensive dynamic purge techniques with dry nitrogen backfill. The residual gas pressure may be further reduced by an order of magnitude whenever such an upgrade is economical. Another concern is that light gasses, like hydrogen, are the hardest for mechanical pumps to compress and evacuate, and the stainless steel walls are constantly releasing hydrogen. While the atmospheric concentration of $\mathrm{H_2}$ is only 0.5 ppm, the major residual gas component in the ultra-high vacuum environment of LIGO is hydrogen ($99\%$) due to the bulk diffusion mechanism \cite{livas1989ligo}. Since the compression ratio ($K_0$) for hydrogen in a lobe pump is roughly $15–20\%$ smaller than that for air \cite{o2003user}, we adopt a conservative hydrogen volume fraction of $1$ ppm. At a total pressure of $1$ Pa, the contribution from bulk diffusion is expected to remain negligible. Nevertheless, a comprehensive study of the residual gas will be required in future work to confirm the assumed mixture and its impact on system performance. 

\textbf{Coating Design.}
The reflection and absorption losses in the AR coating are based on the simulation with optimized 8-layer $\mathrm{Ta_2O_5/SiO_2}$ stacks designed for the telecom band. With the introduction of the steering mirror, we assume the HR coating has a similar performance to the AR one. This assumption is conservative, given that the HR coating transmission loss is less than 5 ppm in LIGO \cite{billingsley2017characterization}, and the dominant contribution to the current steering mirror losses is from angular misalignment rather than coating absorption. For polarization encoding, the crucial point is to ensure that the HR coating maintains its attenuation performance for both polarization modes at the designed incident angle (e.g. $\ang{45}$), as mentioned in Sec.~\ref{sec:polarizationNoise}. Although achieving such a loss requirement for both polarizations is not difficult and has been readily demonstrated with $\mathrm{Ta_2O_5/SiO_2}$ layers \cite{Kraus2013Improving,cui2017simultaneous}, this constraint will complicate the coating design, increase thermal noise, and inflate budgets. In addition, the birefringence phase shift will also need to be compensated by carefully arranging the optic axes of the lenses. For these reasons, time-bin encoding may provide a simpler and more robust alternative. Furthermore, the assumption we made about the material properties for HR/AR coatings when estimating the Brownian noise and thermal-optics noise in Sec.~\ref{subsec:thermalNoise} may not be true if the material properties change significantly during manufacturing, especially after the doping and annealing process \cite{granata2020amorphous}. Therefore, a dedicated experimental characterization of coating losses and noise will be required in future work to validate these assumptions. 

\textbf{Control System Robustness.}
Although a prototype design of the control system is provided, its performance is analyzed only theoretically with reasonable redundancy, incorporating all noise sources of which we are aware. As a result, this assessment should only be taken as an order-of-magnitude demonstration until experimental measurements are performed to confirm the models. The proposed design primarily serves as a proof-of-principle guideline that can be further modified and improved through careful engineering if the model assumptions fail, designs lack a robust duty cycle, or restrictions become unnecessarily limiting. Alternative designs are also suggested at the end of the corresponding subsections. In addition to the rotational RMS noise and the ultra-low drifting issues mentioned earlier, it is noted that the AUX beam jitter~\cite{canuel2014sub} can raise concerns if it is too intense and rapid, drowning out the actual motion signal of the mirrors or lenses, which could cause failure in the ACS system. Encouragingly, reinforcement learning can be integrated into the control system to enhance performance and avoid overshooting due to a suboptimal control strategy~\cite{buchli2025improving}.

\textbf{Protocol Performance.}
The performance of the applications estimated in this work is mainly based on the limitations induced by the optical metrics imposed by the VBG channel, while neglecting other external factors such as the quality of quantum sources, photon detector efficiency, and quantum control ability. Although this approach directly probes the scalability of the protocols and unravels the intrinsic improvements and advantages of the VBG channel, it does not represent the performance that is immediately realizable in current experiments. In practice, these \textit{one-time} losses and noise sources may constitute the dominant constraints on feasibility, particularly in regimes where the VBG channel considerably reduces the quantum communication cost. Nevertheless, we believe that the VBG is an indispensable building block for the protocols mentioned in Fig.~\ref{fig:applicationSummary}, and the analysis presented here reflects the asymptotic performance of these protocols in the near future.

A more detailed analysis of these issues is beyond the scope of this work. However, we look forward to future investigations, especially full simulations and small-scale experimental demonstrations.

% END EXACT SI BODY
\end{document}